\documentclass[aps,floatfix,amsmath,nofootinbib,amssymb,superscriptaddress]{revtex4}

\usepackage{overpic}
\usepackage{amssymb}
\usepackage{indentfirst}
\usepackage{feynmf}   
\usepackage{slashed}  
\usepackage{cases}
\usepackage{color}
\usepackage{multirow}
\usepackage{epstopdf}
\usepackage{graphicx,color,bm}
\usepackage{epstopdf}
\usepackage{amsmath} 

\usepackage[colorlinks,
            citecolor=green,
            anchorcolor=red,
            menucolor=red,
            linkcolor=red,
            filecolor=red,
            runcolor=red,
            urlcolor=blue,
            frenchlinks=red]{hyperref}

\begin{document}

\title{Twist-3 gluon GPDs for the nucleon in a spectator model}

\author{Chentao Tan}\email[]{tanchentao@seu.edu.cn}
\author{Zhun Lu}
\email[]{zhunlu@seu.edu.cn}
\affiliation{School of Physics, Southeast University, Nanjing 211189, China}

\begin{abstract}
	
We investigate the complete set of twist-3 gluon generalized parton distributions (GPDs) of the nucleon at nonzero skewness within a spectator model, in which the nucleon is treated as a two-body system consisting of an active spin-1 gluon and a spin-1/2 spectator particle. The spectator mass is allowed to vary continuously and is described by a spectral function, while the nucleon-gluon-spectator coupling is modeled by an effective vertex containing two form factors. We calculate the twist-3 gluon GPDs over a broad kinematic range and investigate the corresponding impact-parameter-dependent parton distributions, defined as two-dimensional Fourier transforms of the GPDs at zero skewness. In the forward limit, we further obtain the corresponding twist-3 parton distribution functions (PDFs). Our results provide model predictions for twist-3 gluon GPDs and PDFs that can be useful for future phenomenological studies and for investigating observables at current and future collider facilities.

\end{abstract}

\maketitle

\section{Introduction}

Understanding the internal structure of hadrons in terms of their quark and gluon degrees of freedom is one of the central goals of hadronic physics. For decades, parton distribution functions (PDFs) have played a fundamental role in the study of hadron structure by encoding the distributions of longitudinal momentum and polarization carried by partons in fast-moving hadrons~\cite{Collins:1981uw,Martin:1998sq,Gluck:1994uf,Gluck:1998xa}. A more complete description of hadron structure is provided by generalized parton distributions (GPDs)~\cite{Muller:1994ses,Diehl:2003ny,Belitsky:2005qn,Ji:1996ek,Radyushkin:1996nd,Goeke:2001tz,Boffi:2007yc,
Radyushkin:1996ru,Ji:1996nm,Radyushkin:1997ki,Ji:2004gf,Belitsky:2001ns}, which extend ordinary PDFs from the forward to the off-forward kinematic regime.
At zero skewness, GPDs can be Fourier transformed with respect to the transverse momentum transfer to obtain impact-parameter-dependent parton distributions (IPDs), providing information on the transverse spatial distribution of partons~\cite{Burkardt:2000za,Burkardt:2002hr,Burkardt:2002ks,Diehl:2002he}. GPDs can be accessed experimentally through hard exclusive reactions, such as deeply virtual Compton scattering (DVCS)~\cite{Ji:1996nm,Collins:1998be,Ji:1998xh,Blumlein:1999sc,Guidal:2013rya} and hard exclusive meson production~\cite{Goeke:2001tz,Ji:1998pc,Collins:1996fb,Goloskokov:2006hr,Goloskokov:2011rd}. Extensive measurements have been performed by the H1~\cite{H1:2001nez,H1:2005gdw,H1:2007vrx,H1:2009wnw}, ZEUS~\cite{ZEUS:2003pwh,ZEUS:2008hcd,ZEUS:2009uxs}, and HERMES~\cite{HERMES:2012gbh,HERMES:2012idp} experiments at HERA, by COMPASS~\cite{Kumericki:2016ehc}, and by CLAS~\cite{CLAS:2001wjj,CLAS:2007clm,CLAS:2008ahu,Niccolai:2012sq,CLAS:2015bqi,CLAS:2015uuo,CLAS:2018bgk,CLAS:2018ddh,CLAS:2021gwi} and Hall A~\cite{JeffersonLabHallA:2006prd,JeffersonLabHallA:2007jdm,JeffersonLabHallA:2012zwt,Georges:2017xjy,
JeffersonLabHallA:2022pnx} at Jefferson Lab (JLab). These measurements have provided important constraints on quark GPDs. 
More precise data are expected from the JLab 12 GeV program and from future Electron-Ion Collider facilities in the United States (EIC)~\cite{AbdulKhalek:2021gbh,Accardi:2012qut} and China (EicC)~\cite{Anderle:2021wcy}, which will further improve our understanding of hadron structure.

Since the matrix elements of the operators defining GPDs are dominated by leading-twist operators in the Bjorken limit~\cite{Bjorken:1968dy}, most theoretical studies have focused on leading-twist GPDs~\cite{Boffi:2007yc,Kumericki:2009uq,Bhattacharya:2018zxi,Bhattacharya:2019cme,Pasquini:2005dk,
Pasquini:2006dv,Frederico:2009fk,Burkardt:2015qoa,Pasquini:2019evu,Meissner:2007rx,Tan:2023kbl}. 
At the energy scales relevant to present experiments, however, higher-twist effects may also play an important role, particularly those associated with twist-3 GPDs~\cite{Guo:2022cgq,Anikin:2009hk,Freund:2003qs,Kivel:2003jt,Kiptily:2002nx,Radyushkin:2001fc,Belitsky:2001yp,
Kivel:2000fg,Belitsky:2000vk,Radyushkin:2000ap,Kivel:2000cn,Radyushkin:2000jy,Kivel:2000rb,Belitsky:2000vx}. 
In the quark sector, twist-3 GPDs have been investigated using the quark target model~\cite{Aslan:2018tff,Aslan:2018zzk}, spectator diquark models~\cite{Aslan:2018tff,Tan:2024doz}, basis light-front quantization~\cite{Zhang:2023xfe}, and lattice QCD~\cite{Dodson:2021rdq,Bhattacharya:2023nmv}.
By contrast, our knowledge of twist-3 gluon GPDs in the nucleon remains limited, partly because of the more intricate operator structure involving the gluon field-strength tensor and its derivatives~\cite{Thakuria:2025ncs}. Beyond their appearance in operator identities~\cite{Kiptily:2002nx}, corrections to DVCS amplitudes~\cite{Belitsky:2000vx}, and Wandzura--Wilczek-type relations~\cite{Hatta:2012jm}, only the chiral-even twist-3 gluon GPDs at zero skewness have been calculated in a light-front model~\cite{Thakuria:2025ncs}.
A systematic model calculation of the complete set of twist-3 gluon GPDs over a wider kinematic range, including nonzero skewness, is therefore still lacking.

Experimentally, the extraction of GPDs remains challenging~\cite{Bertone:2021yyz,Moffat:2023svr}, particularly for higher-twist distributions. Near-threshold exclusive photoproduction of heavy vector mesons, such as $J/\psi$ and $\Upsilon$, has emerged as a promising channel for probing leading-twist gluon GPDs in the nucleon~\cite{Guo:2023pqw,Guo:2021ibg,Guo:2023qgu,Tan:2026qba}, and may also provide sensitivity to higher-twist contributions. Isolating the contribution of an individual twist-3 gluon GPD, however, remains experimentally difficult. Nevertheless, twist-3 GPDs are of interest for several reasons. First, they provide higher-twist corrections to leading-twist amplitudes in hard exclusive reactions and may therefore be relevant for a more reliable extraction of leading-twist GPDs. Second, model-independent relations connect certain $x$ moments of twist-3 GPDs to partonic kinetic quantities, including orbital angular momentum and spin-orbit correlations~\cite{Zhang:2023xfe,Penttinen:2000dg,Lorce:2014mxa,Bhoonah:2017olu,Thakuria:2025ncs,Tan:2024doz}, thereby providing additional information on the spin structure of the nucleon~\cite{Leader:2013jra,Hatta:2012jm,Deur:2018roz,Aidala:2012mv}. Finally, relations between twist-3 GPDs and generalized transverse-momentum-dependent parton distributions (GTMDs), IPDs, and PDFs~\cite{Meissner:2009ww,Meissner:2008ay,Lorce:2013pza} establish connections among different descriptions of the multidimensional partonic structure. These relations provide access to information on the momentum, angular-momentum, and spatial distributions of partons~\cite{Leader:2013jra,Burkardt:2000za,Burkardt:2002hr,Burkardt:2002ks,Diehl:2002he,Ralston:2001xs,Burkardt:2008ps,Aslan:2019jis}.

In this work, we calculate the twist-3 gluon GPDs at nonzero skewness, together with the corresponding IPDs and PDFs, using a spectator model~\cite{Lu:2016vqu,Bacchetta:2020vty,Bacchetta:2024fci}. In this framework, the nucleon is treated as a two-body system consisting of an active gluon and a spectator particle carrying the quantum numbers of the remaining degrees of freedom. The spectator mass is allowed to vary continuously according to a spectral function, while the nucleon-gluon-spectator vertex is modeled using two dipolar form factors~\cite{Bacchetta:2008af}. The present study complements existing phenomenological analyses and light-front model calculations by providing model predictions for the complete set of twist-3 gluon GPDs at nonzero skewness.

The remainder of this paper is organized as follows. In Sec.~\ref{Sec:2}, we introduce the complete set of twist-3 gluon GPDs in the nucleon. In Sec.~\ref{Sec:3}, we calculate these distributions analytically at nonzero skewness within the spectator model. In Sec.~\ref{Sec:4}, we present numerical results for the GPDs over a broad kinematic range and discuss the corresponding IPDs and PDFs. Finally, Sec.~\ref{Sec:5} summarizes our main findings.

\section{Definitions of twist-3 gluon GPDs}\label{Sec:2}

\begin{figure}
	\centering
	\includegraphics[width=0.48\columnwidth]{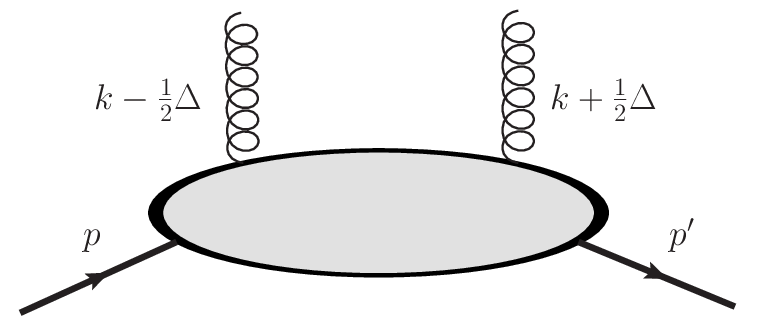}
	\caption{Kinematics for gluon GPDs.}
	\label{fig:gluon}
\end{figure}

GPDs encode the dependence of partonic correlations on the average longitudinal momentum fraction $x=k^+/P^+$ of the active parton, the skewness variable $\xi=-\Delta^+/2P^+$, and the momentum transfer squared $t=\Delta^2$. 
We represent a generic four-momentum $a^\mu$ by its light-front components
$[a^+,a^-,\bm{a}_\perp]=\left[\frac{a^0+a^3}{\sqrt{2}},\frac{a^0-a^3}{\sqrt{2}},a^1,a^2\right]$.
Together, these variables provide multidimensional information on the internal structure of hadrons.
Fig.~\ref{fig:gluon} illustrates the kinematics for gluon GPDs. For spin-$1/2$ hadrons, a complete set of twist-3 gluon GPDs was given in Ref.~\cite{Lorce:2013pza}. At fixed light-front time $z^+=0^+$, the corresponding correlator is defined as the off-forward matrix element of a bilocal gluon field-strength operator connected by two straight Wilson lines:
\begin{align}
	F^{\mu\nu;\rho\sigma}_{\lambda^\prime\lambda}(P,x,\Delta)=\frac{1}{xP^+}\int \frac{dz^-}{2\pi} e^{ik\cdot z} \langle{p^\prime; \lambda^\prime}|2\text{Tr}\left[ F^{\mu\nu}\left(-\frac{z}{2}\right) \mathcal{W} F^{\rho\sigma}\left(\frac{z}{2}\right)\mathcal{W}^\prime\right]|p; \lambda \rangle \big|_{z^+=0^+,\bm{z}_\perp=\bm{0}_\perp},
	\label{eq:correlator}
\end{align}
where the one-particle states are covariantly normalized according to $\langle p^\prime|p\rangle=2p^0(2\pi)^3\delta^{(3)}(\bm{p}^\prime-\bm{p})$. Here, $p$ ($p^\prime$) and $\lambda$ ($\lambda^\prime$) denote the momentum and helicity of the initial (final) nucleon, respectively, $P=(p+p^\prime)/2$ is the average nucleon four-momentum, $\Delta=p^\prime-p$ is the four-momentum transfer to the nucleon, and $k$ denotes the average gluon four-momentum. The gluon field-strength tensor $F_a^{\mu\nu}$ is related to the gluon field $A_a^\mu$ through
$$F^{\mu\nu}_a(x)=\partial^\mu A^\nu_a(x)-\partial^\nu A_a^\mu(x)+gf_{abc}A_b^\mu(x)A_c^\nu(x).$$ 

The correlator in Eq.~(\ref{eq:correlator}) contains two Wilson lines, $\mathcal{W}\equiv\mathcal{W}(-\frac{z}{2};\frac{z}{2})$ and $\mathcal{W}^{\prime}\equiv\mathcal{W}(\frac{z}{2};-\frac{z}{2})$, which ensure color gauge invariance. The Wilson line $\mathcal{W}$ connects the two space-time points $(0^+,-\frac{z^-}{2},\bm{0}_\perp)$ and $(0^+,\frac{z^-}{2},\bm{0}_\perp)$ by a straight path and is given by
\begin{align}
	\mathcal{W}_{ab} \left(-\frac{z}{2};\frac{z}{2} \right) \bigg|_{z^+=0^+,\bm{z}_\perp=\bm{0}_\perp}=\left[0^+,-\frac{z^-}{2},\bm{0}_\perp;0^+,\frac{z^-}{2},\bm{0}_\perp\right]_{ab}=\mathcal{P}\exp\left[-g\int^{\frac{z^-}{2}}_{-\frac{z^-}{2}} dy^- f_{abc} A^+_c(0^+,y^-,\bm{0}_\perp)\right],
\end{align}
and $\mathcal{W}^\prime$ is defined analogously. Here, $\mathcal{P}$ denotes path ordering. In the following, we work in the light-front gauge $A^+=0$, in which the Wilson lines reduce to unity. We choose the symmetric frame in which the average nucleon momentum $P$ has no transverse component~\cite{Bacchetta:2008af}. Thus, the momenta of the initial and final nucleons can be written as
\begin{align} 
	p&=\left((1+\xi)P^+,\frac{M^2+\frac{\bm{\Delta}_\perp^2}{4}}{(1+\xi)P^+},
	{-\frac{\bm{\Delta}_\perp}{2}}\right),\\ p^\prime&=\left((1-\xi)P^+,\frac{M^2+\frac{\bm{\Delta}_\perp^2}{4}}
	{(1-\xi)P^+},{\frac{\bm{\Delta}_\perp}{2}}\right),
\end{align}
where $M$ is the nucleon mass and $p^2=p^{\prime 2}=M^2$. 

At twist-3, there are sixteen gluon GPDs for the nucleon in total, which appear in the parametrization of the correlator in Eq.~(\ref{eq:correlator}) according to~\cite{Lorce:2013pza}
\begin{align}
	F^{\{+-;+R\}}_{\lambda^\prime\lambda}=& \frac{M}{2(P^+)^2}\bar{u}(p^\prime,\lambda^\prime)
	\bigg[i\sigma^{+R}H^{g}_{2T}(x,\xi,t)+\frac{\gamma^+\Delta_R-\Delta^+\gamma_R}{2M}E_{2T}^{g}(x,\xi,t)\nonumber\\	&+\frac{P^+\Delta_R-\Delta^+P_R}{M^2}\tilde{H}^{g}_{2T}(x,\xi,t)+\frac{\gamma^+P_R-P^+\gamma_R}{M}\tilde{E}_{2T}^{g}(x,\xi,t)\bigg]u(p,\lambda),\label{eq:p1}\\
	F^{[+-;+R]}_{\lambda^\prime\lambda}=& \frac{M}{2(P^+)^2}\bar{u}(p^\prime,\lambda^\prime) \bigg[i\sigma^{+R}\bar{H}^{g}_{2T}(x,\xi,t)+\frac{\gamma^+\Delta_R-\Delta^+\gamma_R}{2M}\bar{E}_{2T}^{g}(x,\xi,t)\nonumber\\ &+\frac{P^+\Delta_R-\Delta^+P_R}{M^2}\tilde{\bar{H}}^{g}_{2T}(x,\xi,t)+\frac{\gamma^+P_R-P^+\gamma_R}{M}\tilde{\bar{E}}_{2T}^{g}(x,\xi,t)\bigg]u(p,\lambda),\label{eq:p2}\\
	\frac{1}{2}F^{[LR;+R]}_{\lambda^\prime\lambda}=& \frac{M}{2(P^+)^2}\bar{u}(p^\prime,\lambda^\prime)
	\bigg[i\sigma^{+R}H^{\prime g}_{2T}(x,\xi,t)+\frac{\gamma^+\Delta_R-\Delta^+\gamma_R}{2M}E_{2T}^{\prime g}(x,\xi,t)\nonumber\\
	&+\frac{P^+\Delta_R-\Delta^+P_R}{M^2}\tilde{H}^{\prime g}_{2T}(x,\xi,t)+\frac{\gamma^+P_R-P^+\gamma_R}{M}\tilde{E}_{2T}^{\prime g}(x,\xi,t)\bigg]u(p,\lambda),\label{eq:p3}\\
	\frac{1}{2}F^{\{LR;+R\}}_{\lambda^\prime\lambda}=& \frac{M}{(P^+)^2}\bar{u}(p^\prime,\lambda^\prime)
	\bigg[i\sigma^{+R}\bar{H}^{\prime g}_{2T}(x,\xi,t)+\frac{\gamma^+\Delta_R-\Delta^+\gamma_R}{2M}\bar{E}_{2T}^{\prime g}(x,\xi,t)\nonumber\\
	&+\frac{P^+\Delta_R-\Delta^+P_R}{M^2}\tilde{\bar{H}}^{\prime g}_{2T}(x,\xi,t)+\frac{\gamma^+P_R-P^+\gamma_R}{M}\tilde{\bar{E}}_{2T}^{\prime g}(x,\xi,t)\bigg]u(p,\lambda)\label{eq:p4},
\end{align}	
where
\begin{align}
	F^{\{\mu\nu;\rho\sigma\}}&=\frac{1}{2}(F^{\mu\nu;\rho\sigma}+F^{\rho\sigma;\mu\nu}),\\
	F^{[\mu\nu;\rho\sigma]}&=\frac{1}{2}(F^{\mu\nu;\rho\sigma}-F^{\rho\sigma;\mu\nu}),
\end{align}
$\sigma^{\mu\nu}=\frac{i}{2}[\gamma^\mu,\gamma^\nu]$, and the transverse polar combinations are defined as $a_{R/L}=a^1\pm ia^2$.  
A different parametrization, in which the gluon operators are separated according to their polarization properties, was introduced in Ref.~\cite{Thakuria:2025ncs}.

Since GPDs at negative $\xi$ are difficult to access directly in known processes, we restrict our discussion to the region $0\leq\xi\leq1$. 
The support region for $x$ in GPDs is $-1 \leq x \leq 1$.
In the region $\xi < x \leq 1$, a gluon is emitted and reabsorbed with longitudinal momentum fractions $x+\xi$ and $x-\xi$, respectively. 
In the region $-\xi\leq x\leq\xi$, the correlator describes the emission of a gluon and an antigluon with momentum fractions $x+\xi$ and $\xi-x$, respectively. 
Finally, in the region $-1\leq x<-\xi$, an antigluon is emitted and reabsorbed with momentum fractions $\xi-x$ and $-\xi-x$, respectively. 
The first and third regions are commonly referred to as the Dokshitzer-Gribov-Lipatov-Altarelli-Parisi (DGLAP) regions~\cite{Gribov:1972ri,Lipatov:1974qm,Borah:2012ey}, while the second is the Efremov-Radyushkin-Brodsky-Lepage (ERBL) region~\cite{Efremov:1978rn,Lepage:1980fj}. Since gluons are their own antiparticles, the corresponding distributions satisfy Bose-symmetry relations~\cite{Diehl:2003ny,Ji:1998pc}. We therefore restrict the numerical analysis to $0\leq x\leq1$, where the twist-3 gluon GPDs are evaluated separately in the two regions $0\leq x\leq\xi$ and $\xi<x\leq1$.

\section{Gluon GPDs in the Spectator Model}\label{Sec:3}

In this section, we present an analytical calculation of the twist-3 gluon GPDs in Eqs.~(\ref{eq:p1})--(\ref{eq:p4}) at $\xi\neq0$ within the spectator model. This model has previously been used to calculate leading-twist gluon transverse-momentum-dependent distributions (TMDs)~\cite{Lu:2016vqu,Bacchetta:2020vty,Bacchetta:2024fci}. A related version without a spectral function for the spectator mass has also been employed to calculate leading-twist gluon GPDs~\cite{Tan:2026qba}. For later convenience, we first introduce the polarization state of the nucleon in a generic direction~\cite{Diehl:2005jf}:
\begin{align} 	
|p;S\rangle=\cos(\theta/2)|p;+\rangle+\sin(\theta/2)e^{i\phi}|p;-\rangle.	
\end{align}
An analogous superposition applies to the final nucleon state $\langle p^\prime;S|$. 
In the nucleon rest frame, the state $|p;S\rangle$ describes a particle with the three-dimensional spin vector $\bm{S}=(S_\perp^1,S_\perp^2,\lambda)=(\sin\theta \cos\phi,\sin \theta \sin\phi,\cos \theta)$. 
Replacing the helicity states $\langle p^\prime;\lambda^\prime|$ and $|p;\lambda\rangle$ in Eq.~(\ref{eq:correlator}) by $\langle p^\prime;S|$ and $|p;S\rangle$, respectively, gives the relation between the correlator $F(x,\Delta;S)$ for an arbitrary spin orientation and the correlators $F(x,\Delta;\lambda,\lambda^\prime)$ for the possible helicity combinations~\cite{Meissner:2007rx}:
\begin{align}	
	F(x,\Delta;S)=&\frac{1}{2}[F(x,\Delta;+,+)+F(x,\Delta;-,-)]
	+\frac{1}{2}\lambda[F(x,\Delta;+,+)-F(x,\Delta;-,-)]\nonumber\\ &+\frac{1}{2}S_\perp^1[F(x,\Delta;-,+)+F(x,\Delta;+,-)]
	+\frac{i}{2}S_\perp^2[F(x,\Delta;-,+)-F(x,\Delta;+,-)].
\end{align}
Following the convention in Ref.~\cite{Lepage:1980fj}, the explicit expressions of the light-front helicity spinors for the nucleon are given by
\begin{align}
	u(p,+)=\frac{1}{\sqrt{2^{3/2}p^+}}\begin{pmatrix}
		\sqrt{2}p^++M\\
		p_\perp^1+ip_\perp^2\\
		\sqrt{2}p^+-M\\
		p_\perp^1+ip_\perp^2
	\end{pmatrix},~~~~
	u(p,-)=\frac{1}{\sqrt{2^{3/2}p^+}}\begin{pmatrix}
		-p_\perp^1+ip_\perp^2\\
		\sqrt{2}p^++M\\
		p_\perp^1-ip_\perp^2\\
		-\sqrt{2}p^++M
	\end{pmatrix}.
\end{align}

\begin{figure}
	\centering
	\includegraphics[width=0.45\columnwidth]{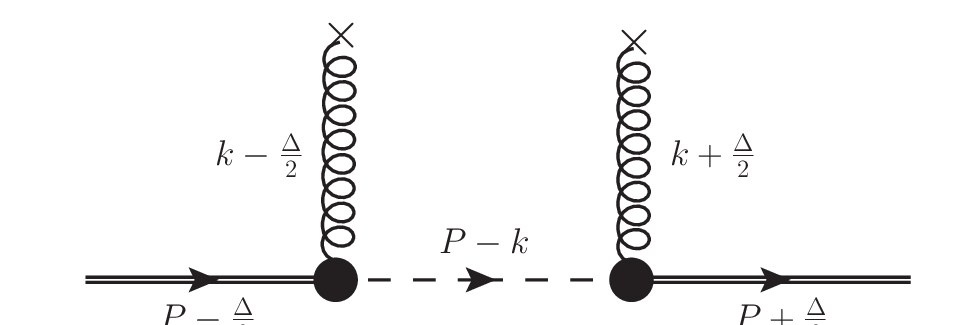}
	\caption{Kinematics for gluon GPDs in the spectator model. The gluon lines with crosses correspond to the specific Feynman rules for the gluon field strength tensor.
	}
	\label{fig:GPD}
\end{figure}

Fig.~\ref{fig:GPD} shows the leading-order Feynman diagram used to calculate the gluon GPDs. Within the spectator approximation, the nucleon state $|p;S\rangle$ is assumed to split into an active spin-1 gluon with momentum $k-\frac{\Delta}{2}$ and a spin-1/2 spectator particle with momentum $P-k$ and mass $M_X$. The tree-level scattering amplitude $\mathcal{M}_a^{\mu\nu}(S)$ is given by~\cite{Bacchetta:2020vty}
\begin{align}
	\mathcal{M}_a^{\mu\nu}(S)&=\langle P-k|F_a^{\mu\nu}|p;S\rangle\nonumber\\ &=\bar{u}_c(P-k)\frac{G_{ab}^{\mu\nu\alpha}(k-\frac{\Delta}{2})}{(k-\frac{\Delta}{2})^2}\mathcal{Y}^-_{\alpha,bc}u(p,S),
\end{align}	
where $\bar{u}_c(P-k)$ is the spectator spinor with the color index $c$, and
\begin{align} G_{ab}^{\mu\nu\alpha}\left(k-\frac{\Delta}{2}\right)=-i\left[\left(k-\frac{\Delta}{2}\right)^\mu g^{\nu\alpha}-\left(k-\frac{\Delta}{2}\right)^\nu g^{\mu\alpha}\right]\delta_{ab}
\end{align}	
denotes the Feynman rule associated with the field-strength tensor in the definition of the correlator~\cite{Goeke:2006ef,Collins:2011zzd}. In general, the nucleon-gluon-spectator vertex $\mathcal{Y}^-_{\alpha,bc}$ may contain several Dirac structures. Following Ref.~\cite{Bacchetta:2020vty}, we model it in analogy with the conserved electromagnetic current of a free nucleon as obtained from the Gordon decomposition:
\begin{align}
	\mathcal{Y}^-_{\alpha,bc}=\delta_{bc}\bigg[g_1^-\gamma_\alpha+g_2^-\frac{i}{2M}\sigma_{\alpha\beta}p^\beta\bigg].
	\label{eq:vertex}
\end{align}	
The model-dependent form factors $g_{1,2}^-$ play a role analogous to that of Dirac and Pauli form factors, although they should not be identified with electromagnetic form factors because the vertex in Eq.~(\ref{eq:vertex}) describes a nonperturbative color interaction.

To parameterize the nonperturbative interaction, we adopt the dipolar form factors~\cite{Bacchetta:2008af}:
\begin{align}
	g_{1,2}^\pm\equiv g_{1,2}\left(\bm{k}_\perp\pm\frac{\bm{\Delta}_\perp}{2}\right)=\kappa_{1,2} \frac{(k\pm\frac{\Delta}{2})^2}{|(k\pm\frac{\Delta}{2})^2-\Lambda_X^2|^2}
=-\kappa_{1,2}\frac{(1-x)[(\bm{k}_\perp\pm\frac{\bm{\Delta}_\perp}{2})^2+L_X^2(0)]}
{[(\bm{k}_\perp\pm\frac{\bm{\Delta}_\perp}{2})^2+L_X^2(\Lambda_X^2)]^2},
\end{align}
where the superscript $+$ ($-$) denotes the outgoing (incoming) gluon vertex, $\kappa_{1,2}$ are normalization parameters controlling the strength of the vertex, $\Lambda_X$ is a cutoff parameter, and the gluon mass is set to $M_g=0$ GeV. Here,
\begin{align}
	L_X^2(0)&=xM_X^2-x(1-x)M^2,\nonumber\\
	L_X^2(\Lambda_X^2)&=xM_X^2+(1-x)\Lambda_X^2-x(1-x)M^2.
\end{align}
With the above assumptions, the spectator is treated as an on-shell spin-1/2 particle. The on-shell condition $(P-k)^2=M_X^2$ then implies that the active gluon is off shell:
\begin{align} \left(k\pm\frac{\Delta}{2}\right)^2&\equiv\tau\left(x,\left(\bm{k}_\perp\pm\frac{\bm{\Delta}_\perp}{2}\right)^2\right)
=-\frac{(\bm{k}_\perp\pm\frac{\bm{\Delta}_\perp}{2})^2+L_X^2(0)}{1-x}.
\end{align} 
The introduction of dipolar form factors cancels the singular behavior associated with the gluon propagators and renders the transverse-momentum integrals finite.

According to Fig.~\ref{fig:GPD}, the correlator in Eq.~(\ref{eq:correlator}) at tree level in the spectator model reads
\begin{align}
	F^{\mu\nu;\rho\sigma}_{\lambda^\prime\lambda}(P,x,\Delta)=\int \frac{dk^-d^2\bm{k}_\perp}{(2\pi)^4}\frac{\bar{u}(p^\prime,\lambda^\prime)G^{\rho\sigma\beta\ast}_{ab^\prime}(k+\frac{\Delta}{2})\mathcal{Y}^{+\ast}_{\beta,b^\prime c^\prime}(\slashed{P}-\slashed{k}+M_X)_{cc^\prime}G^{\mu\nu\alpha}_{ab}(k-\frac{\Delta}{2})\mathcal{Y}^-_{\alpha,bc}u(p,\lambda)}{D_{\text{SM}}}
	\label{eq:F},
\end{align}
where
\begin{align} D_{\text{SM}}=\bigg[\left(k+\frac{\Delta}{2}\right)^2+i\epsilon\bigg]
	\bigg[\left(k-\frac{\Delta}{2}\right)^2+i\epsilon\bigg]
	\bigg[(P-k)^2-M_X^2+i\epsilon\bigg],
\end{align}
and all color and spinor indices are contracted.

In the calculation, two types of $k^-$ integrals arise:
\begin{align}
	I &=\int^{+\infty}_{-\infty}dk^-\frac{1}{D_{\text{GPD}}}=
	\frac{1}{C}\int^{+\infty}_{-\infty}dk^-\frac{1}{(k^--k_1^-)(k^--k_2^-)(k^--k_3^-)},
	\label{eq:I}\\
	I_k&=\int^{+\infty}_{-\infty}dk^-\frac{k^-}{D_{\text{GPD}}}=
	\frac{1}{C}\int^{+\infty}_{-\infty}dk^-\frac{k^-}{(k^--k_1^-)(k^--k_2^-)(k^--k_3^-)}	,
	\label{eq:Ik}
\end{align}
where
\begin{align}
	C=-8(x+\xi)(x-\xi)(1-x)(P^+)^3.
\end{align}
As discussed in Sec.~\ref{Sec:2}, the range of $x$ is restricted to $0\leq x\leq1$. Consequently, only the poles at $k_2^-$ and $k_3^-$ in Eqs.~(\ref{eq:I})--(\ref{eq:Ik}) contribute:
\begin{align}  
	    k_2^-&=\frac{\Delta^-}{2}+\frac{(\bm{k}_\perp
		-\frac{\bm{\Delta}_\perp}{2})^2-i\epsilon}{2(x+\xi)P^+},\\
	k_3^-&=P^--\frac{\bm{k}_\perp^2+M_X^2-i\epsilon}{2(1-x)P^+},
\end{align}
corresponding to the incoming gluon propagator $(k-\frac{\Delta}{2})$ and the spectator propagator $(P-k)$, respectively. Since the locations of the poles depend on $x$, the $k^-$ integration is evaluated by contour integration separately in the two kinematic regions. The corresponding propagators are replaced by
\begin{align}
	\frac{1}{\left(k-\frac{\Delta}{2}\right)^2+i\epsilon}&\rightarrow \frac{-2\pi i}{2(x+\xi)P^+}\delta\left(k^--k_2^-\right),\\
	\frac{1}{(P-k)^2-M_X^2+i\epsilon}&\rightarrow \frac{-2\pi i}{2(1-x)P^+}\delta\left(k^--k_3^-\right).
\end{align}
The resulting integrals are
\begin{align}
	I&= 
	\begin{cases}
		-\frac{2\pi i}{C}\frac{1}{(k_2^--k_1^-)(k_2^--k_3^-)}, &  0 \leq  x\leq \xi,\\
		\frac{2\pi i}{C}\frac{1}{(k_3^--k_1^-)(k_3^--k_2^-)}, &  \xi< x \leq 1, 
	\end{cases}\\
    I_k &= 
    \begin{cases}
	    -\frac{2\pi i}{C}\frac{k_2^-}{(k_2^--k_1^-)(k_2^--k_3^-)}, &  0 \leq  x\leq \xi,\\
    	\frac{2\pi i}{C}\frac{k_3^-}{(k_3^--k_1^-)(k_3^--k_2^-)}, &  \xi< x \leq 1, 
    \end{cases}
\end{align}
where the region $0\leq x\leq\xi$ corresponds to the contribution from the incoming-gluon pole, while the region $\xi<x\leq1$ corresponds to the contribution from the spectator pole.

After selecting the appropriate helicity combinations and performing the $k^-$ integrations in Eq.~(\ref{eq:F}), the twist-3 gluon GPDs can be written in the generic form
\begin{align}
	\hat{H}^g(x,\xi,t;M_X) &= 
	\begin{cases}
		\int d^2\bm{k}_\perp \frac{N_H^{0 \leq  x\leq \xi}}{D_\text{SM}^{0 \leq  x\leq \xi}}, &  0 \leq  x\leq \xi,\\
        \int d^2\bm{k}_\perp \frac{N_H^{\xi< x \leq 1}}{D_\text{SM}^{\xi< x \leq 1}}, &  \xi< x \leq 1.
	\end{cases}
	\label{eq:GPDH}
\end{align}
Because the explicit expressions for the numerators $N_H^{0 \leq x\leq\xi}$ and $N_H^{\xi<x\leq1}$ are lengthy, we do not display them here. The corresponding numerators and denominators at $\xi=0$ are collected in the Appendix.

\section{Numerical results and discussion}\label{Sec:4}

In this section, we present numerical results for the twist-3 gluon GPDs at $\xi\neq0$ obtained within the spectator model, and discuss the corresponding twist-3 IPDs and PDFs. The GPDs in Eq.~(\ref{eq:GPDH}) depend explicitly on the spectator mass $M_X$, which is not treated as a single-valued parameter as in Ref.~\cite{Bacchetta:2020vty}. Instead, $M_X$ is integrated over a continuous range according to the spectral function
\begin{align}
	\rho_X(M_X)=\mu^{2a}\left[\frac{A}{B+\mu^{2b}}+\frac{C}{\pi\sigma}e^{-\frac{(M_X-D)^2}{\sigma^2}}\right],
	\label{eq:rhox}
\end{align}
where $\mu^2=M_X^2-M^2$ and $\{A,B,a,b,C,D,\sigma\}$ are parameters. Then the actual model results for GPDs are obtained by weighting the analytic expressions in Eq.~(\ref{eq:GPDH}) with the spectral function $\rho_X$:
\begin{align}
	H^g(x,\xi,t)=\int^\infty_M dM_X\,\rho_X(M_X) \hat{H}^g(x,\xi,t;M_X).
\end{align}

The model contains seven parameters associated with the spectral function, $\{A,B,a,b,C,D,\sigma\}$, and three parameters associated with the dipolar form factors, $\{\kappa_1,\kappa_2,\Lambda_X\}$. As shown in Table~\ref{tab1}, we adopt the parameter values and uncertainties obtained in Ref.~\cite{Bacchetta:2020vty}, while fixing $B=2.1$ because the corresponding fit is relatively insensitive to this parameter.

\begin{table}[htbp]
	\centering
	\caption{Model parameters with uncertainties~\cite{Bacchetta:2020vty}}
	\label{tab1}
	\begin{tabular}{lcc} 
		\hline   
		Parameter & Central value & Uncertainty \\ 
		\hline 
		~~~~~$A$ & 6.1  & $\pm2.3$    \\
		~~~~~$a$ & 0.82 & $\pm0.21$   \\
		~~~~~$b$ & 1.43  & $\pm0.23$   \\
		~~~~~$C$ & 371  & $\pm58$    \\
		~~~~~$D$ [GeV] & 0.548 & $\pm0.081$    \\
		~~~~~$\sigma$ [GeV] & 0.52  & $\pm0.14$   \\
		~~~~~$\Lambda_X$ [GeV] & 0.472 & $\pm0.058$   \\
		~~~~~$\kappa_1$ [$\text{GeV}^2$] & 1.51  & $\pm0.16$   \\
		~~~~~$\kappa_2$ [$\text{GeV}^2$] & 0.414 & $\pm0.036$   \\
		\hline 
	\end{tabular}
\end{table} 

\subsection{twist-3 GPDs}

\begin{figure}
	\centering
	\includegraphics[width=0.32\columnwidth]{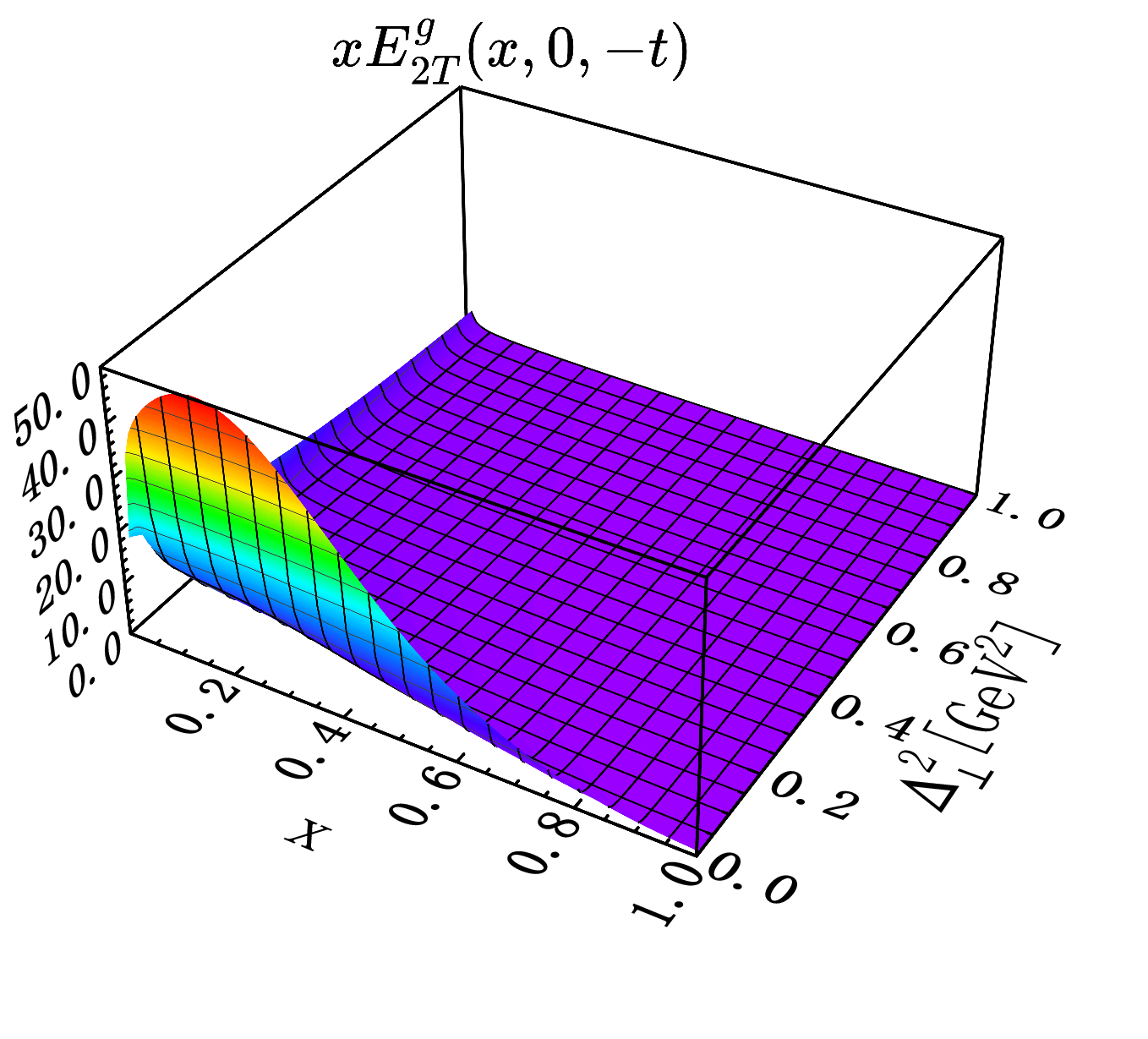}
	\includegraphics[width=0.32\columnwidth]{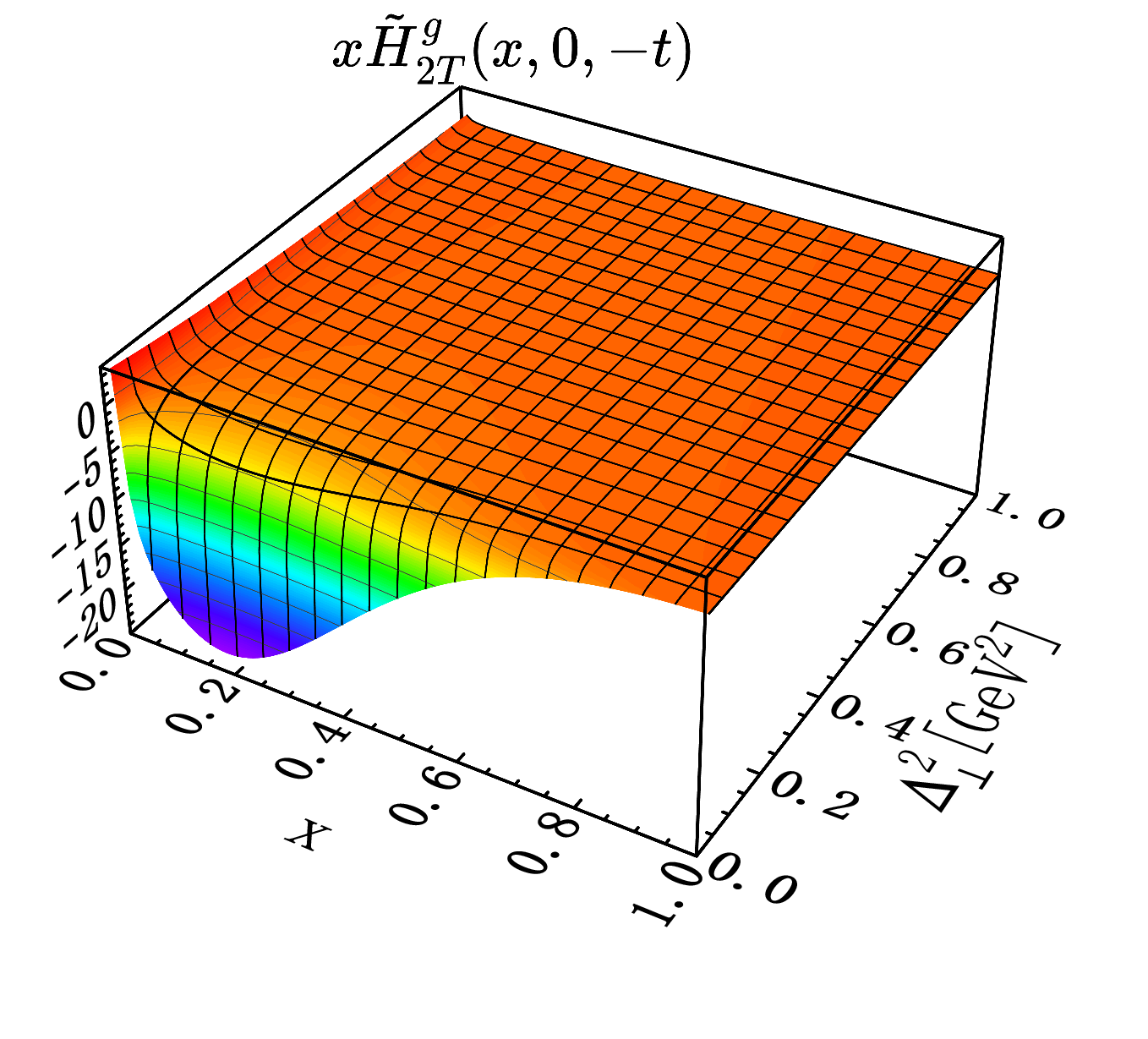}
	\includegraphics[width=0.32\columnwidth]{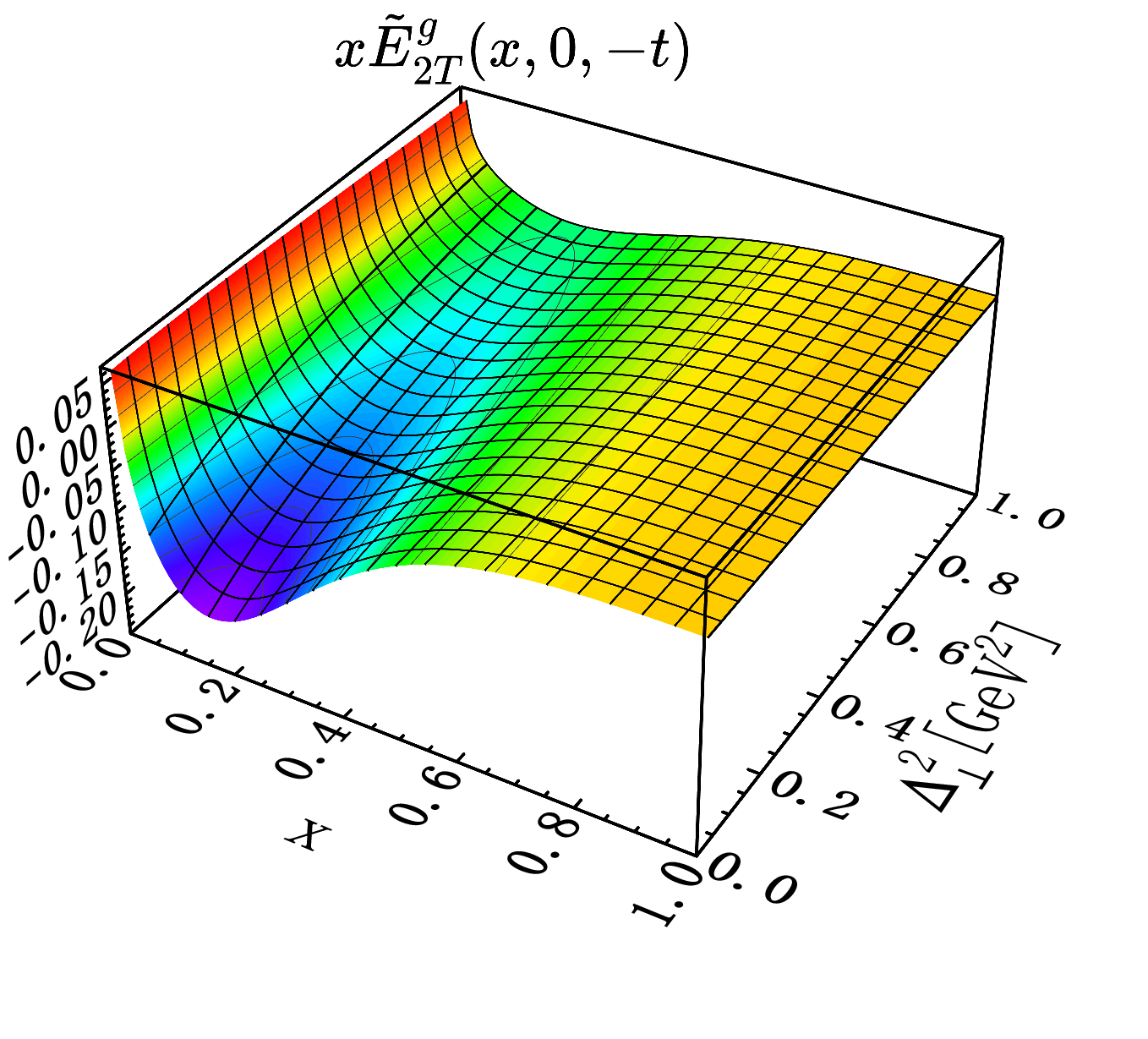}\\
	\includegraphics[width=0.32\columnwidth]{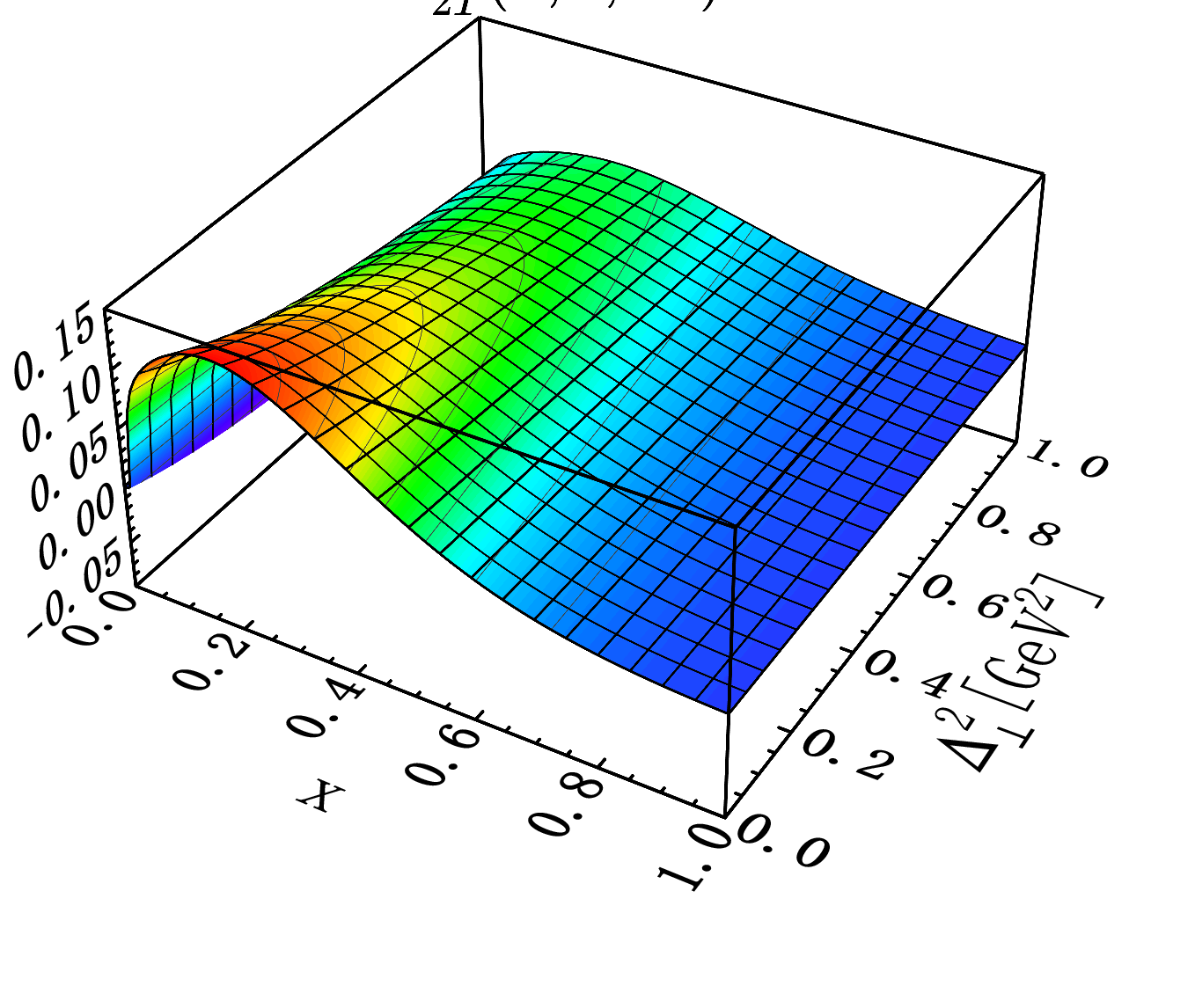}
	\includegraphics[width=0.32\columnwidth]{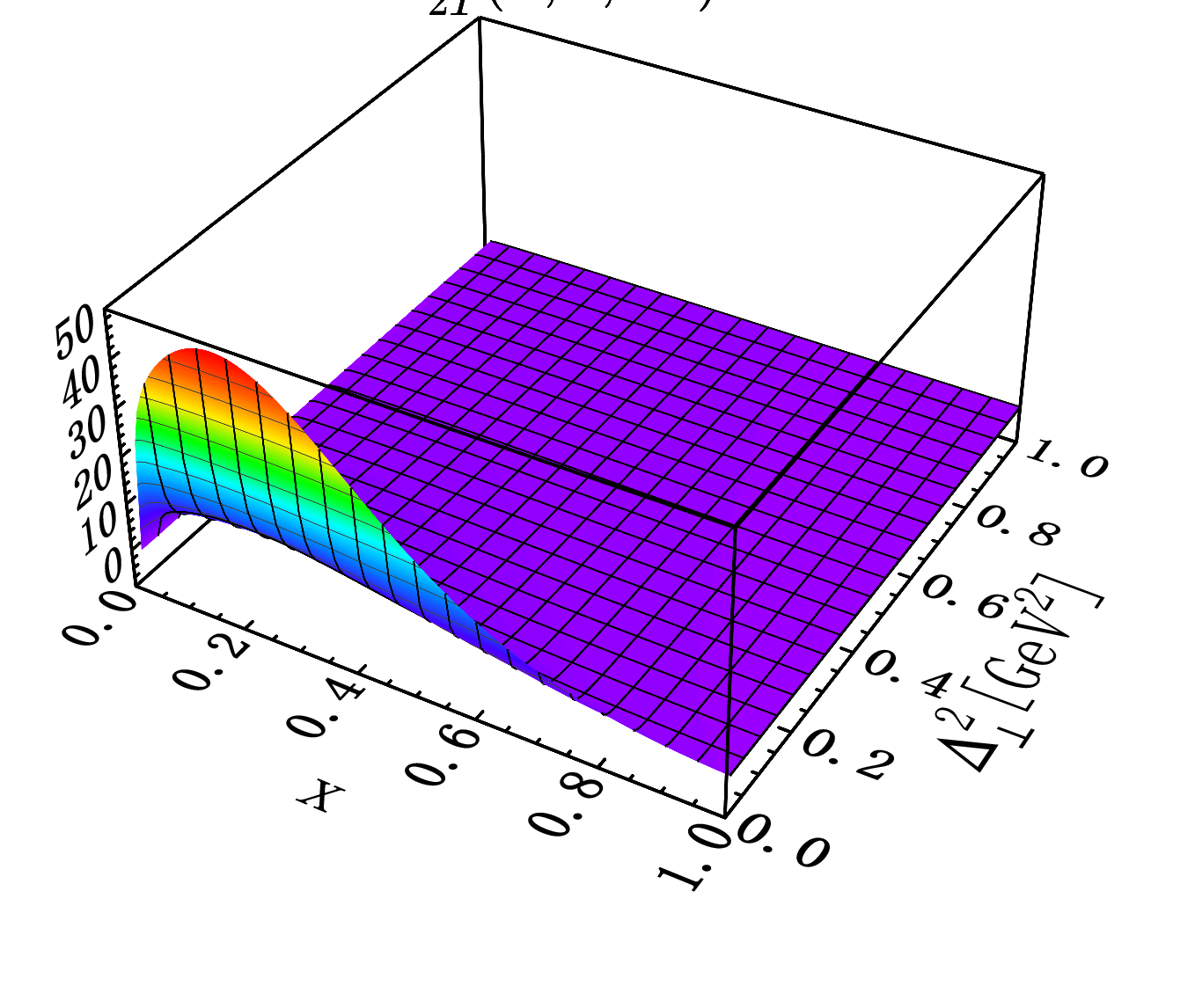}
	\includegraphics[width=0.32\columnwidth]{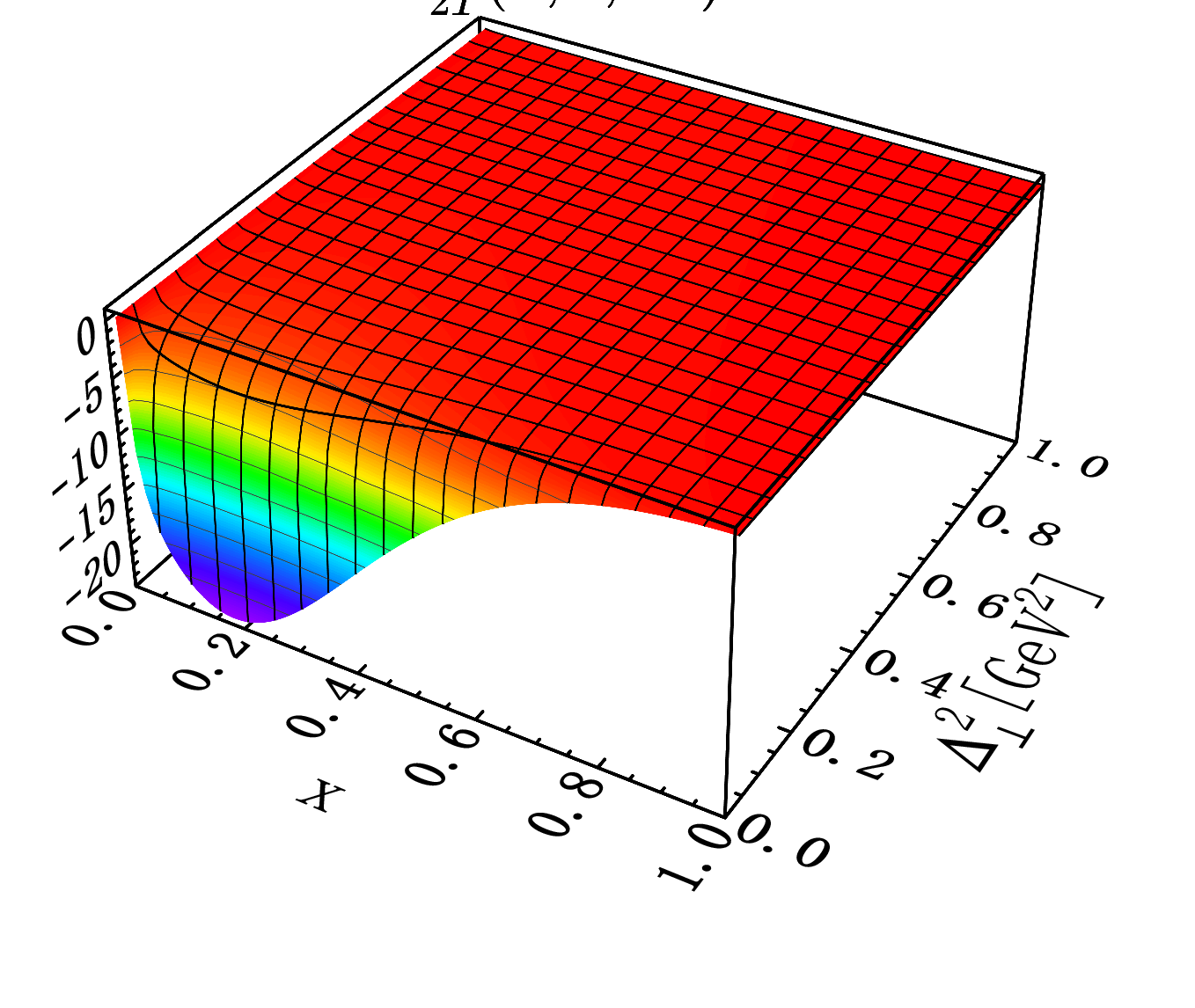}\\
	\includegraphics[width=0.32\columnwidth]{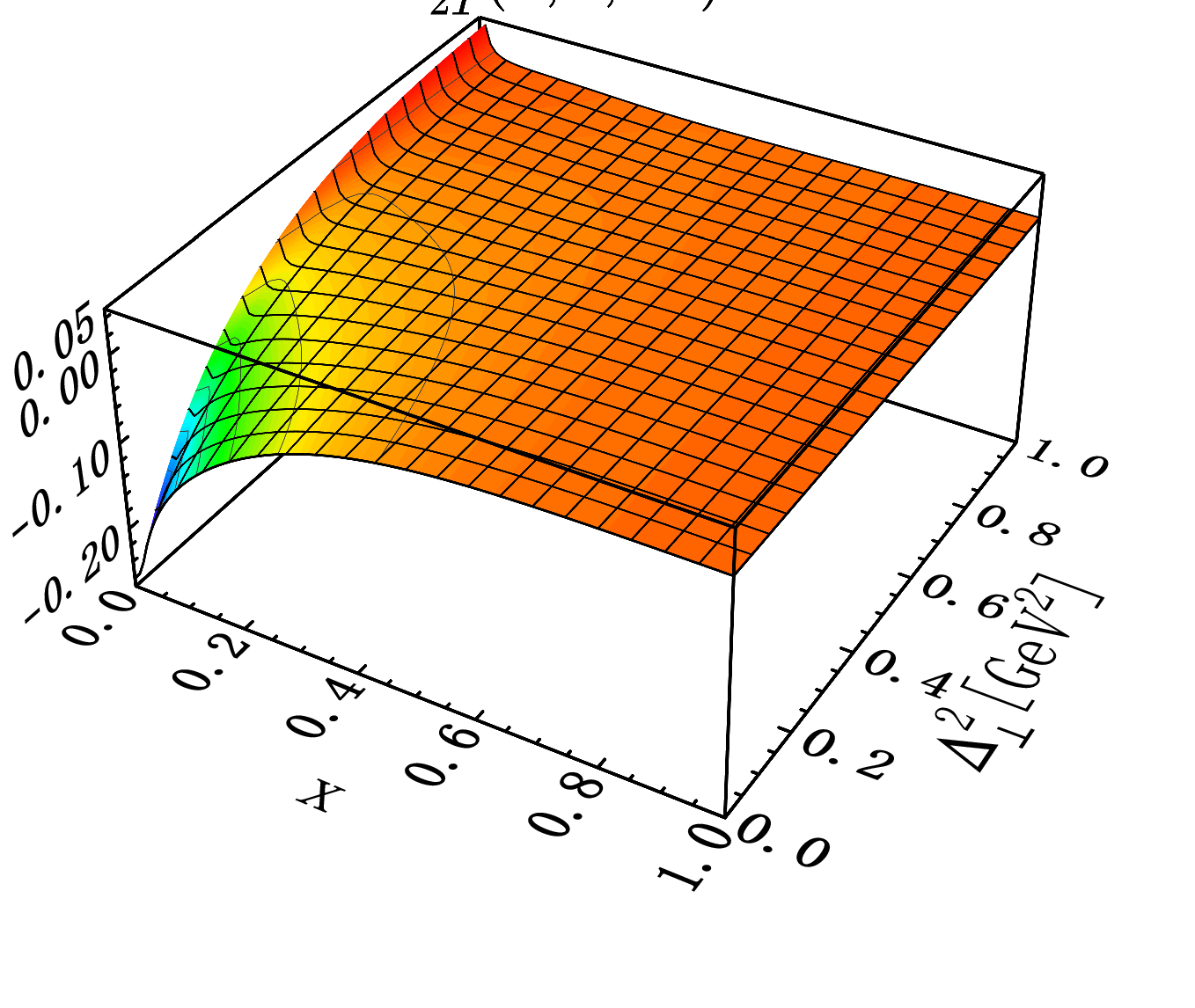}
	\includegraphics[width=0.32\columnwidth]{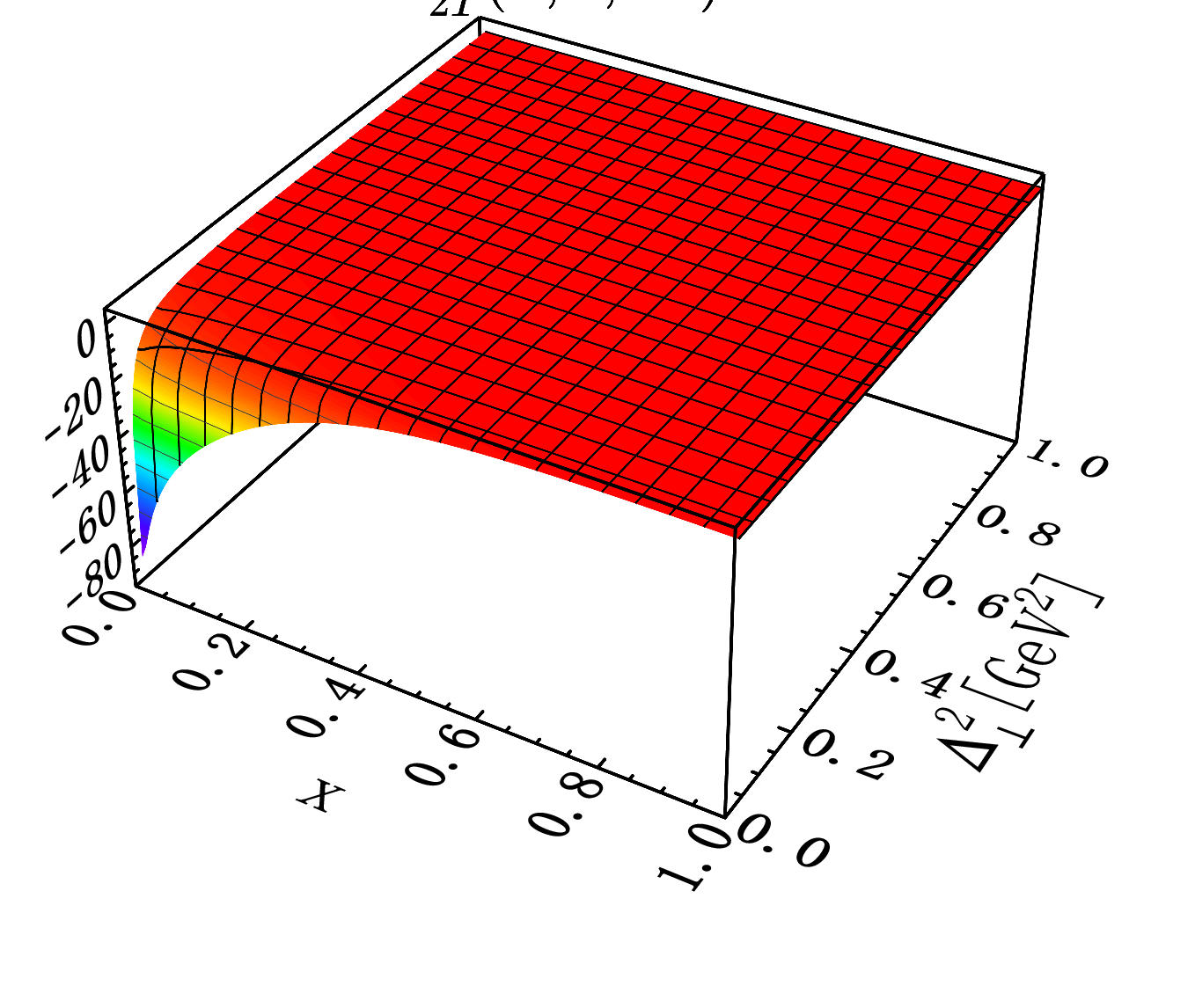}
	\includegraphics[width=0.32\columnwidth]{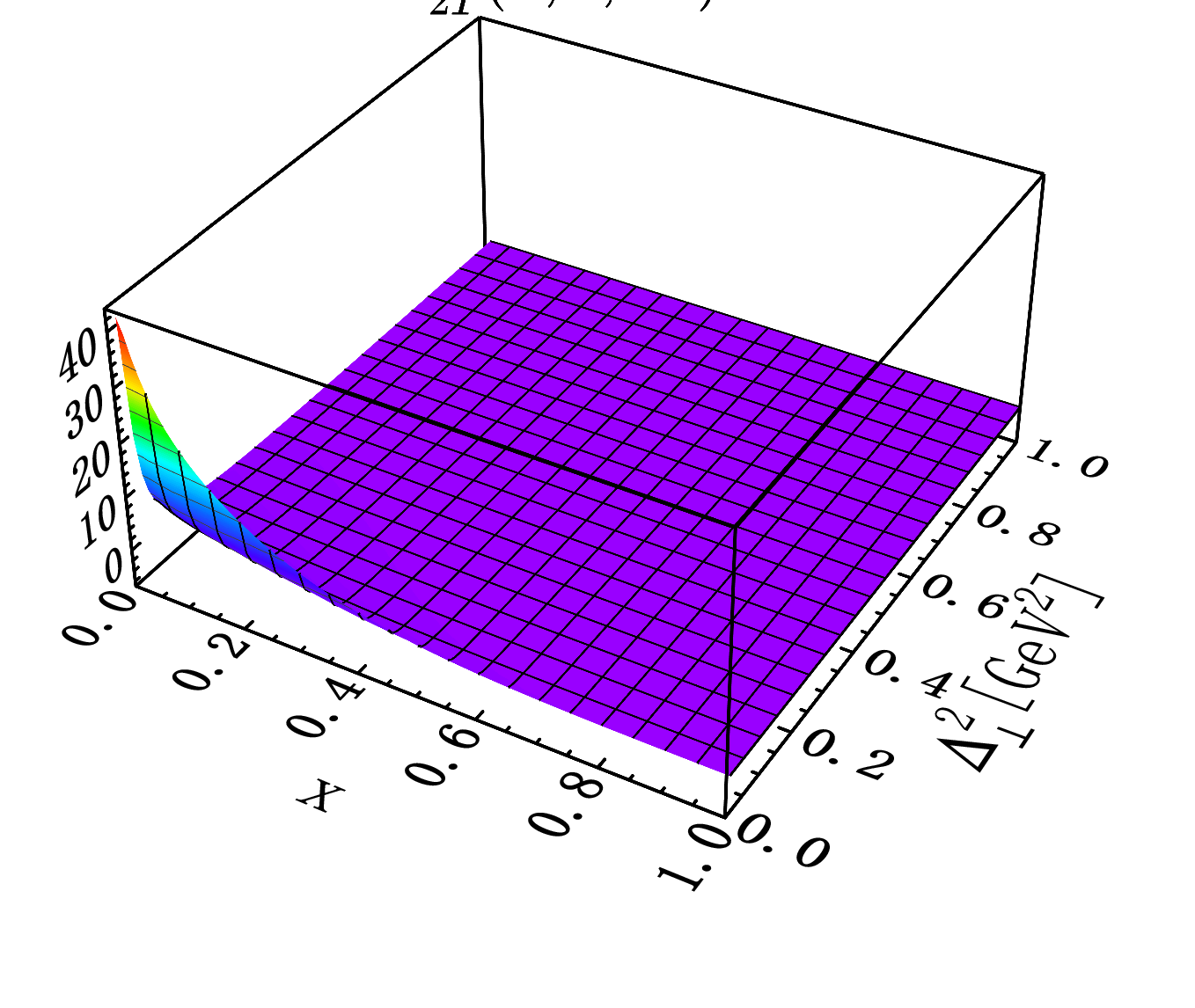}\\
	\includegraphics[width=0.32\columnwidth]{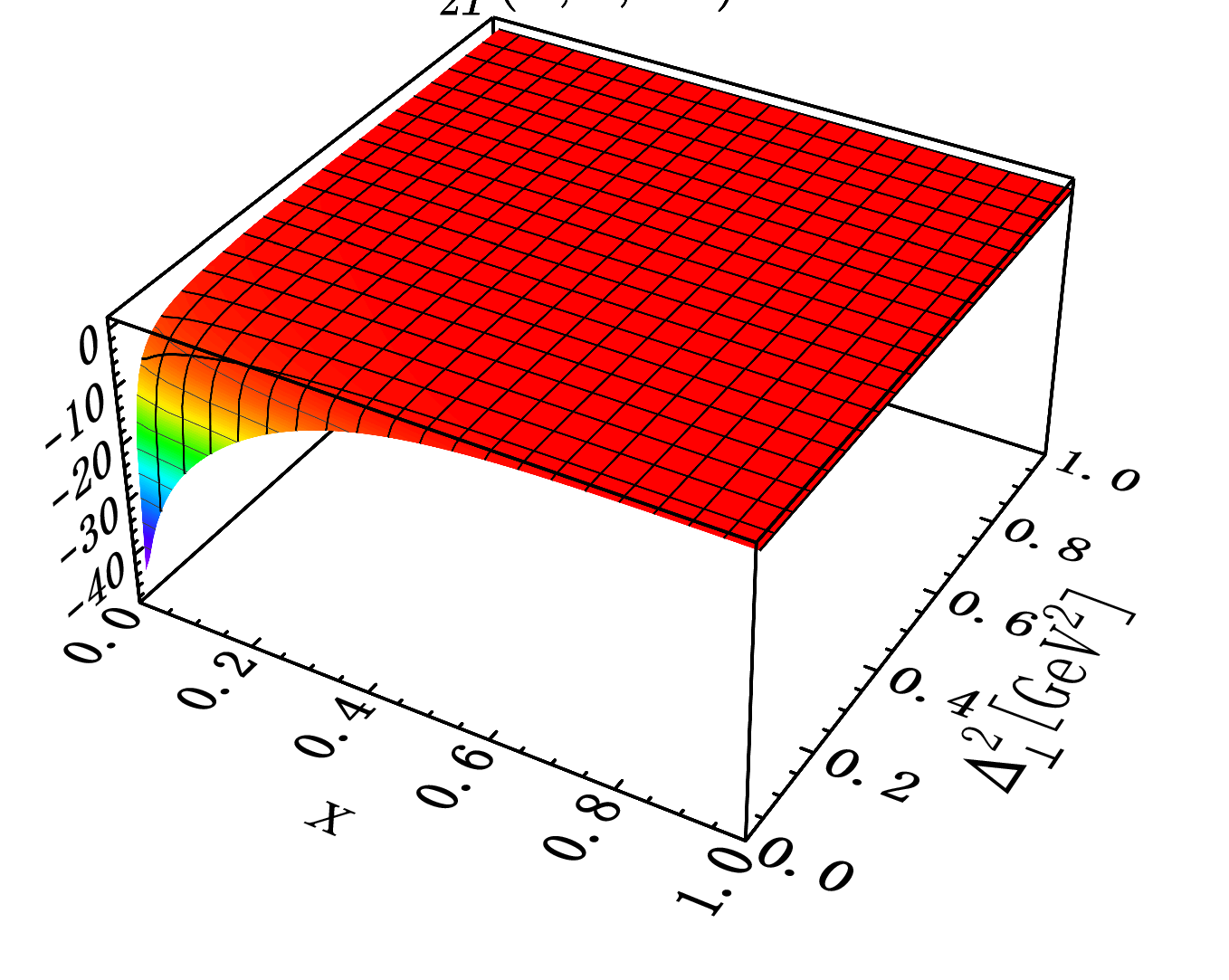}
	\includegraphics[width=0.32\columnwidth]{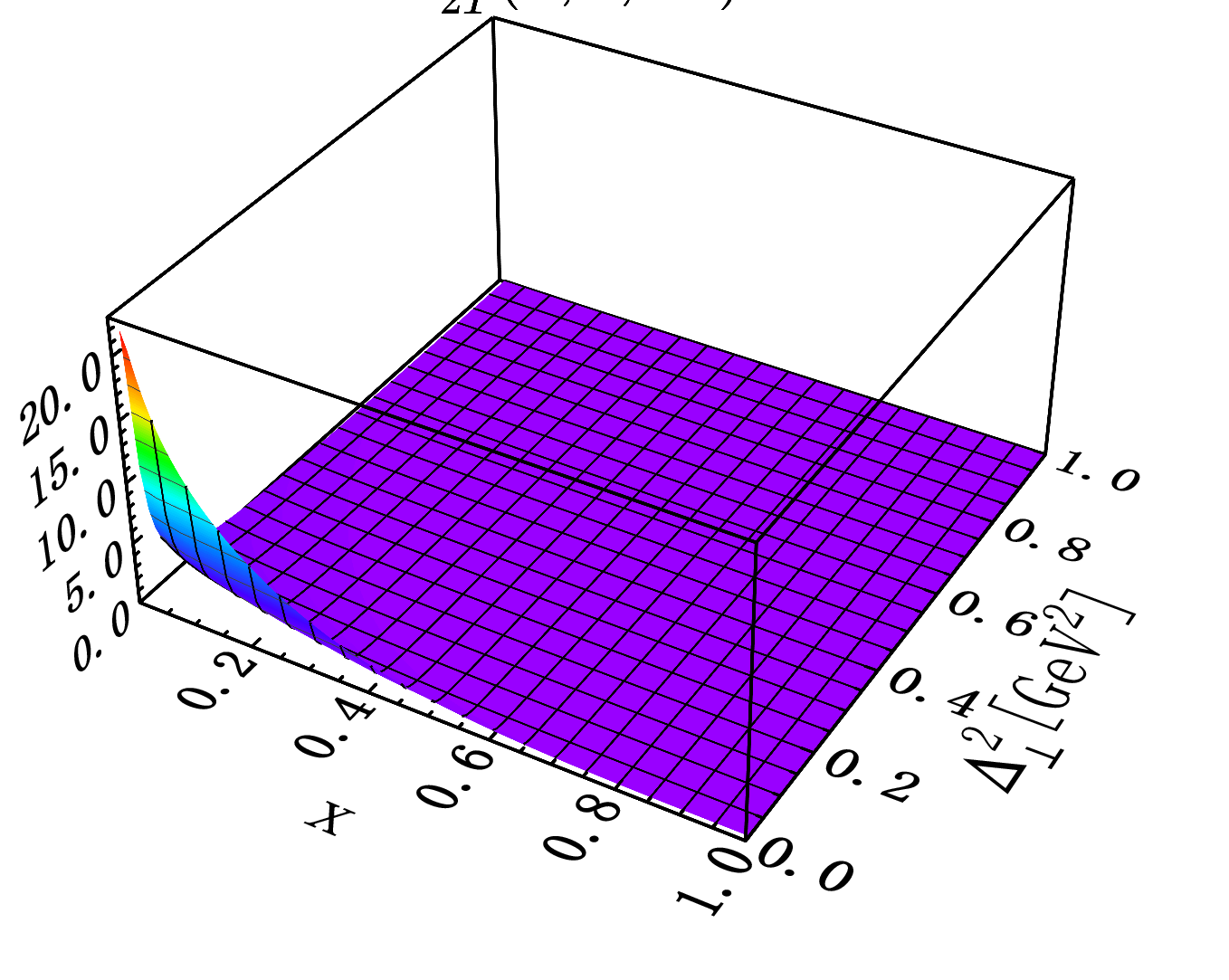}
	\includegraphics[width=0.32\columnwidth]{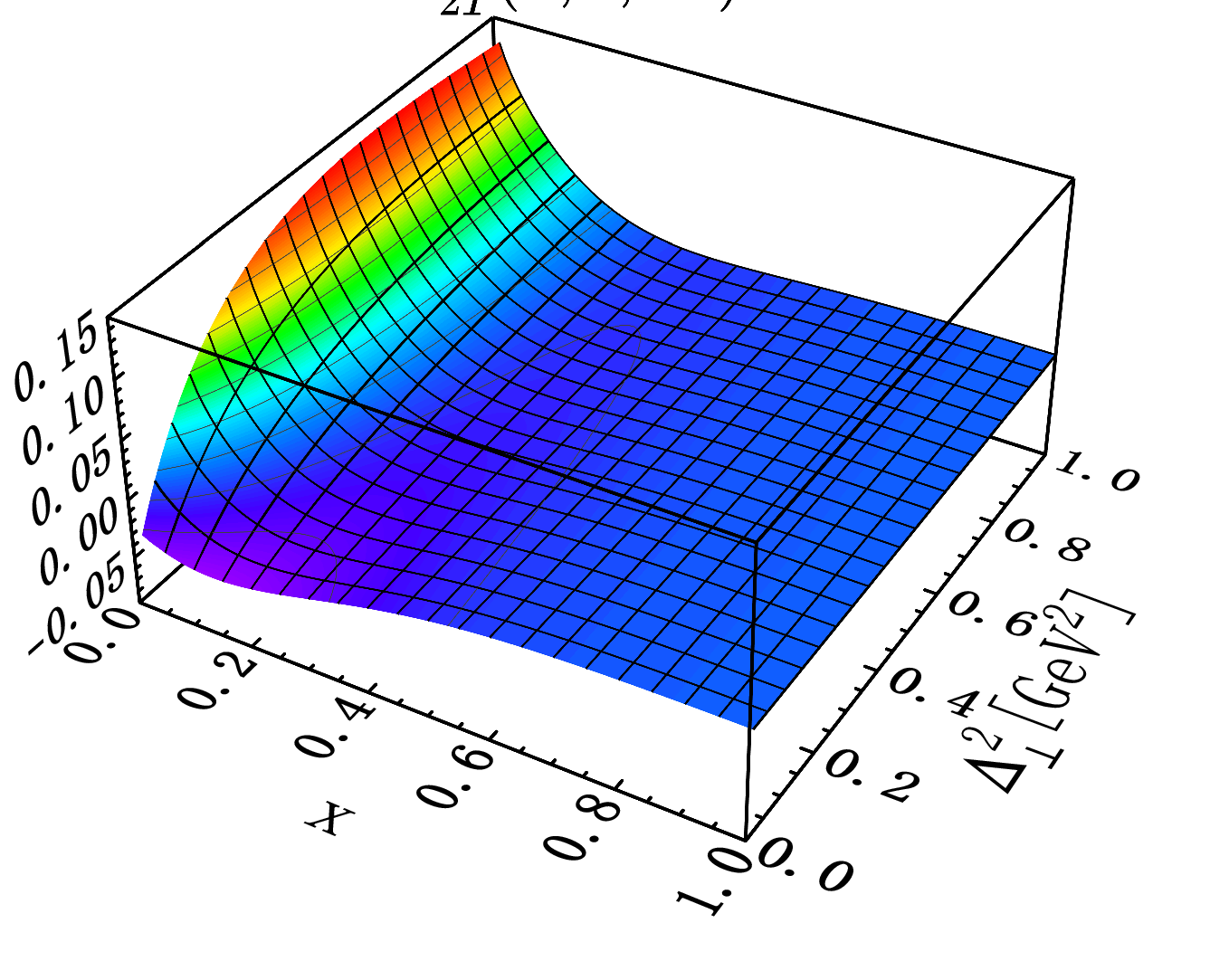}
	\caption{The twist-3 GPDs are plotted with respect to $x$ and $\bm{\Delta}_\perp^2$ in the kinematic range $x\in [0,1]$ and $\bm{\Delta}_\perp^2\in [0,1]\,\text{GeV}^2$.}
	\label{fig:GPD1}
\end{figure}

At $\xi=0$, corresponding to $\Delta^+=0$, the right-hand sides of Eqs.~(\ref{eq:p1}-\ref{eq:p4}) are considerably simplified. Each correlator is then parameterized by three GPDs, while $H_{2T}^g,\tilde{\bar{E}}_{2T}^g,\tilde{E}^{\prime g}_{2T}$, and $\bar{H}_{2T}^{\prime g}$ do not contribute. Fig.~\ref{fig:GPD1} presents the twist-3 gluon GPDs as functions of $x$ and $\bm{\Delta}_\perp^2=-t$. Most GPDs exhibit a pronounced rise toward $\bm{\Delta}_\perp^2\to0$, followed by a monotonic decrease with increasing $\bm{\Delta}_\perp^2$, while $H_{2T}^{\prime g}$ decreases much more slowly. The peak magnitudes generally decrease, and the peak positions shift toward larger $x$, as $\bm{\Delta}_\perp^2$ increases. The GPDs are concentrated predominantly in the small-$x$ region and smoothly approach zero as $x\to1$ due to phase-space suppression, consistent with the behavior commonly found for gluon PDFs, TMDs, GTMDs, and leading-twist GPDs~\cite{Tan:2023kbl,Bacchetta:2020vty,Lyubovitskij:2020xqj,Tan:2024dmz}. Certain GPDs peak around $x=0.2$, whereas others increase toward $x\to0$. GPDs with similar shapes generally have comparable magnitudes, whereas $\tilde{E}_{2T}^g, \bar{H}_{2T}^g, H_{2T}^{\prime g}$, and $\tilde{\bar{E}}_{2T}^{\prime g}$ are numerically suppressed. Despite their smaller magnitudes, these distributions are broader in $\bm{\Delta}_\perp^2$, suggesting a more diffuse transverse spatial distribution. These $(x,\bm{\Delta}_\perp^2)$ slices thus illustrate the interplay between longitudinal momentum and transverse spatial structure in twist-3 gluon dynamics.

\begin{figure}
	\centering
	\includegraphics[width=0.24\columnwidth]{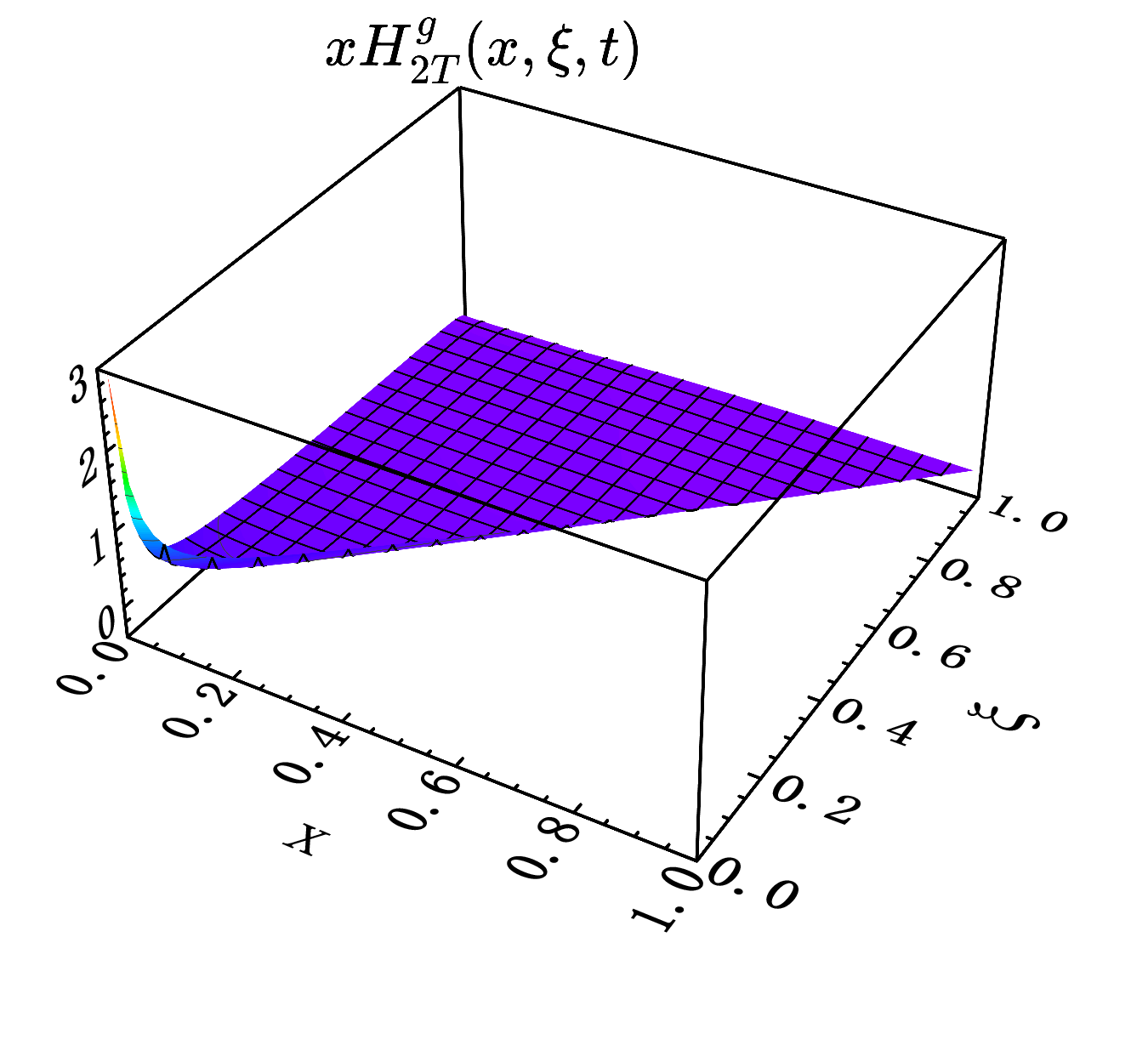}
	\includegraphics[width=0.24\columnwidth]{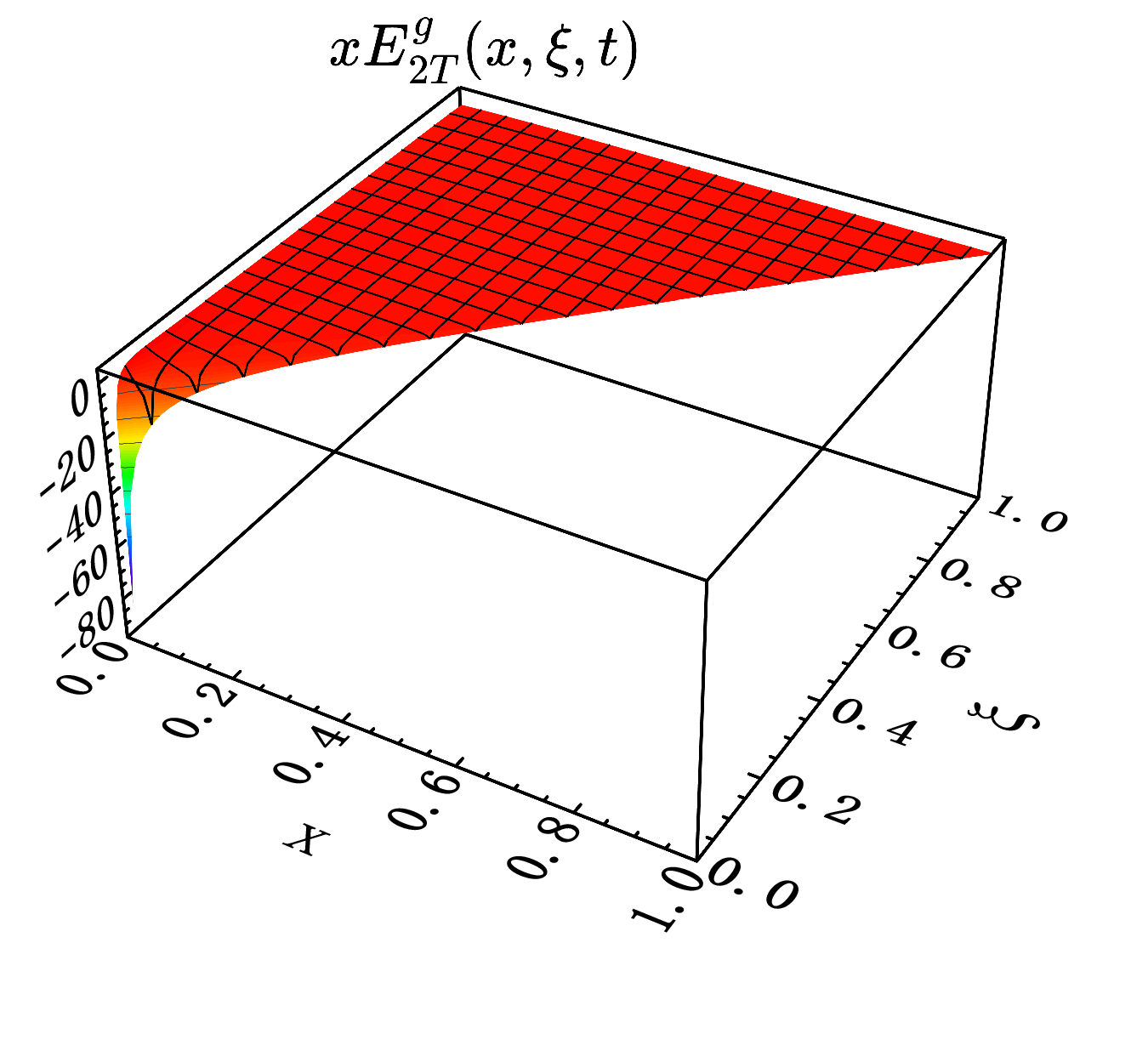}
	\includegraphics[width=0.24\columnwidth]{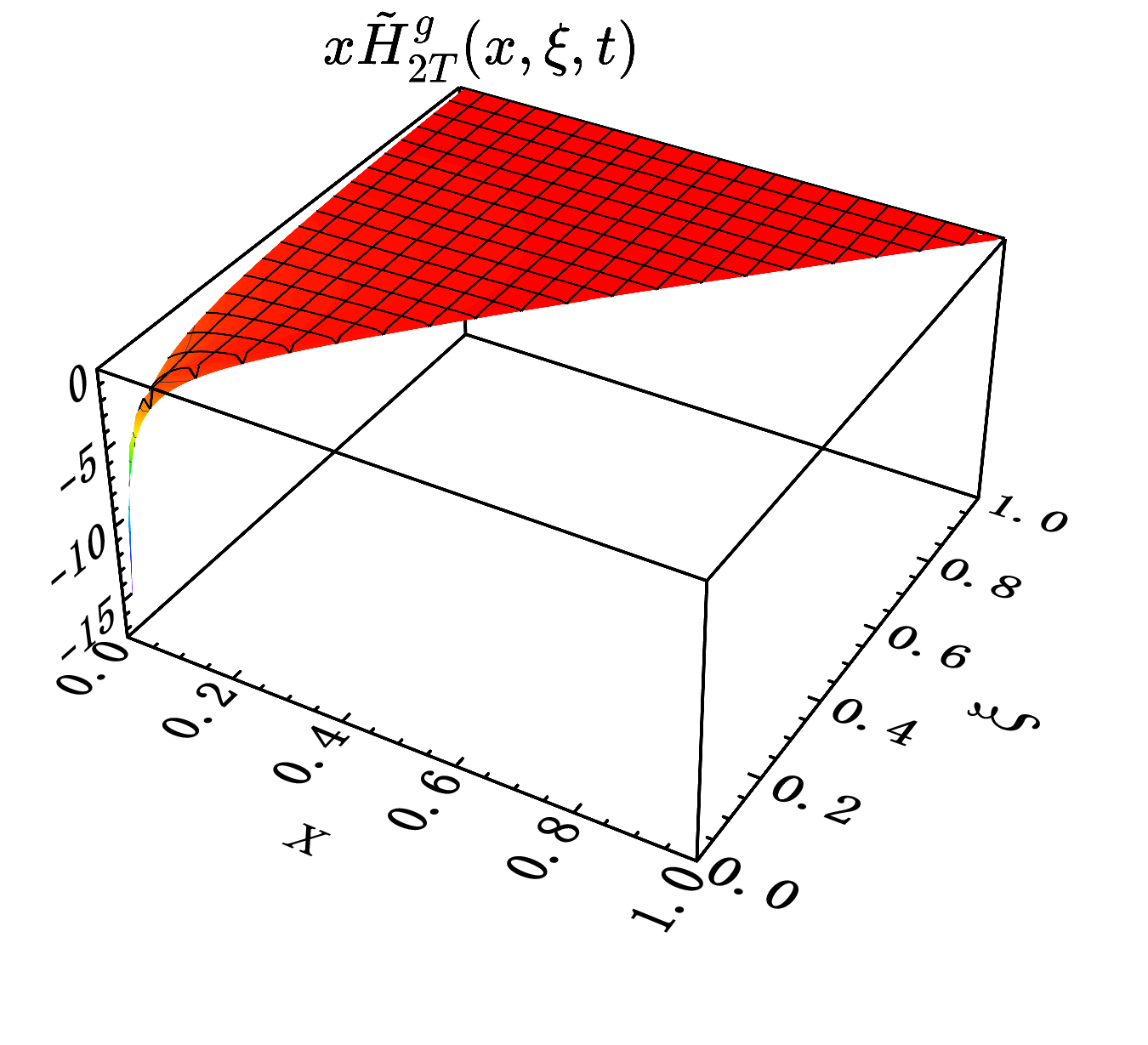}
	\includegraphics[width=0.24\columnwidth]{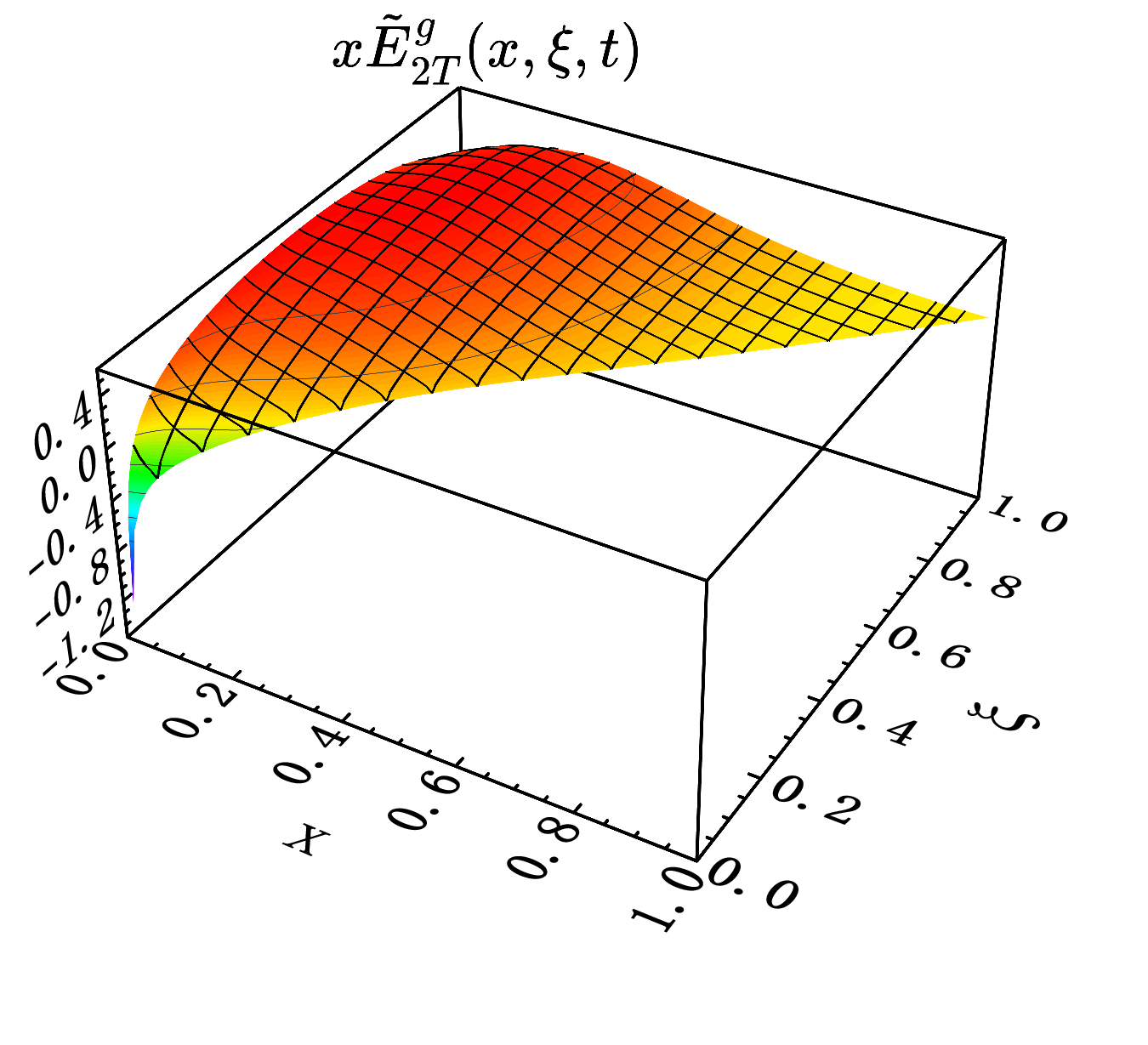}\\
	\includegraphics[width=0.24\columnwidth]{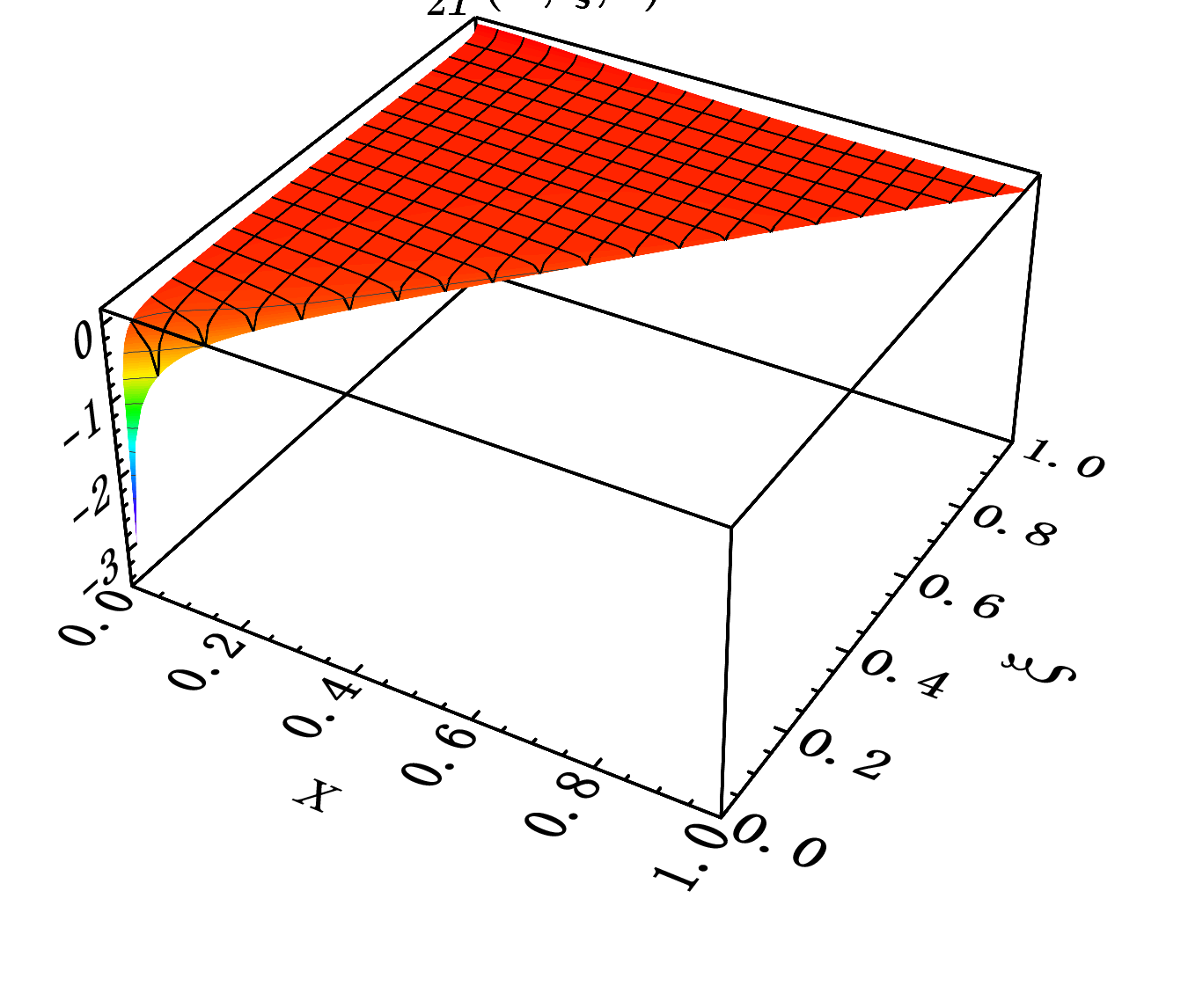}
	\includegraphics[width=0.24\columnwidth]{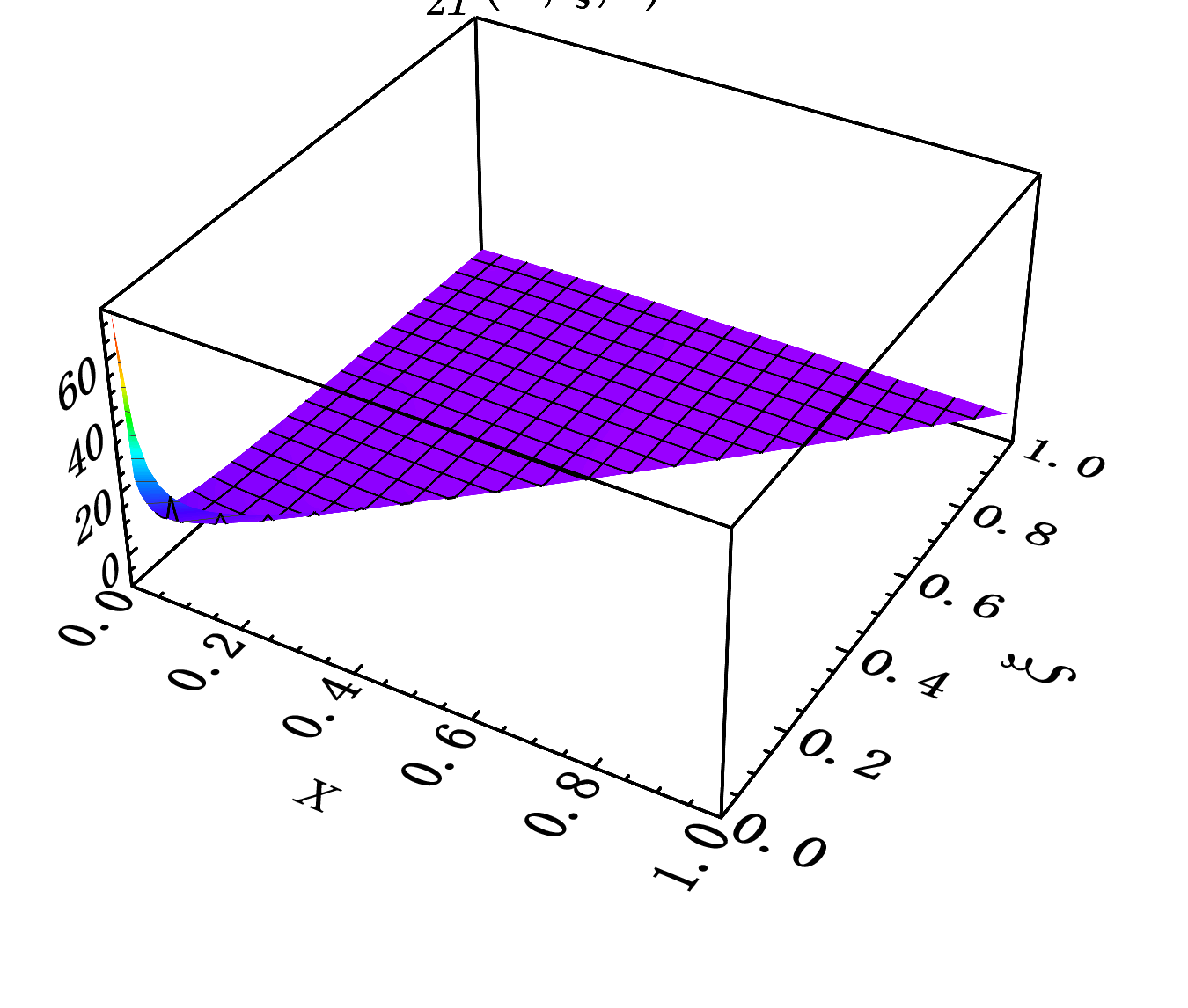}
	\includegraphics[width=0.24\columnwidth]{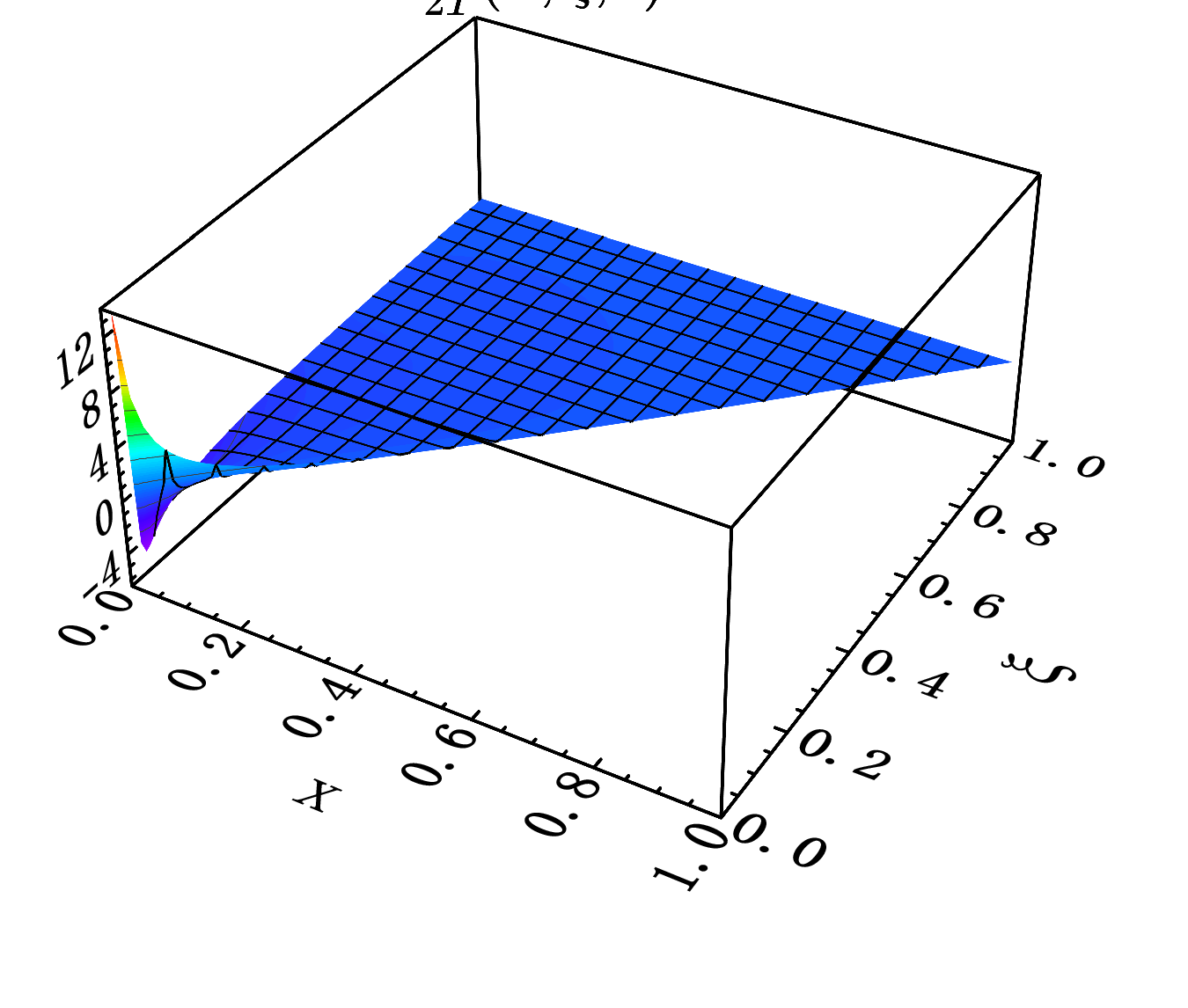}
	\includegraphics[width=0.24\columnwidth]{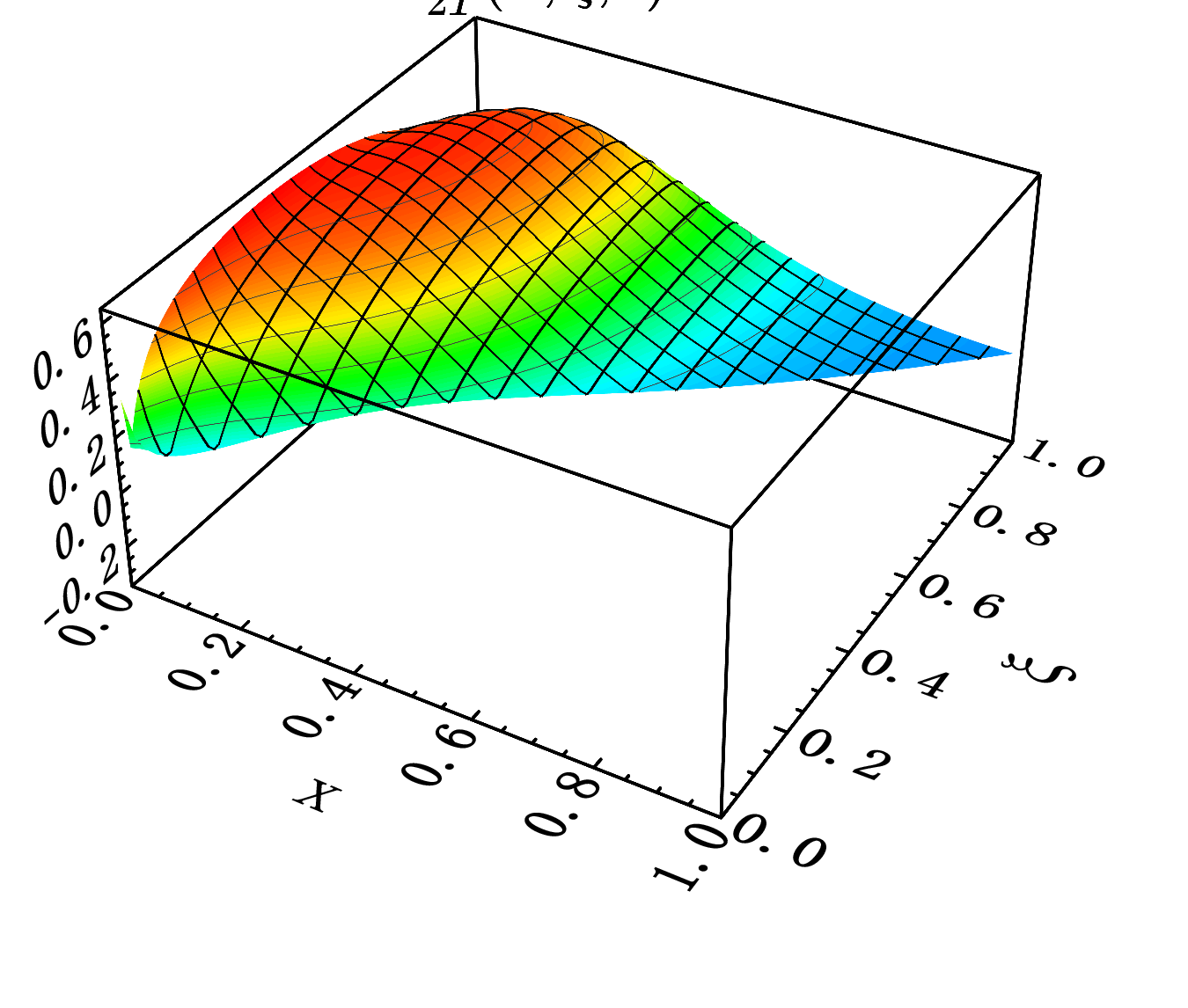}\\
	\includegraphics[width=0.24\columnwidth]{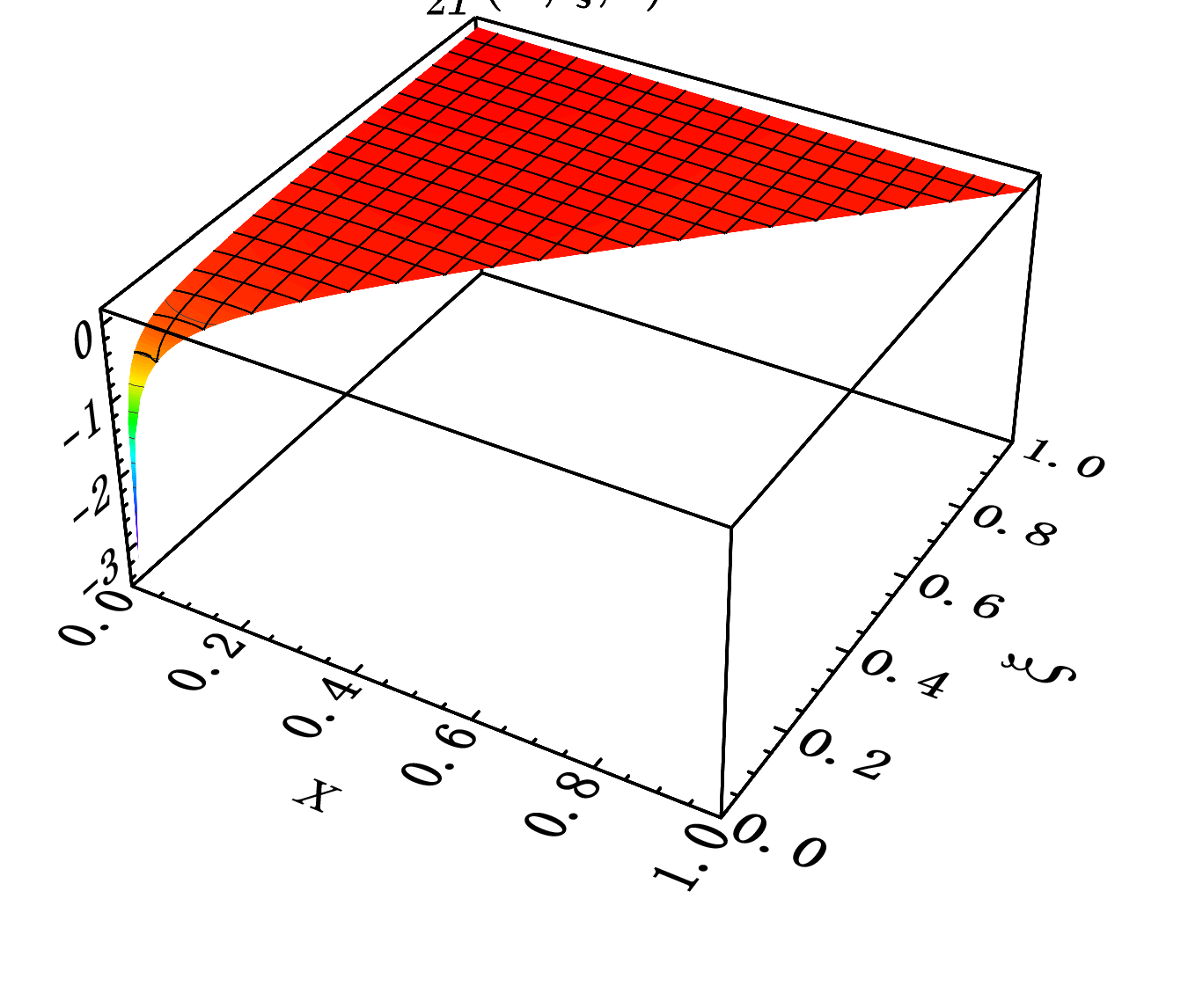}
	\includegraphics[width=0.24\columnwidth]{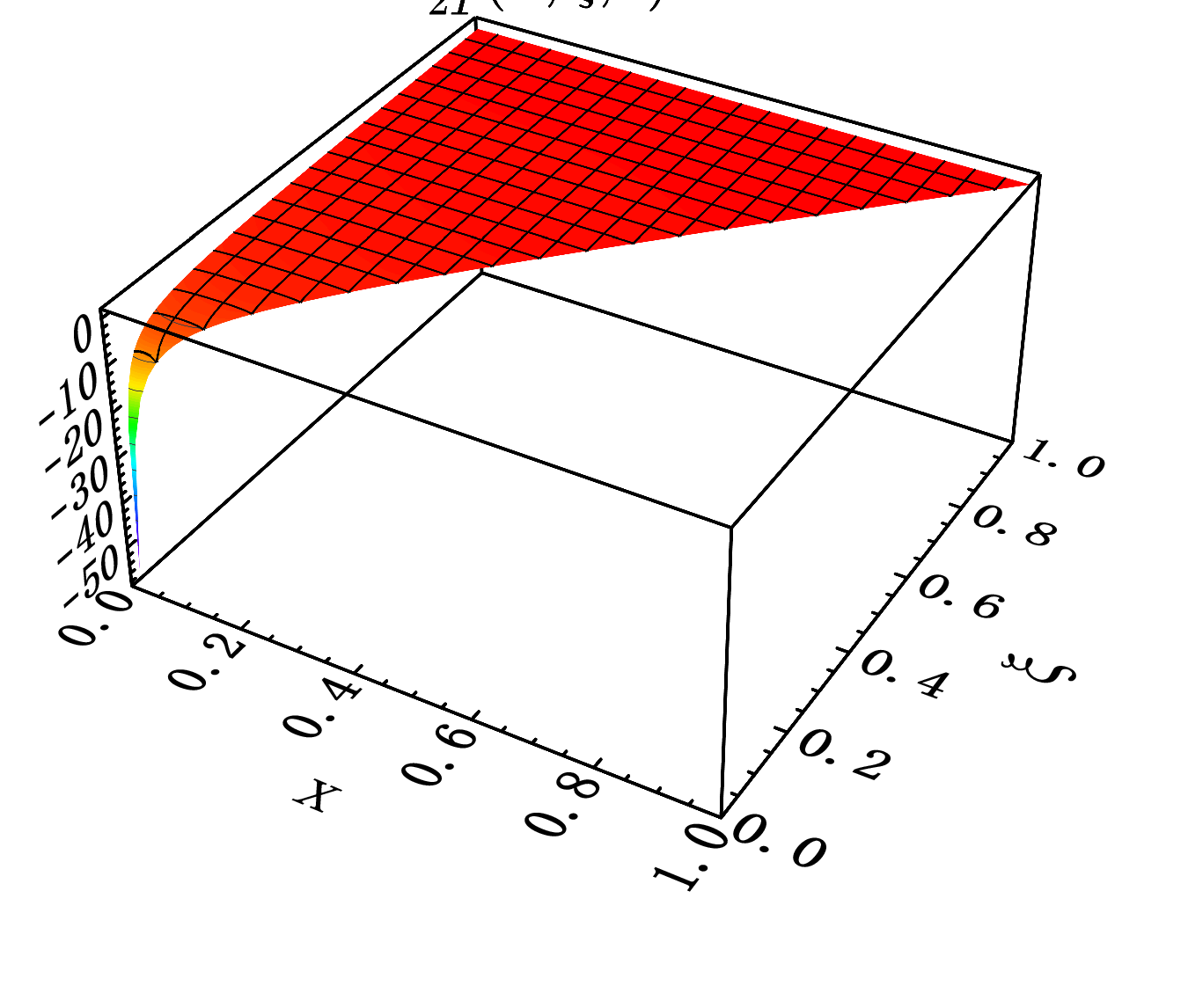}
	\includegraphics[width=0.24\columnwidth]{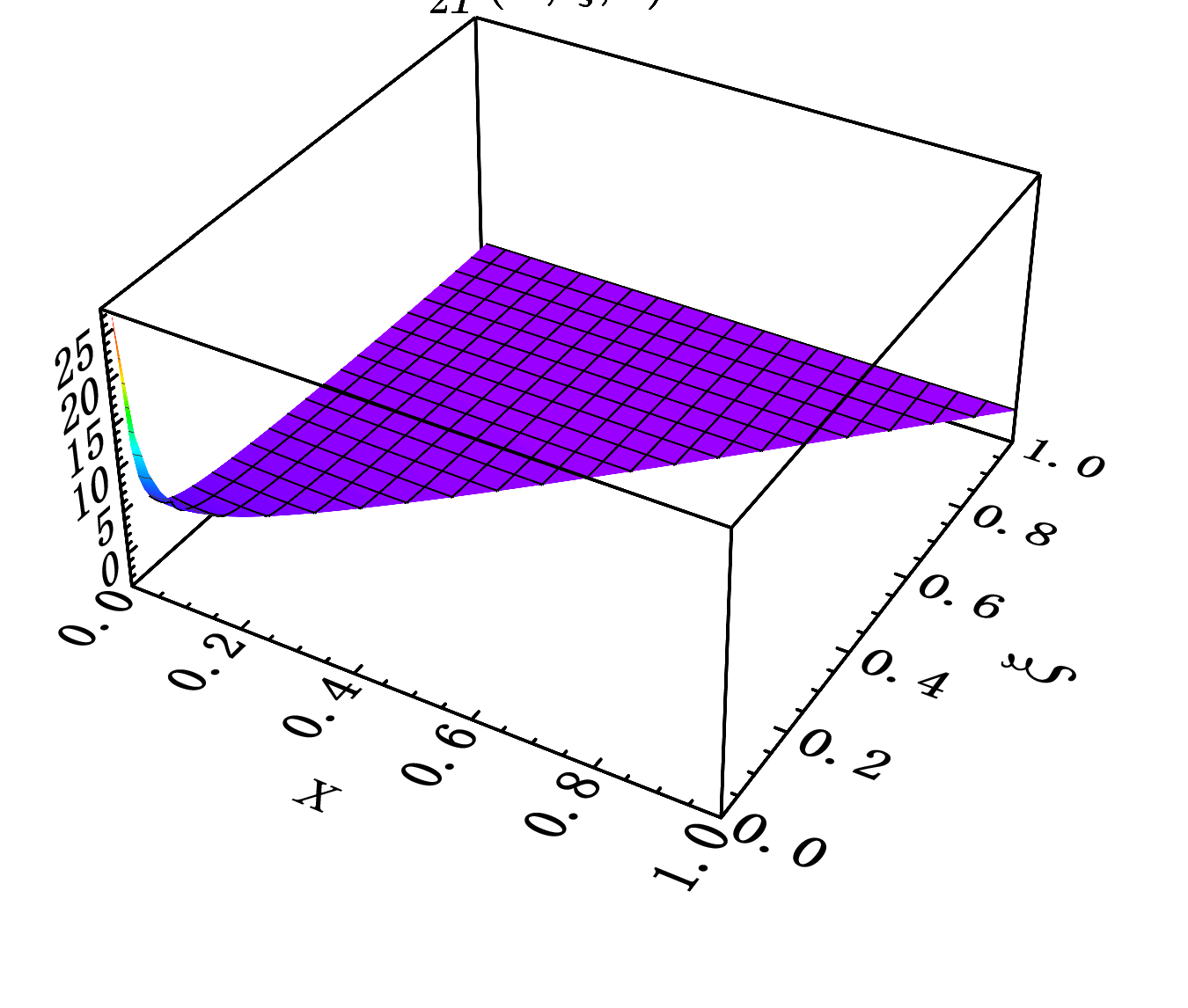}
	\includegraphics[width=0.24\columnwidth]{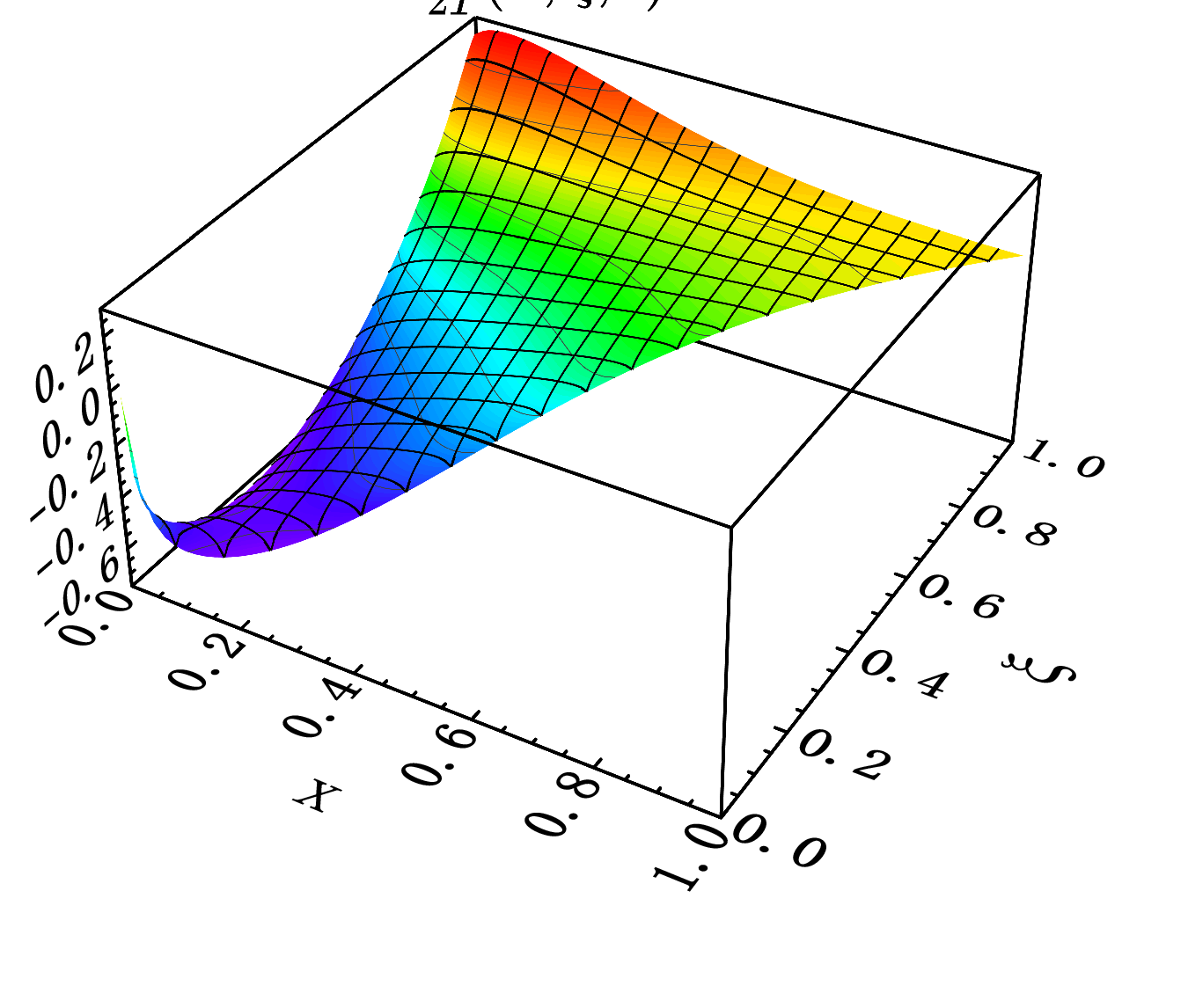}\\
	\includegraphics[width=0.24\columnwidth]{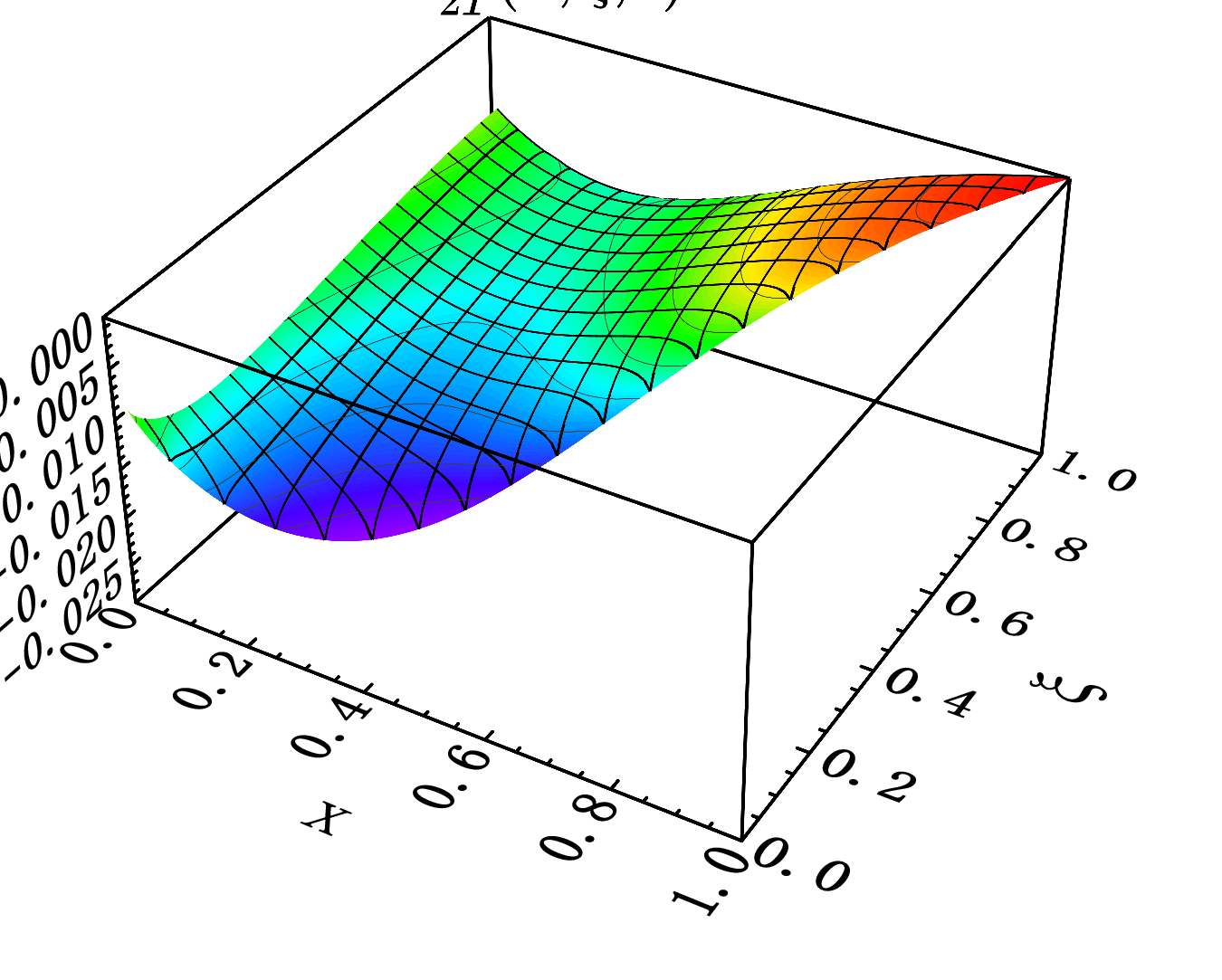}
	\includegraphics[width=0.24\columnwidth]{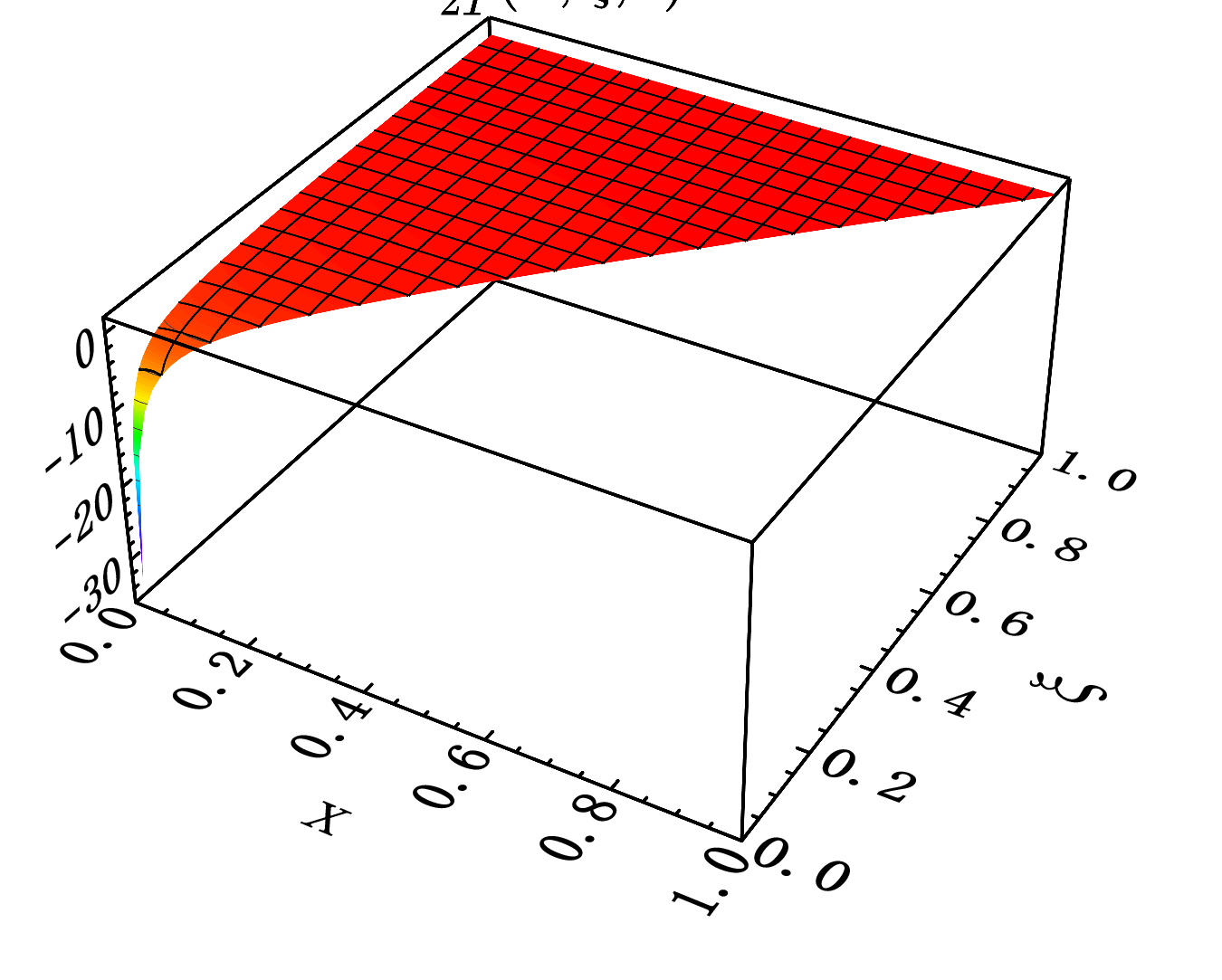}
	\includegraphics[width=0.24\columnwidth]{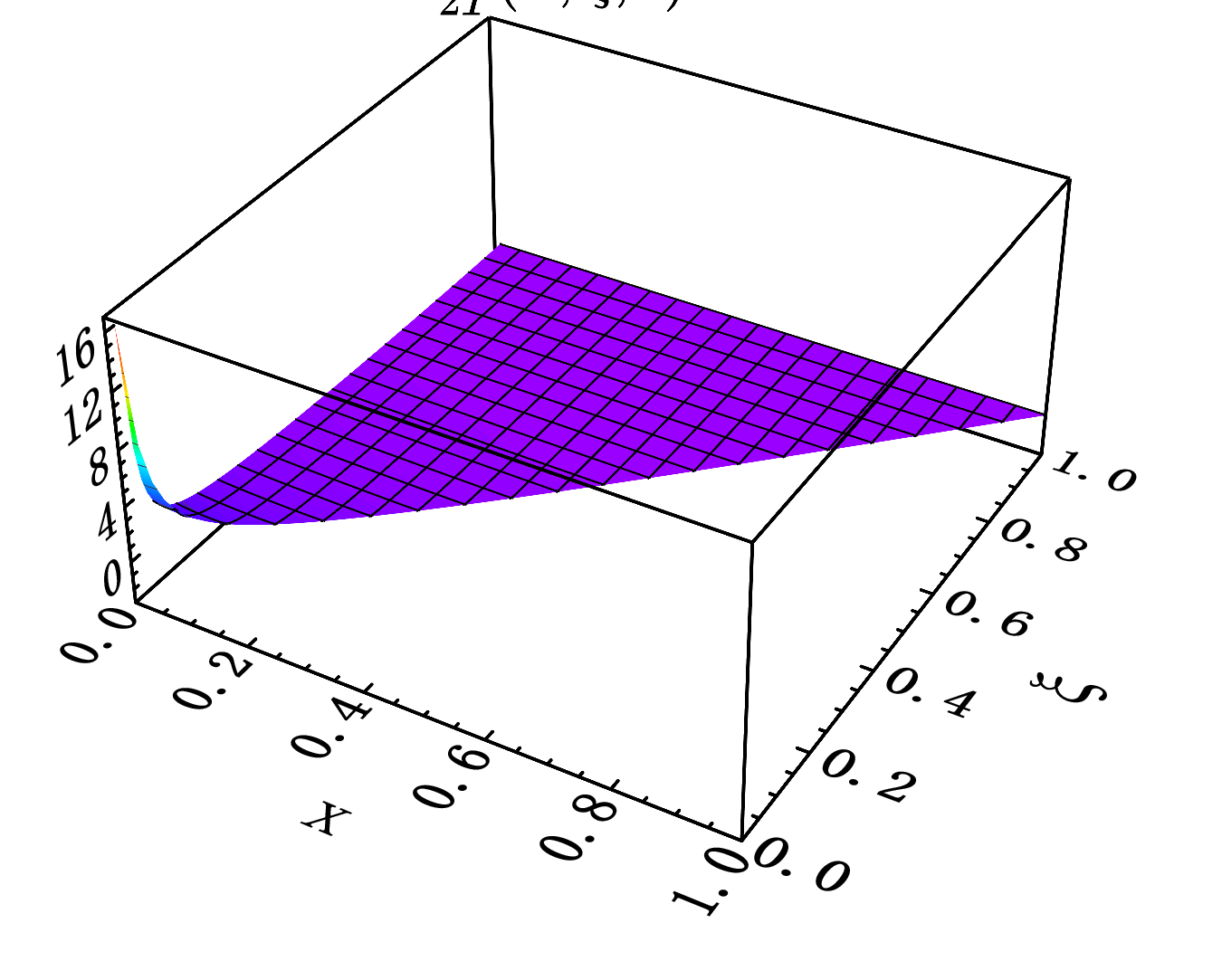}
	\includegraphics[width=0.24\columnwidth]{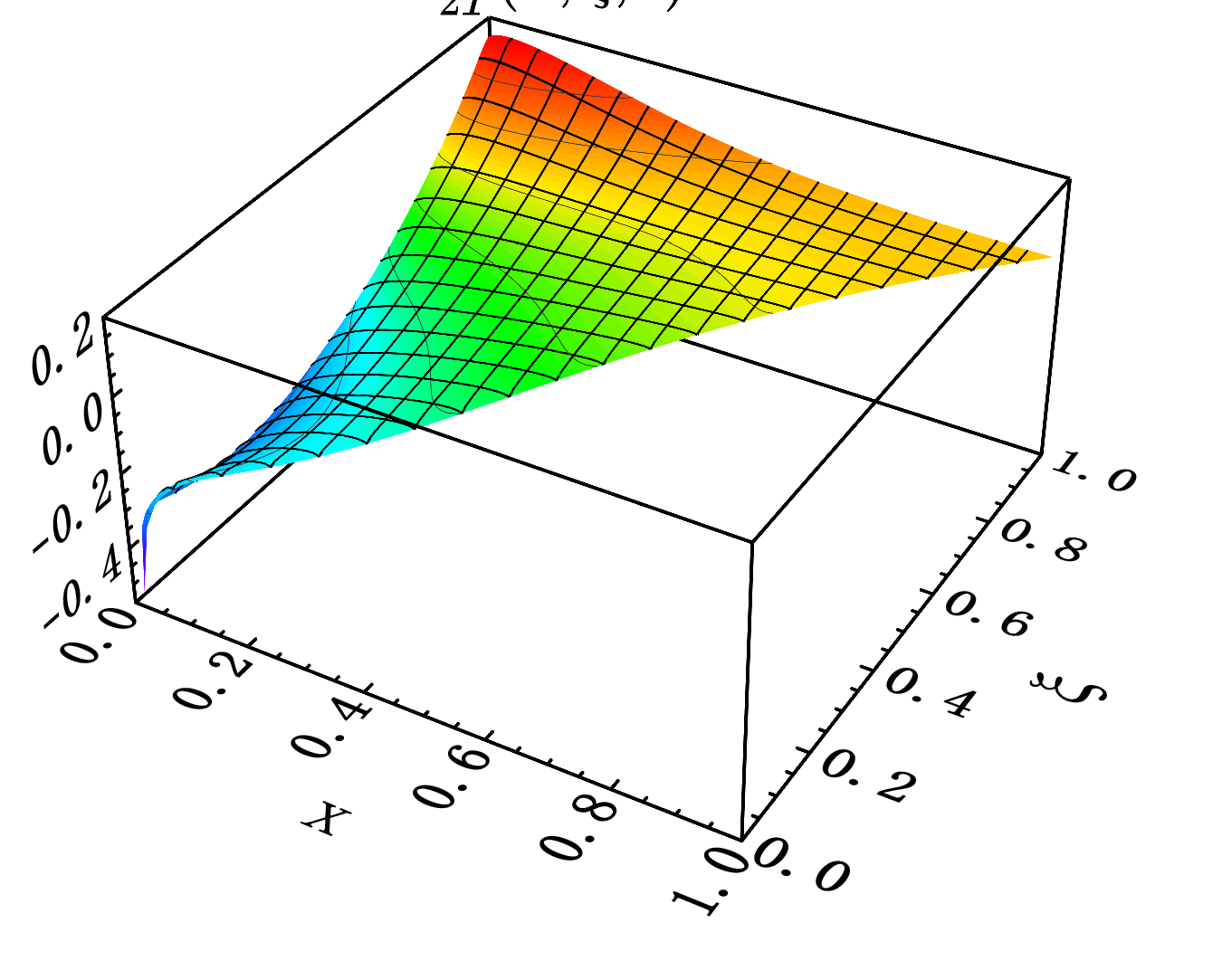}
	\caption{The twist-3 GPDs are plotted with respect to $x$ and $\xi$ in the kinematic range $x\in [0,\xi]$ and $\xi\in [0,1]$ at fixed $\bm{\Delta}_\perp^2=0.2\,\text{GeV}^2$.}
	\label{fig:GPD2}
\end{figure}

Fig.~\ref{fig:GPD2} shows the GPDs in the ERBL region at fixed $\bm{\Delta}_\perp^2=0.2\,\text{GeV}^2$ as functions of $x$ and $\xi$. The distributions are defined for $0\leq x\leq\xi$ and are concentrated in the small-$x$ and low-$\xi$ region. Most GPDs increase strongly toward $x,\xi\to0$, while several distributions exhibit a weaker $\xi$ dependence. The GPDs with larger magnitudes generally have similar shapes, whereas a subset is suppressed by one to two orders of magnitude. The suppressed distributions are nevertheless broader over the kinematic domain, indicating a distinct dependence on the model kinematics.

\begin{figure}
	\centering
	\includegraphics[width=0.24\columnwidth]{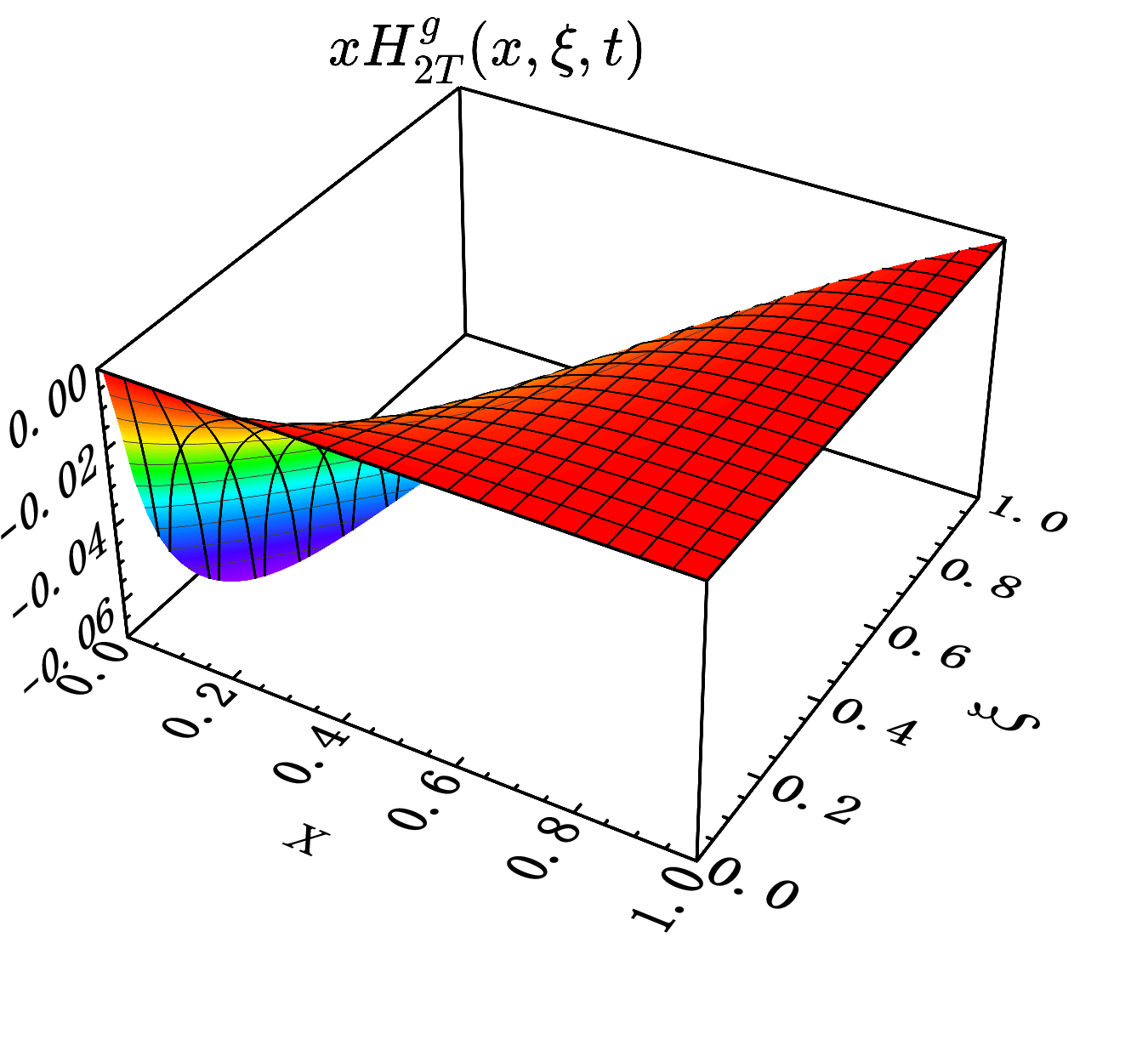}
	\includegraphics[width=0.24\columnwidth]{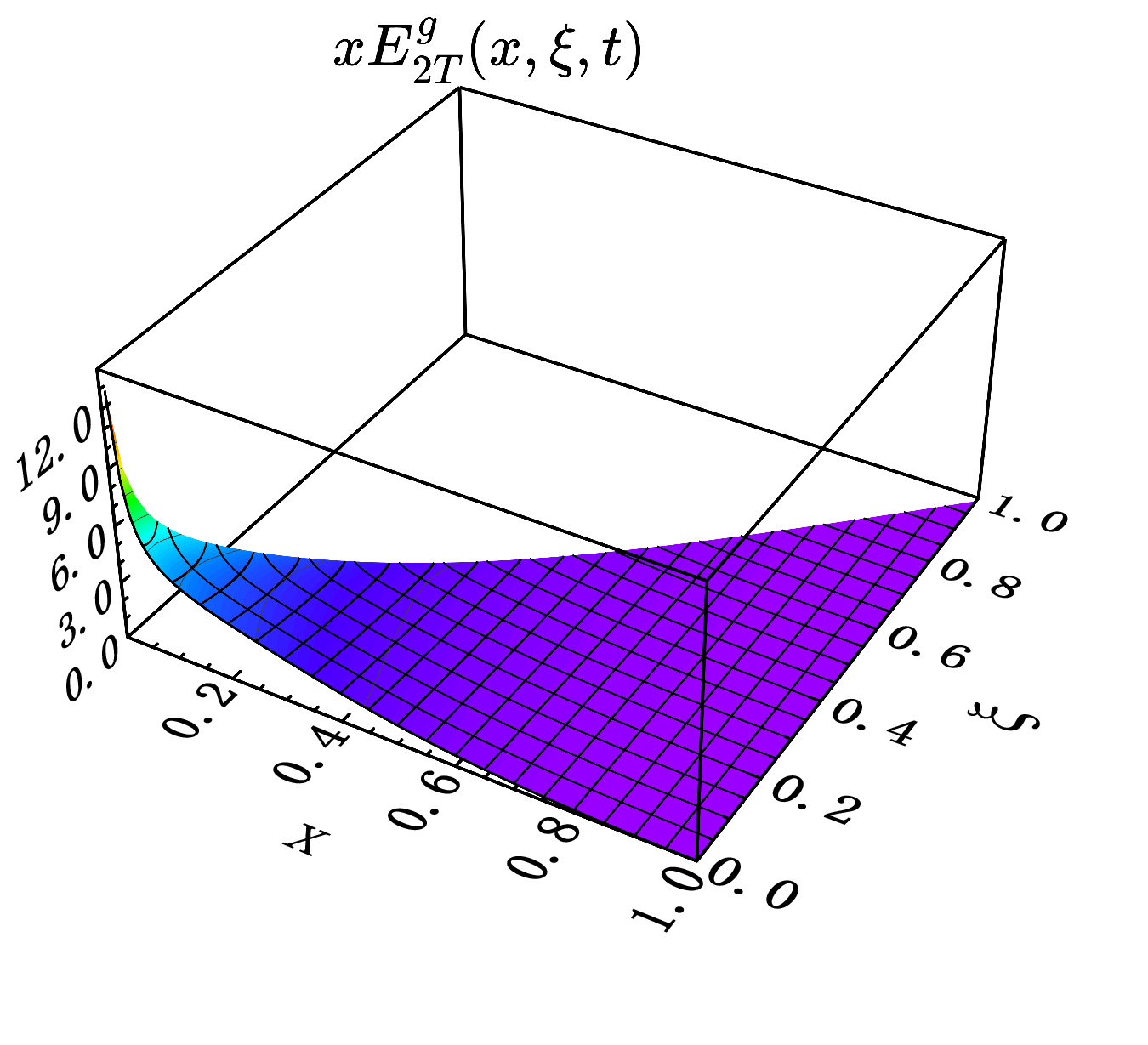}
	\includegraphics[width=0.24\columnwidth]{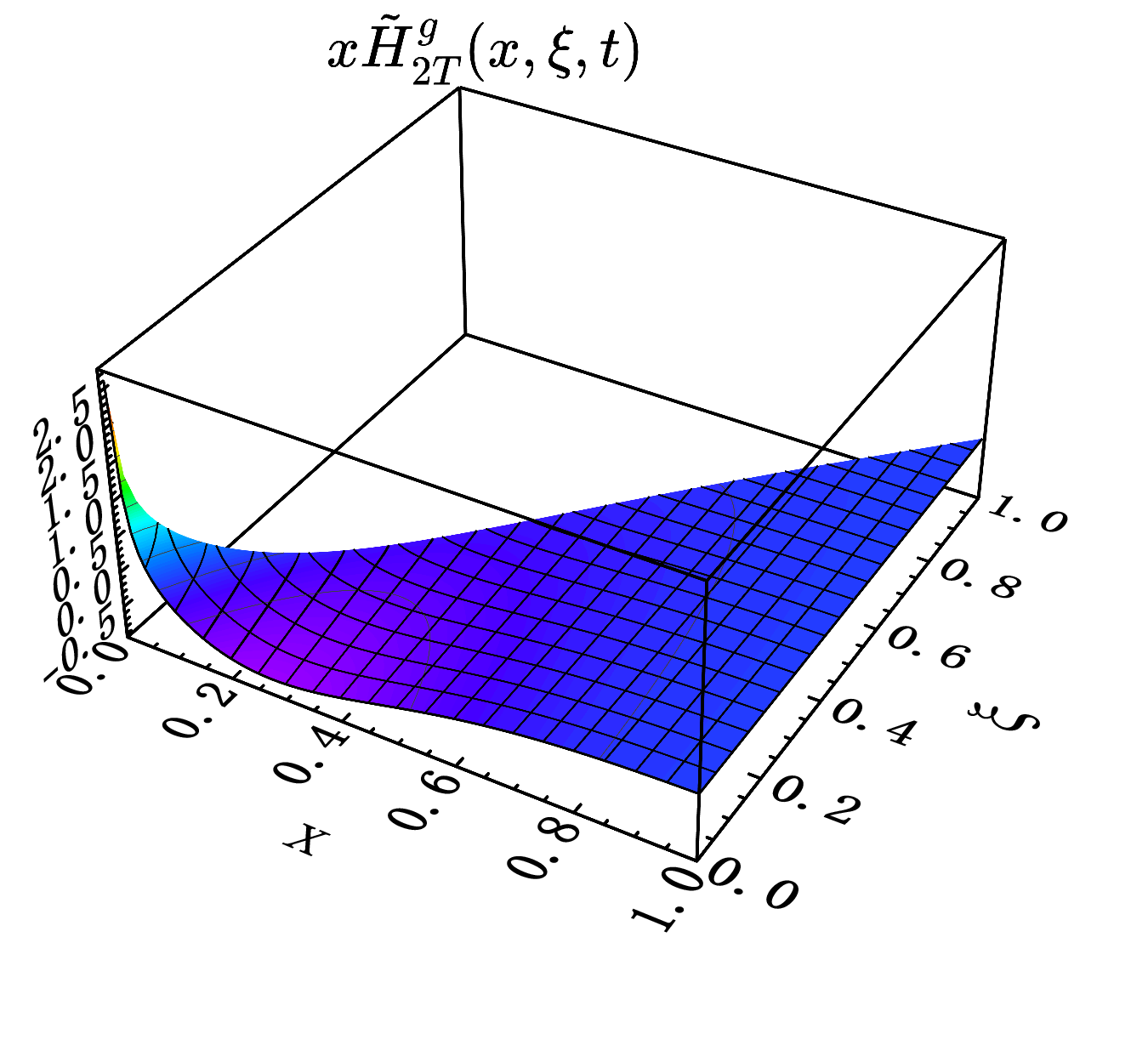}
	\includegraphics[width=0.24\columnwidth]{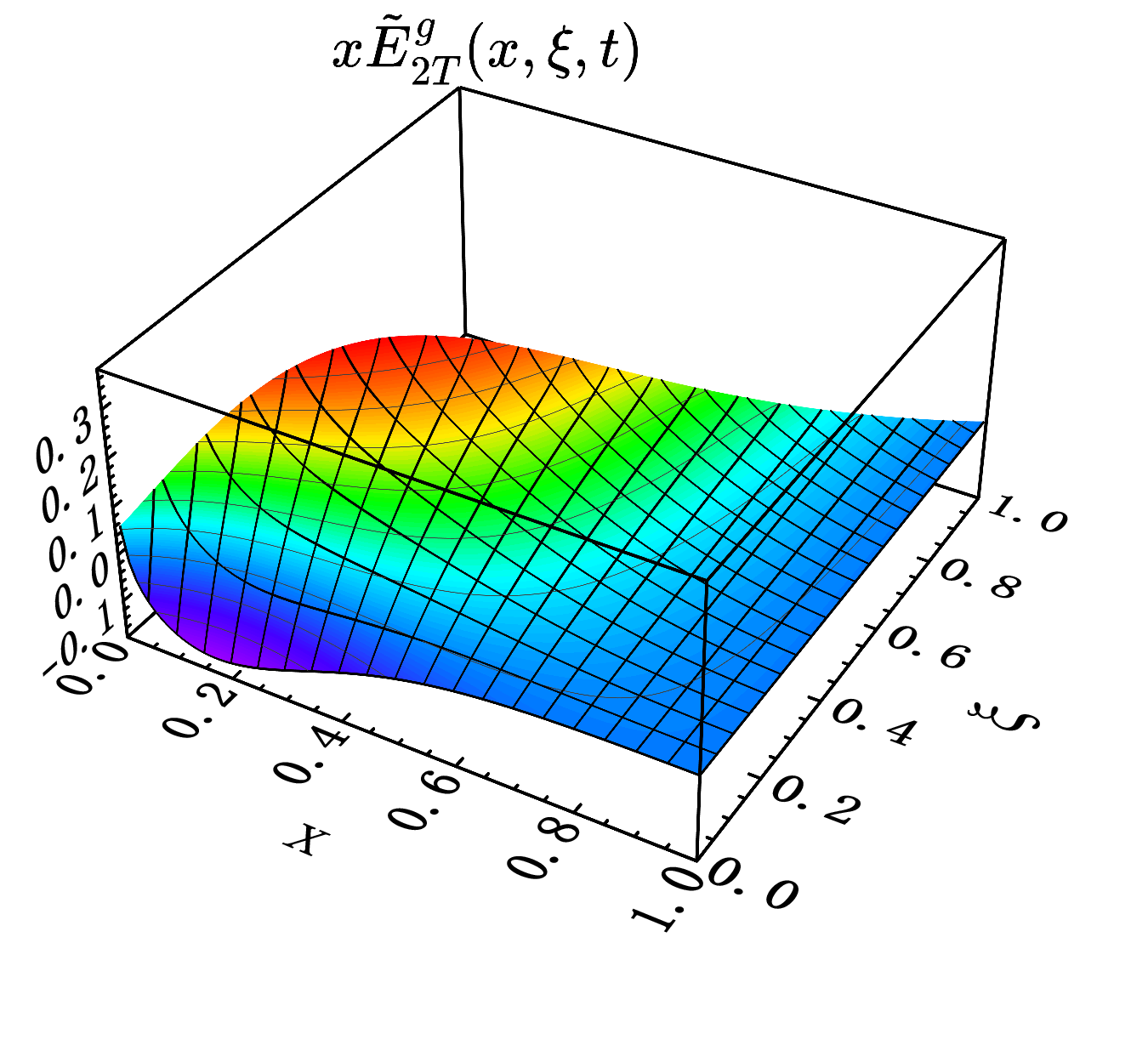}\\
	\includegraphics[width=0.24\columnwidth]{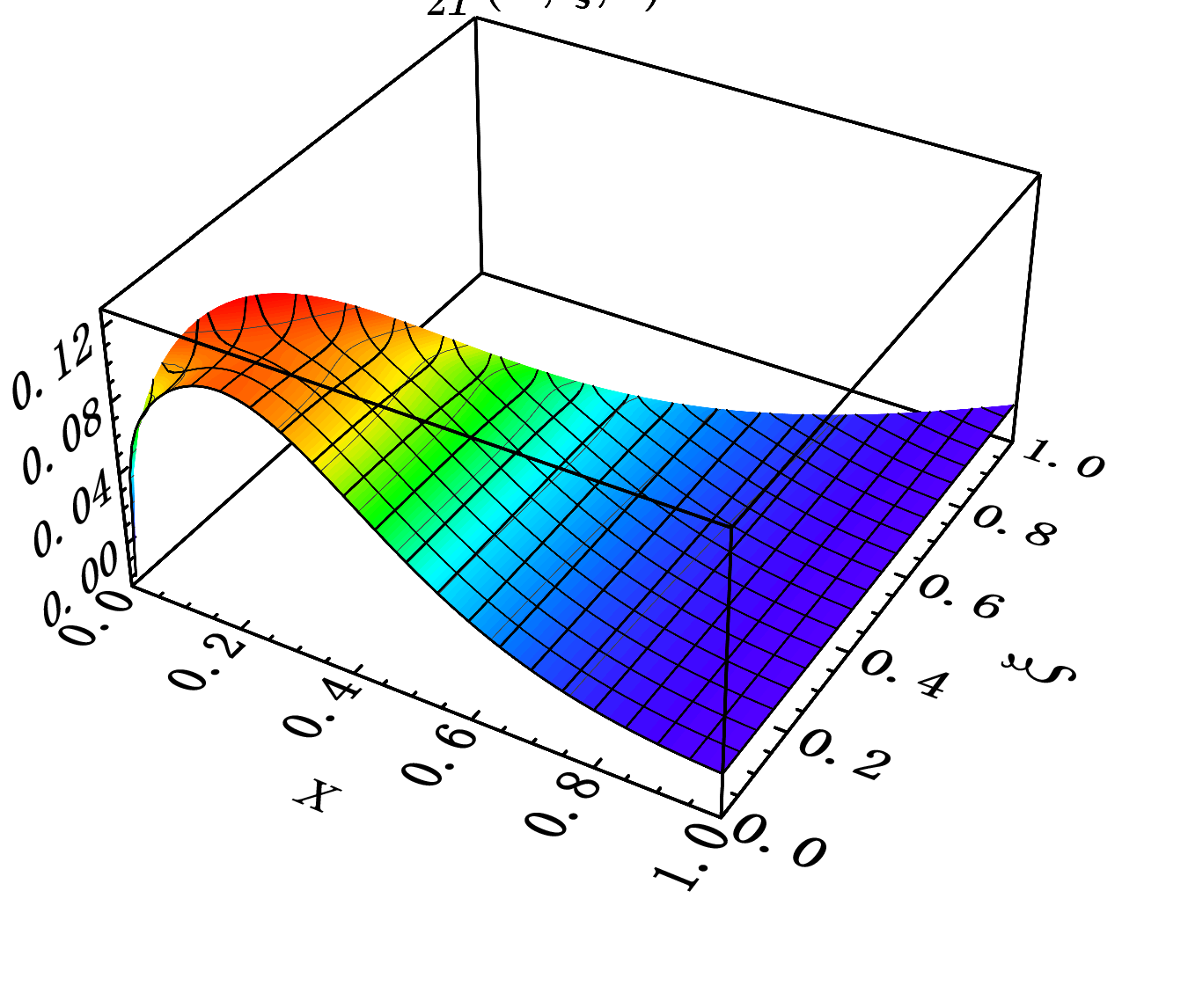}
	\includegraphics[width=0.24\columnwidth]{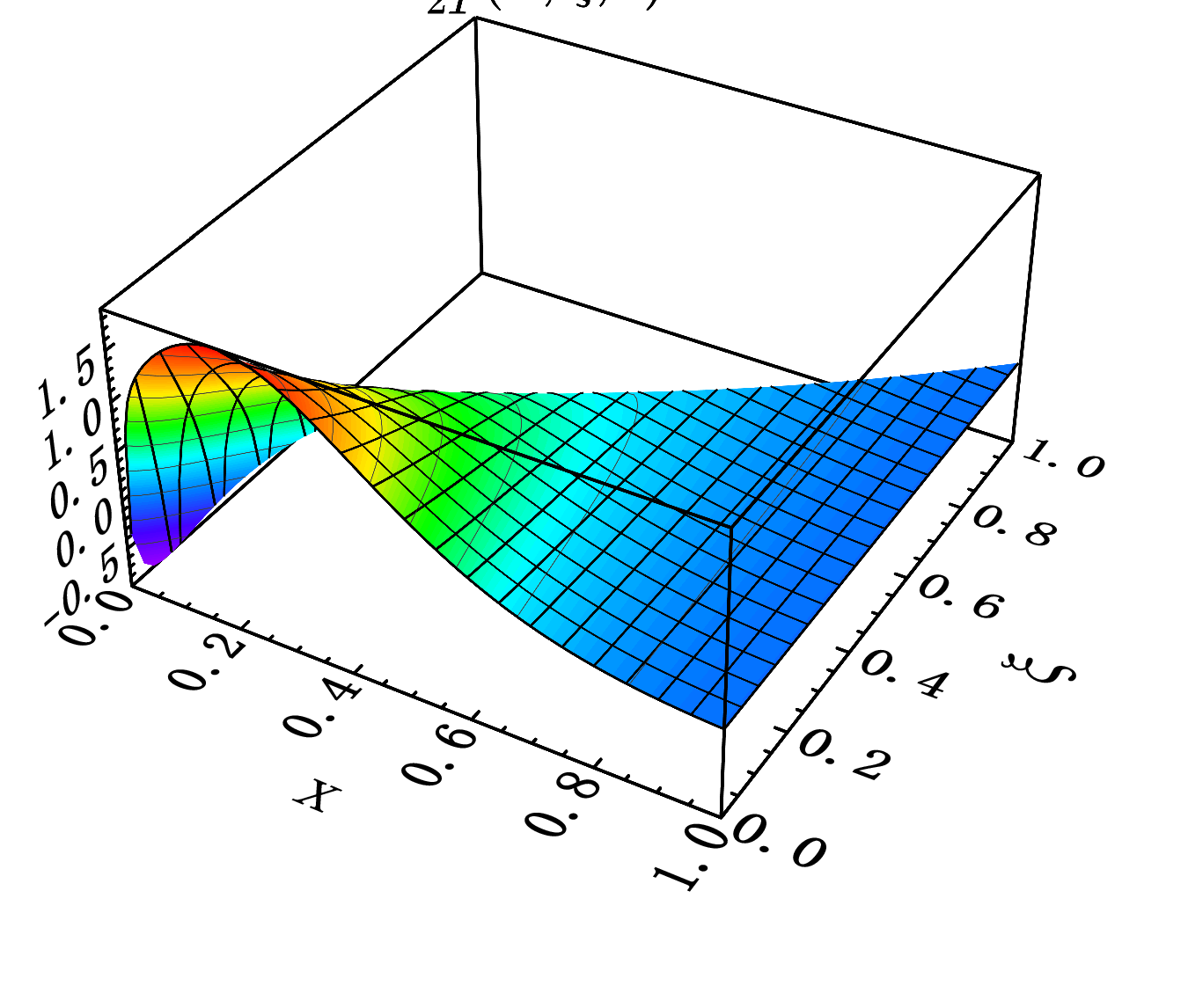}
	\includegraphics[width=0.24\columnwidth]{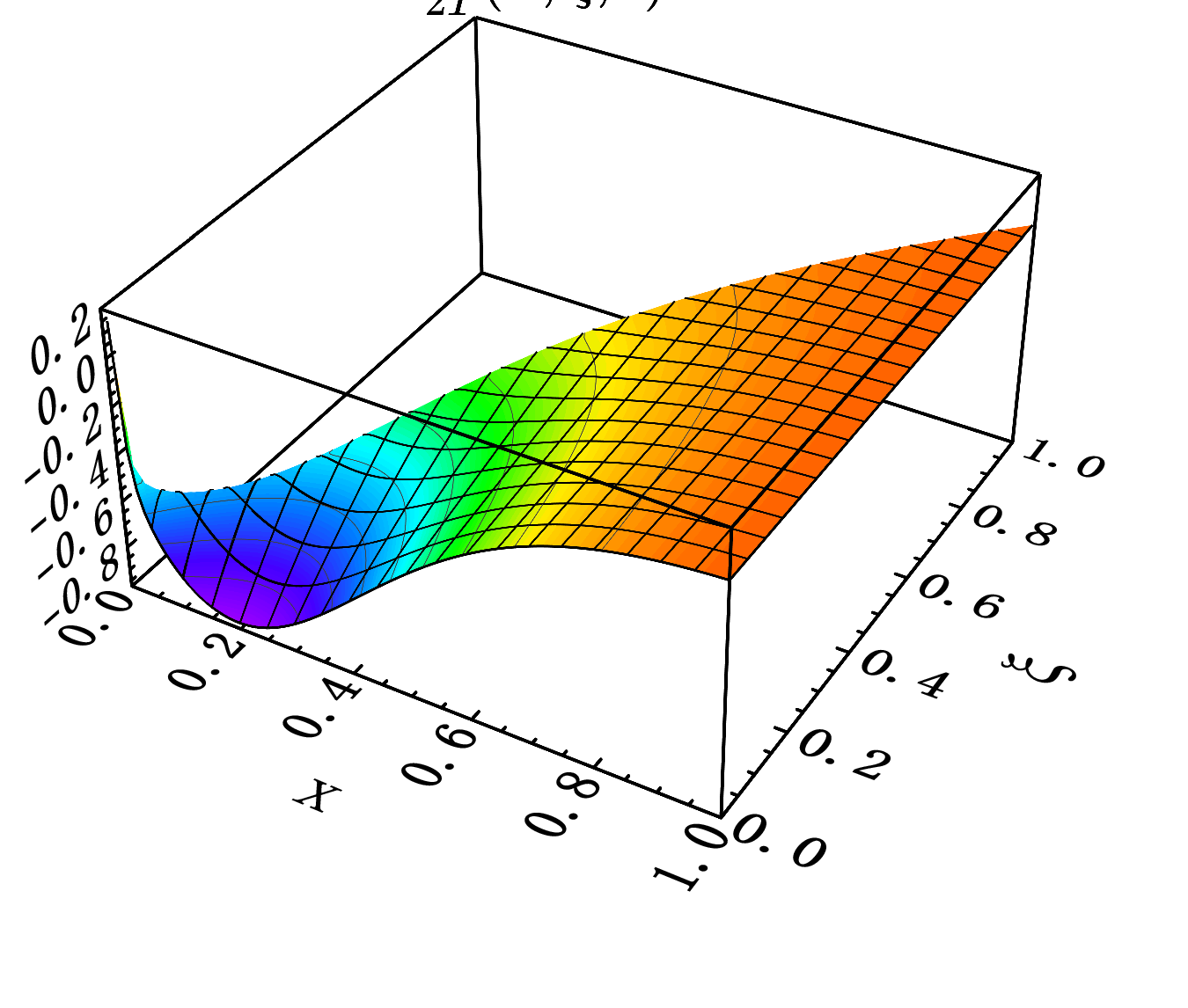}
	\includegraphics[width=0.24\columnwidth]{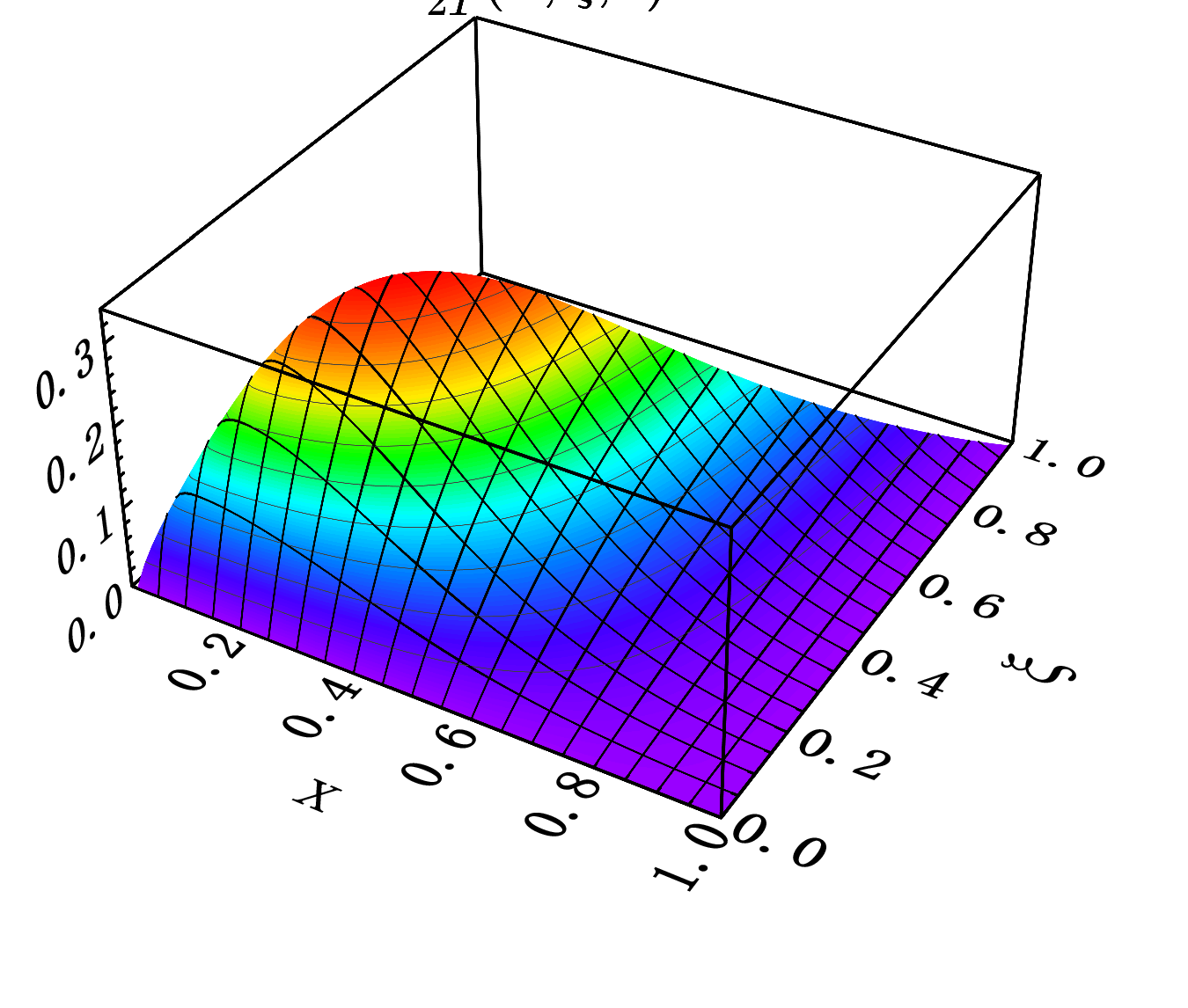}\\
	\includegraphics[width=0.24\columnwidth]{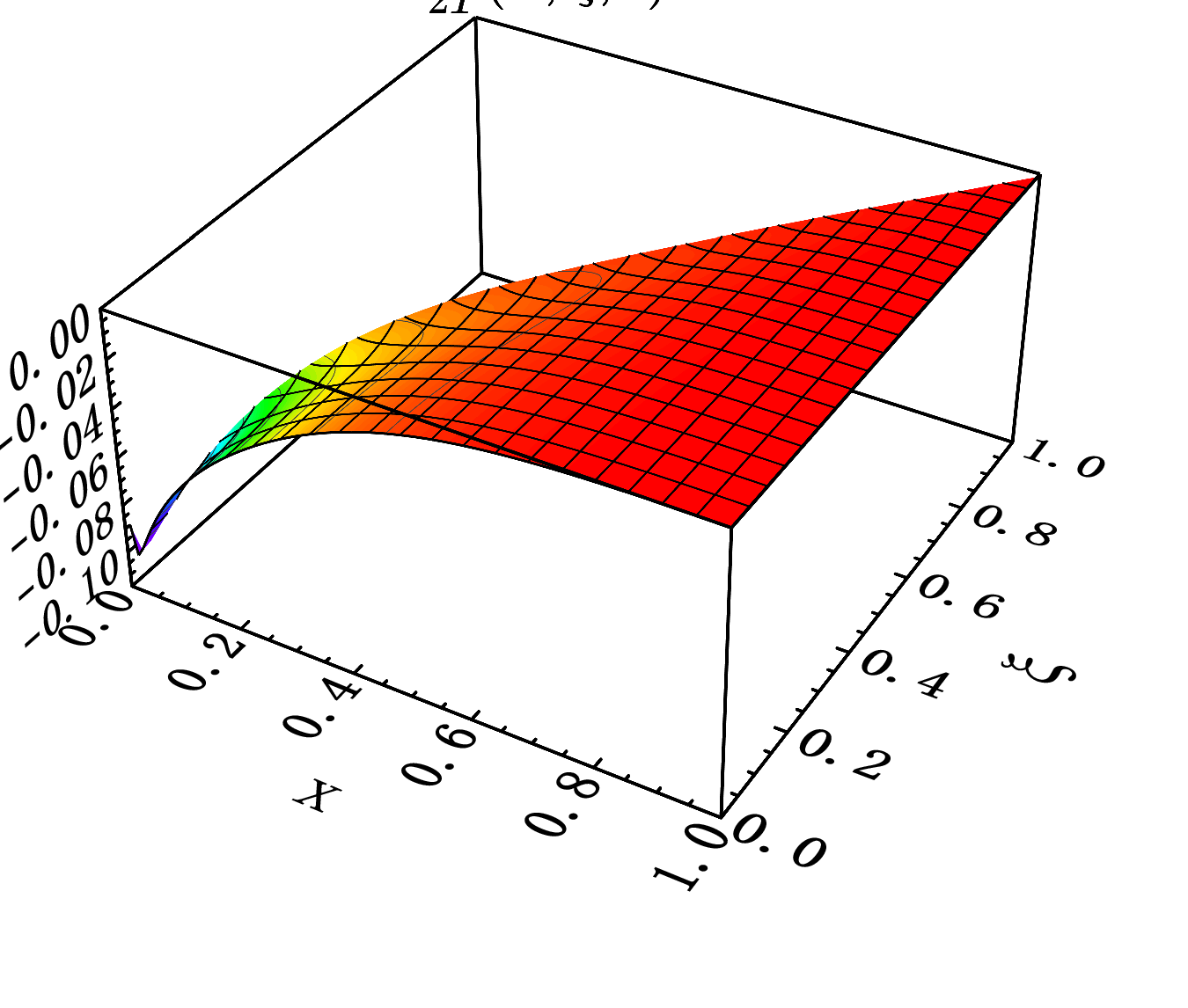}
	\includegraphics[width=0.24\columnwidth]{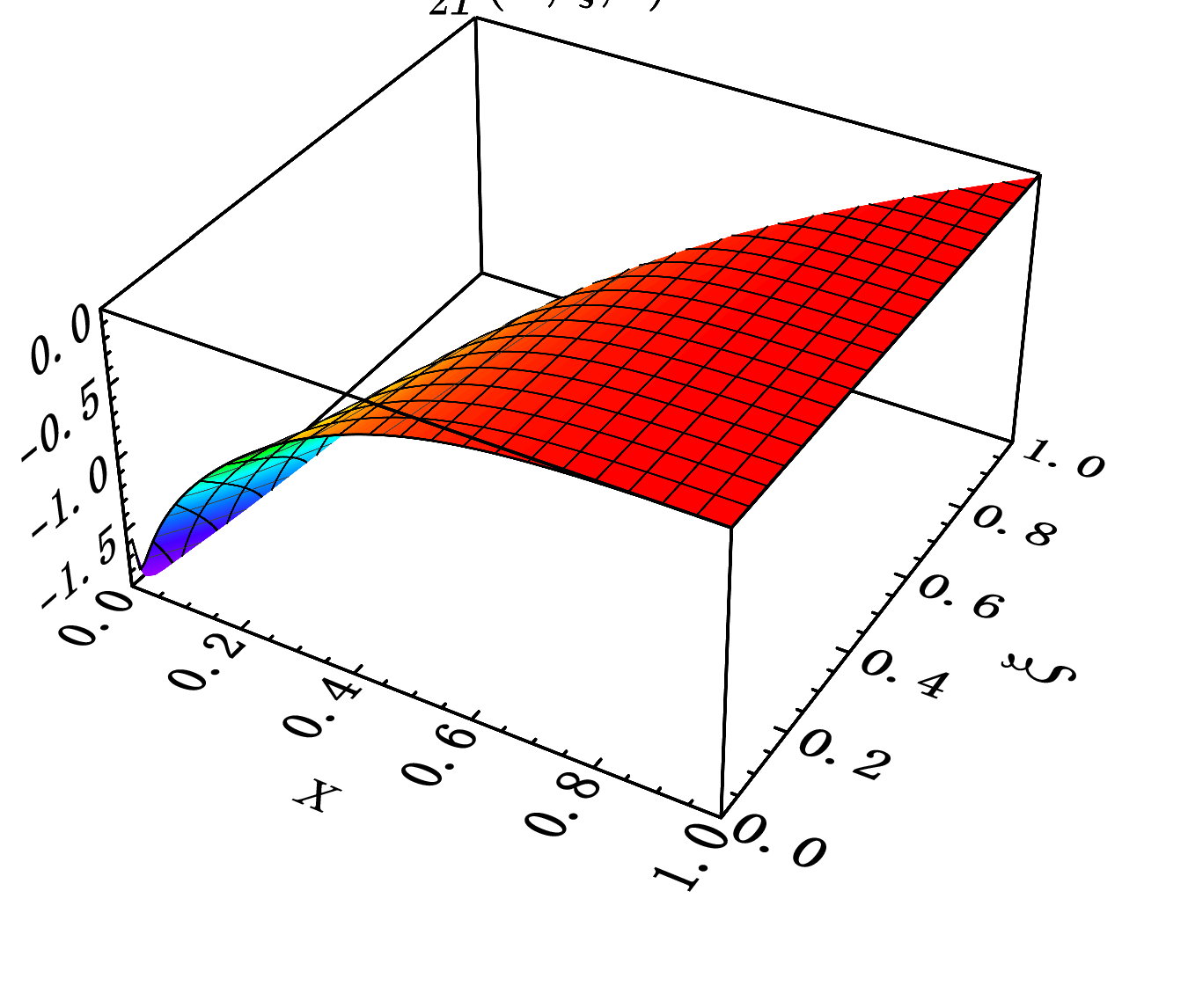}
	\includegraphics[width=0.24\columnwidth]{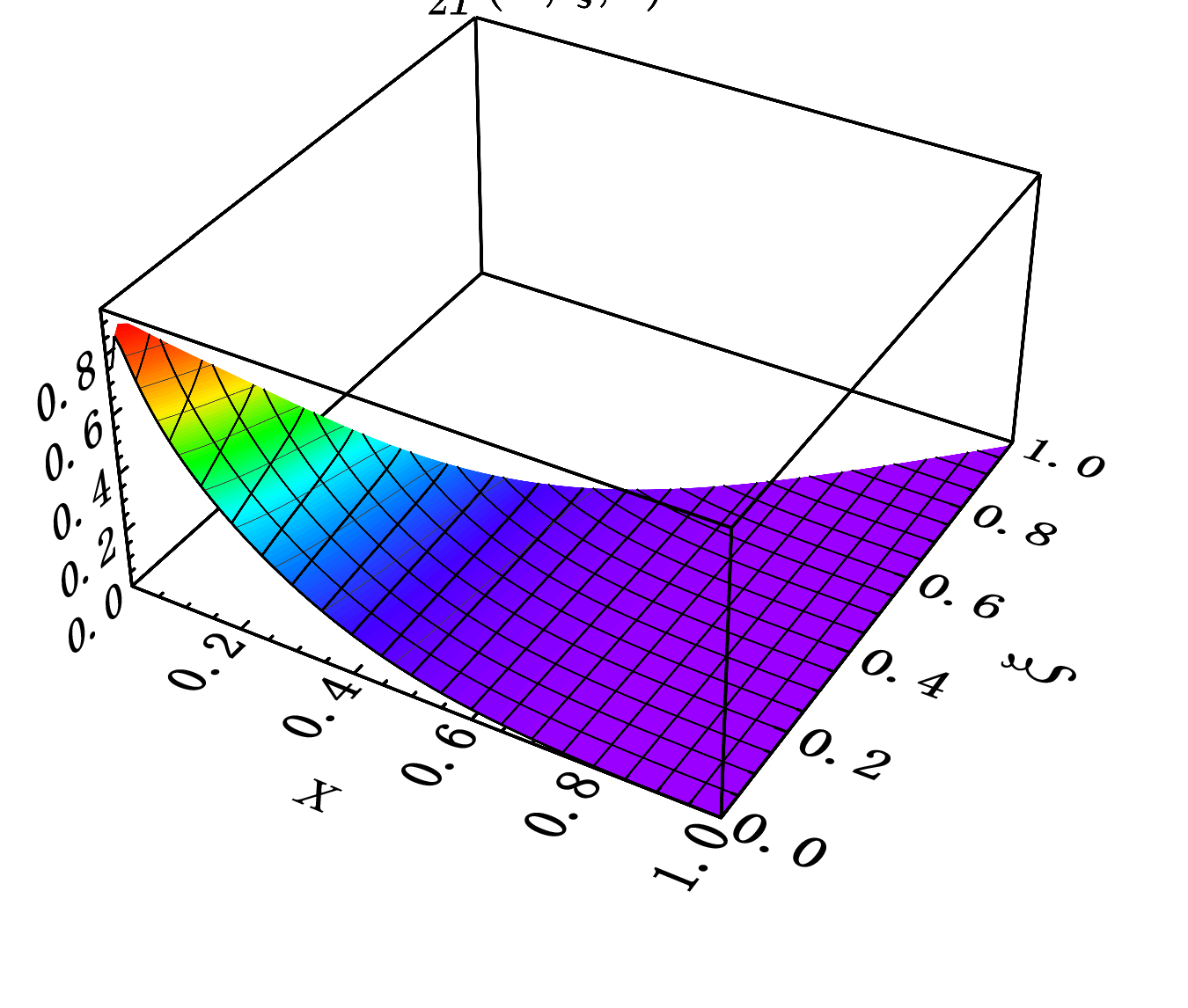}
	\includegraphics[width=0.24\columnwidth]{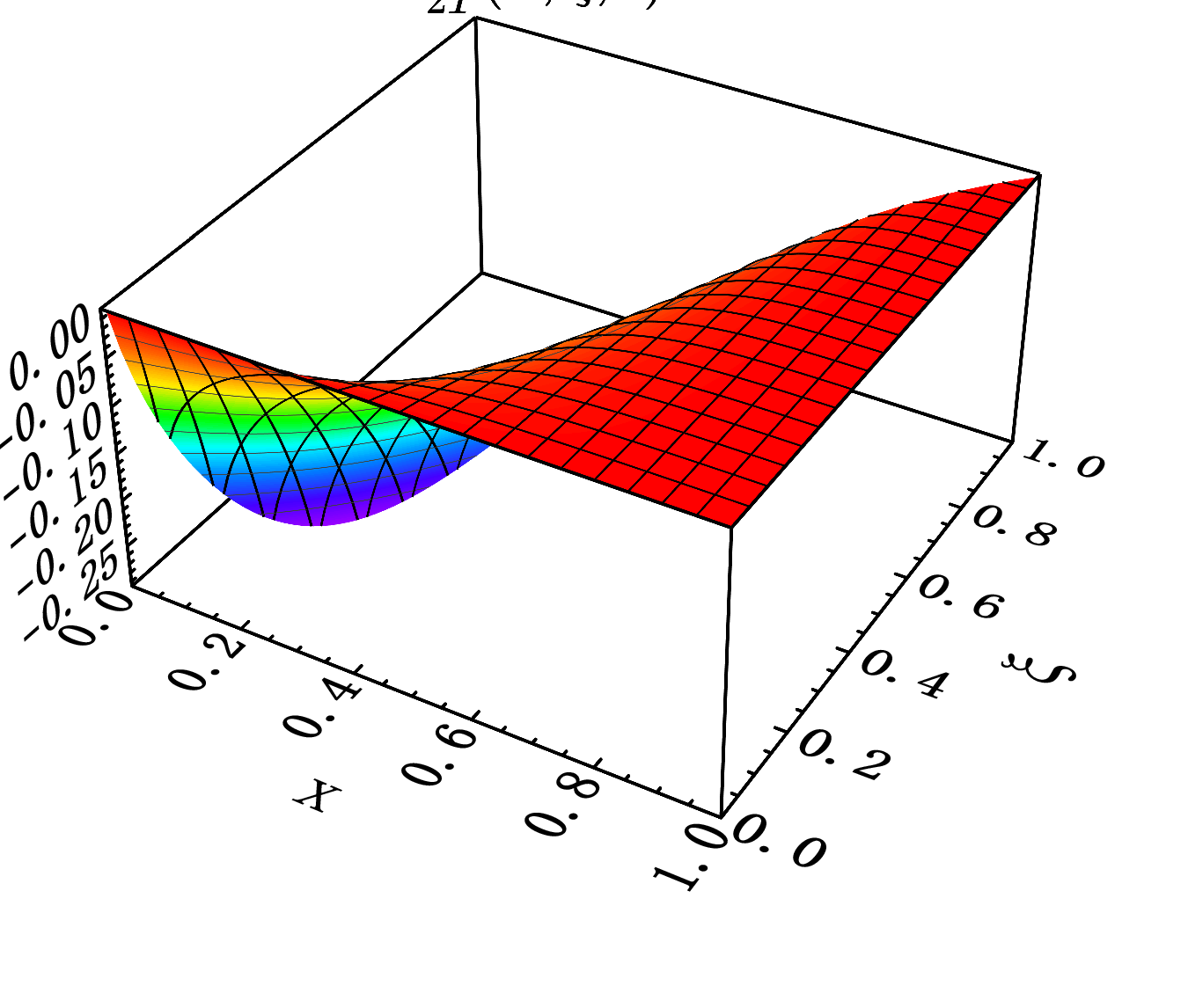}\\
	\includegraphics[width=0.24\columnwidth]{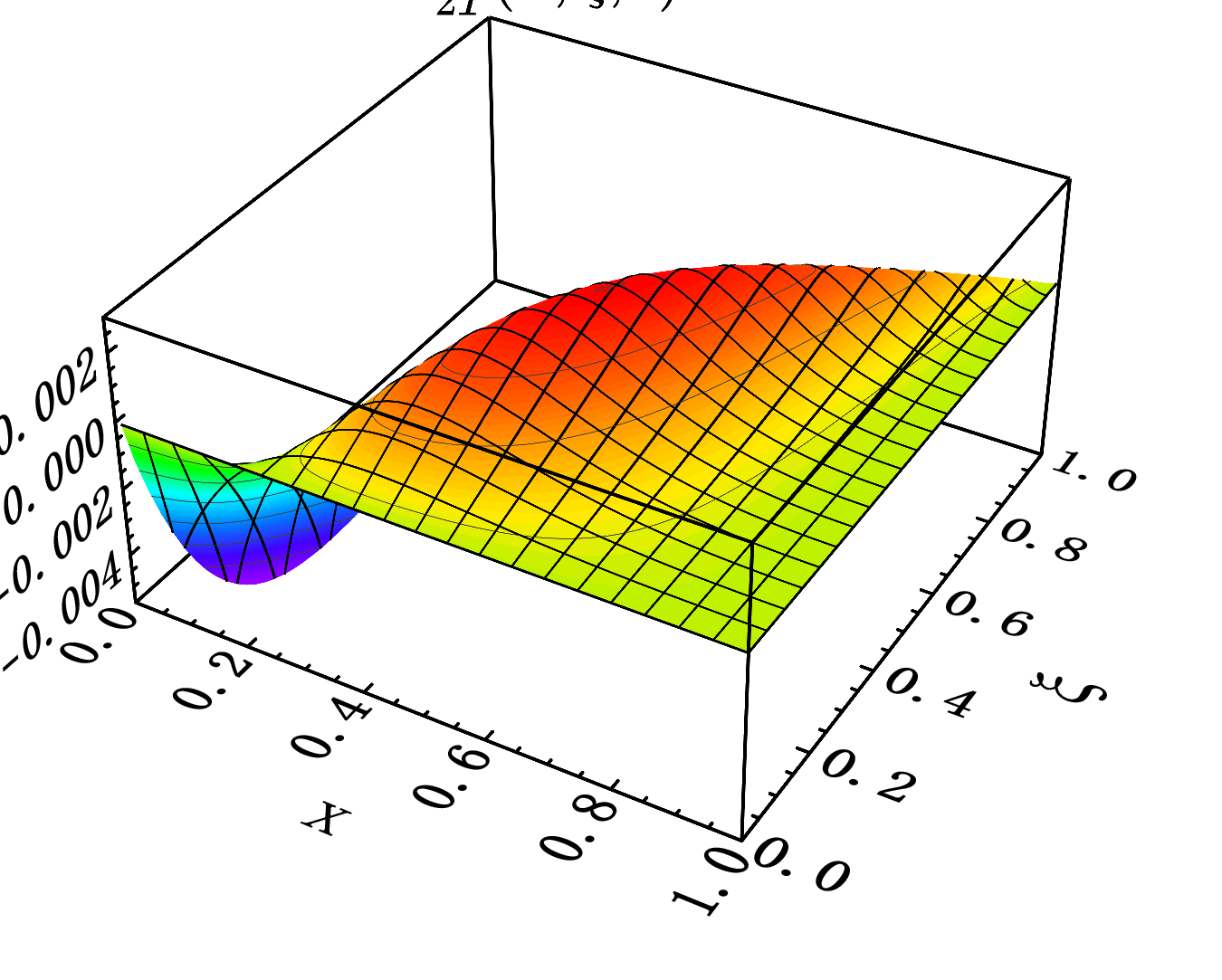}
	\includegraphics[width=0.24\columnwidth]{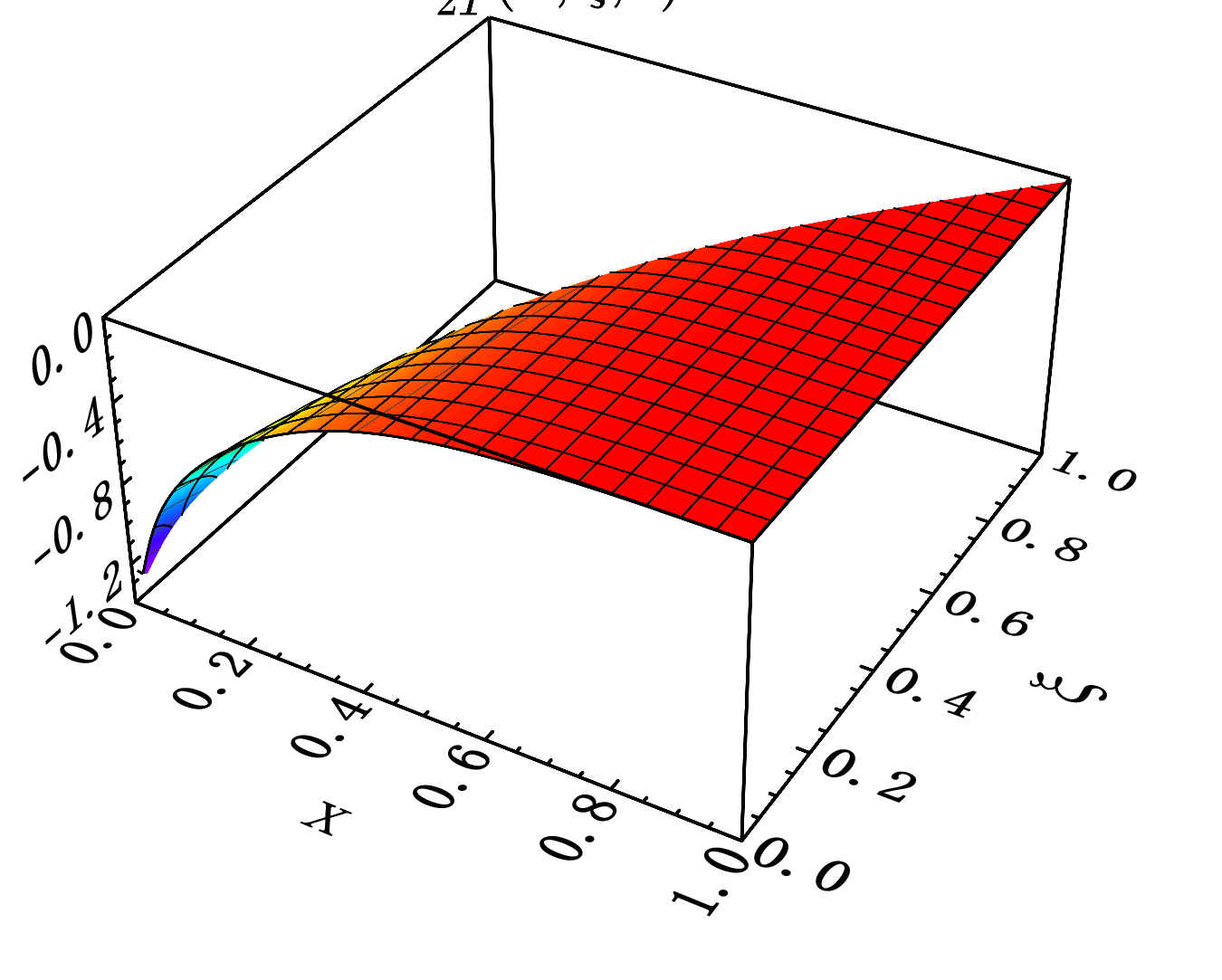}
	\includegraphics[width=0.24\columnwidth]{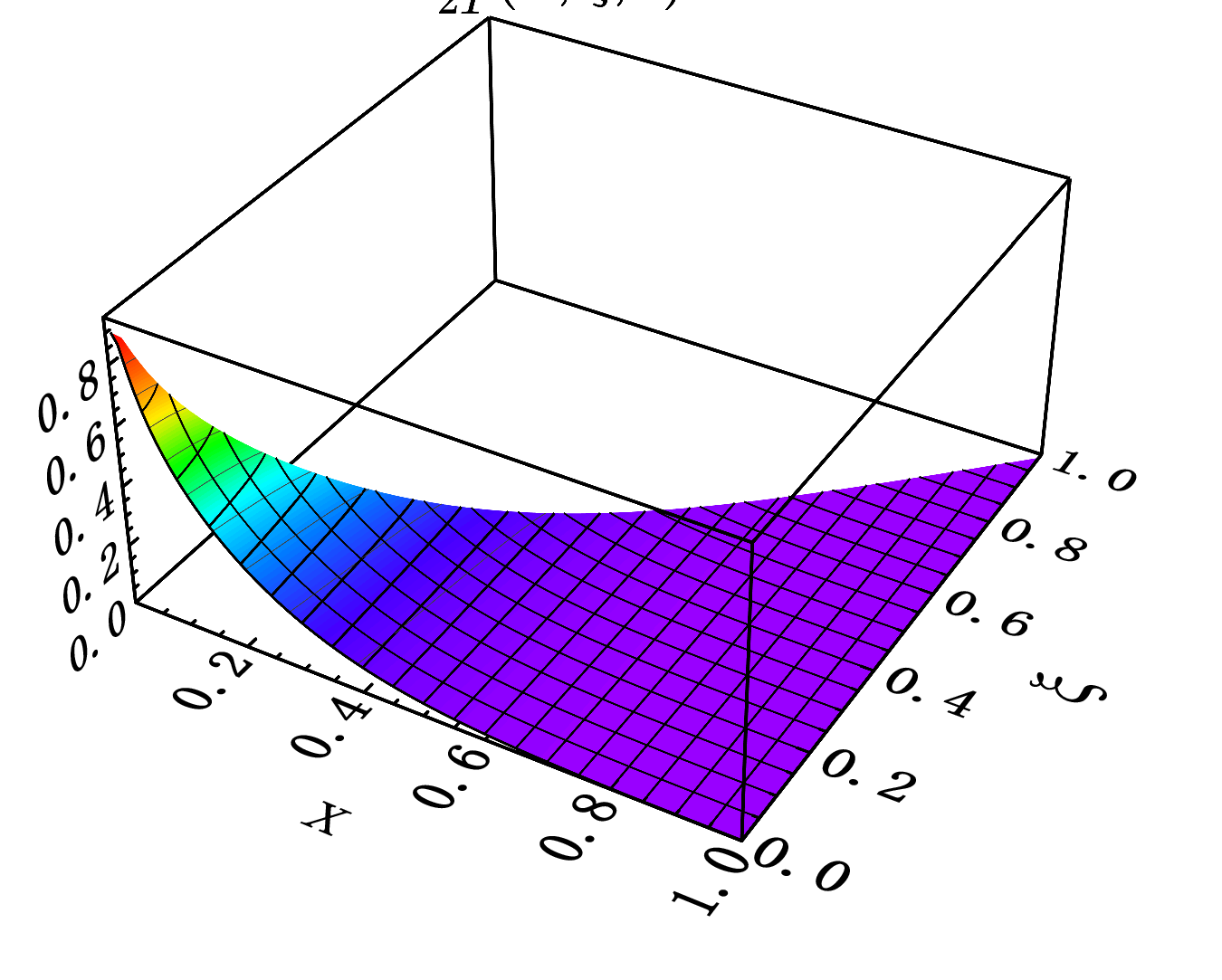}
	\includegraphics[width=0.24\columnwidth]{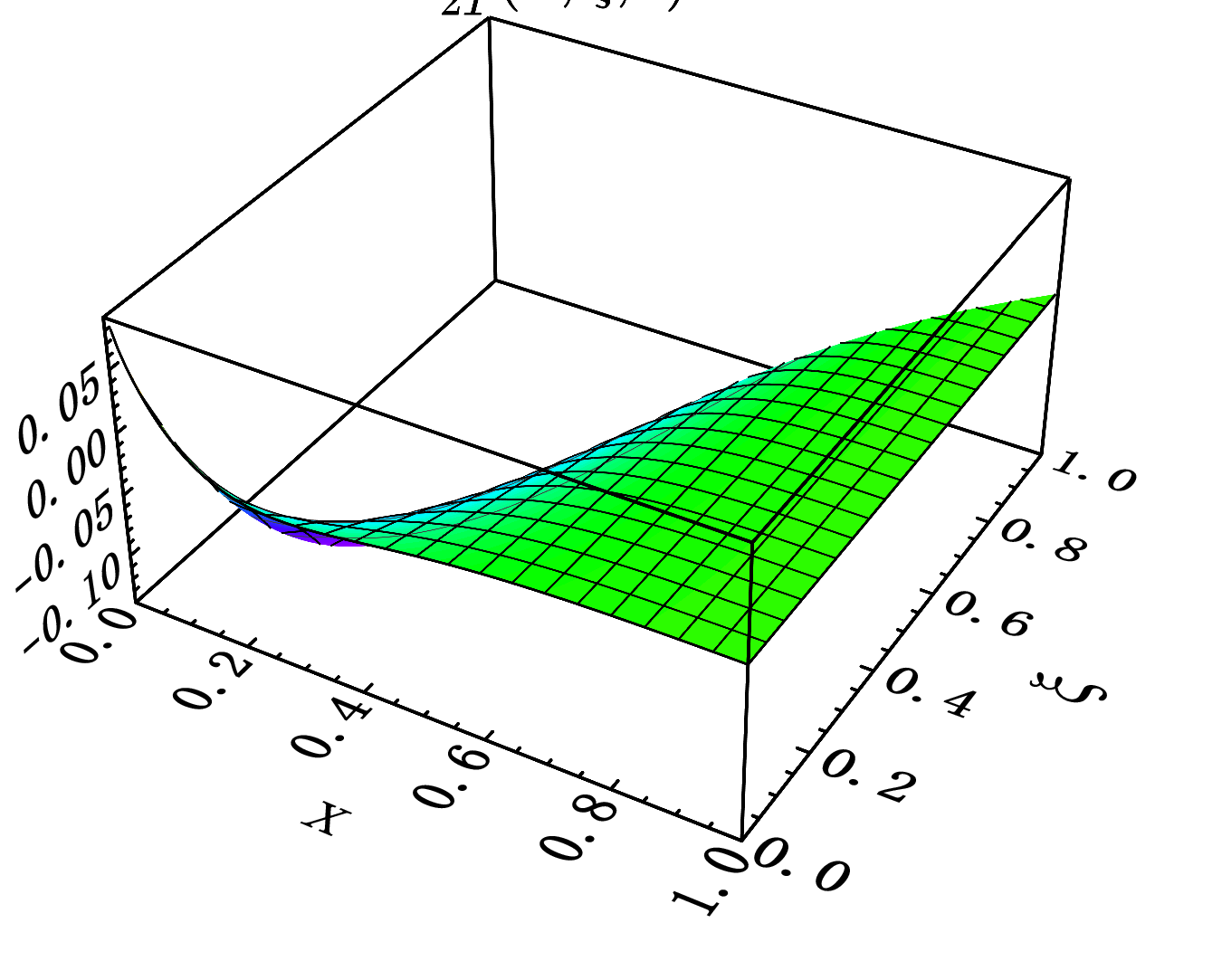}
	\caption{The twist-3 GPDs are plotted with respect to $x$ and $\xi$ in the kinematic range $x\in (\xi,1]$ and $\xi\in [0,1]$ at fixed $\bm{\Delta}_\perp^2=0.2\,\text{GeV}^2$.}
	\label{fig:GPD3}
\end{figure}

Fig.~\ref{fig:GPD3} presents the corresponding results in the DGLAP region, $\xi<x\leq1$. The GPDs exhibit the same general behavior as in the ERBL region as $x\to1$ and remain predominantly concentrated at small $x$. The model results exhibit a significant change across $x=\xi$, with the magnitudes generally larger on the ERBL side, similar to the quark case~\cite{Tan:2024doz}. Most distributions occur in sign-opposite pairs with similar shapes, while $E_{2T}^g$ is numerically enhanced by one to two orders of magnitude relative to the others.

\begin{figure}
	\centering
	\includegraphics[width=0.24\columnwidth]{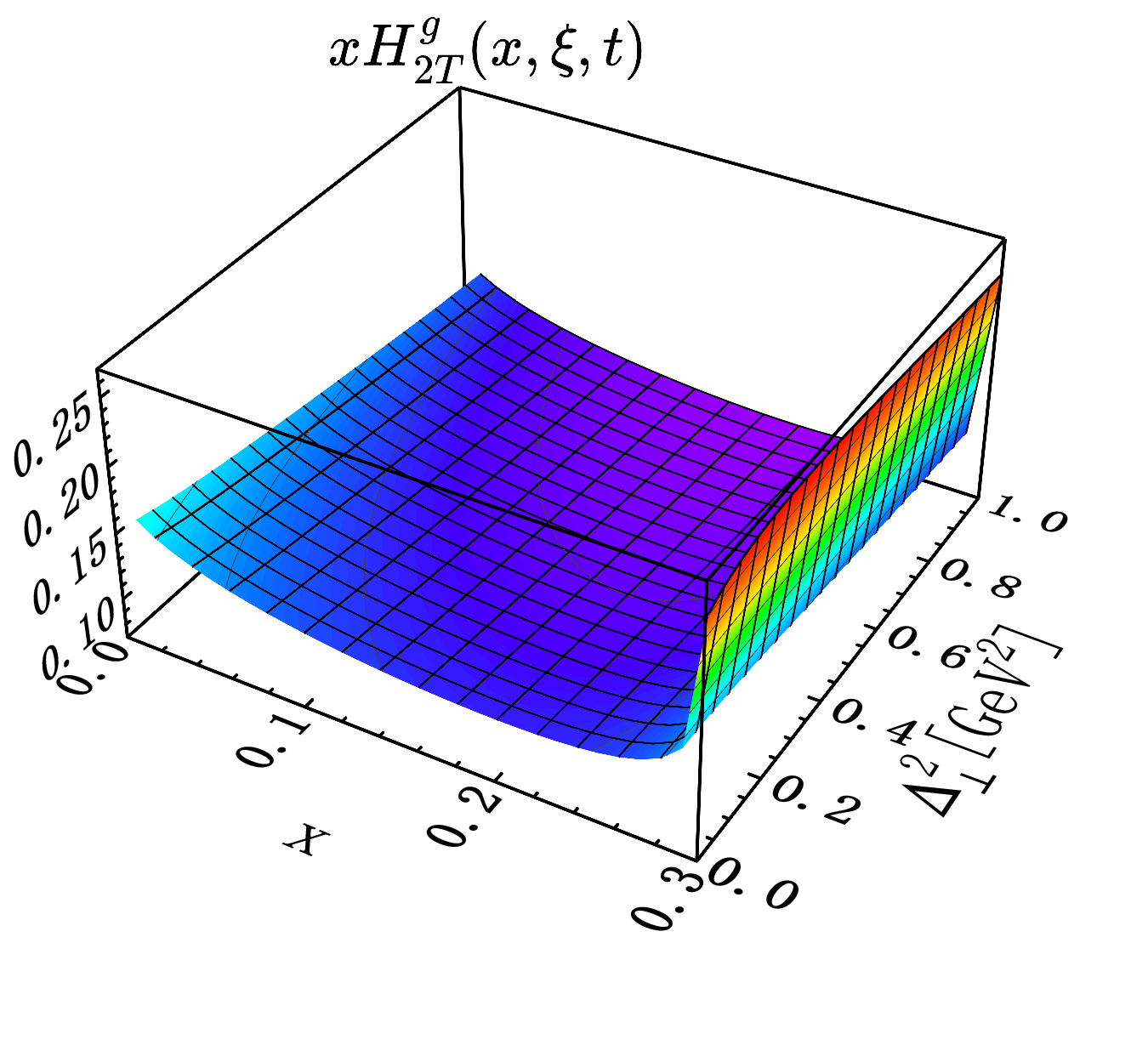}
	\includegraphics[width=0.24\columnwidth]{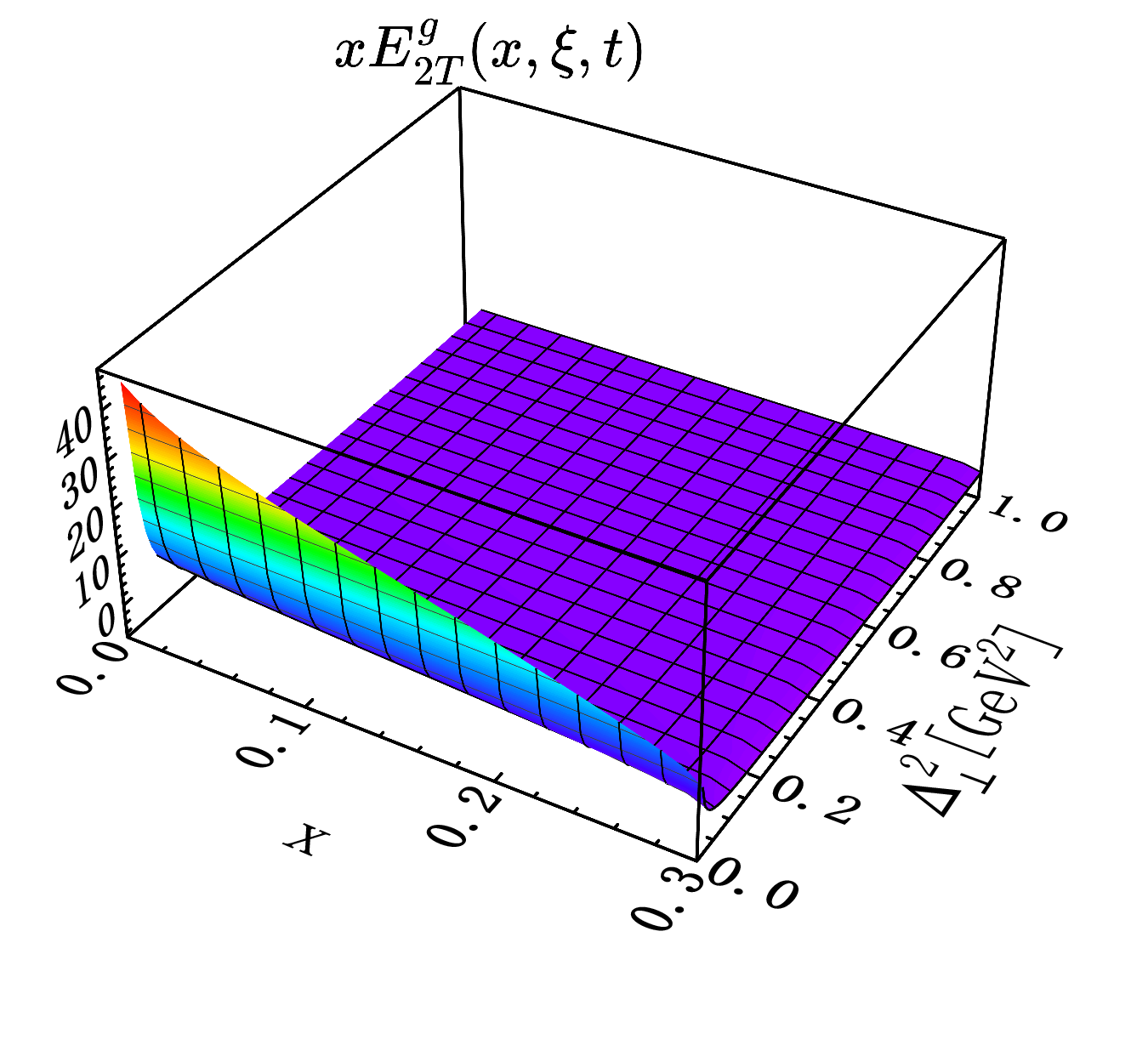}
	\includegraphics[width=0.24\columnwidth]{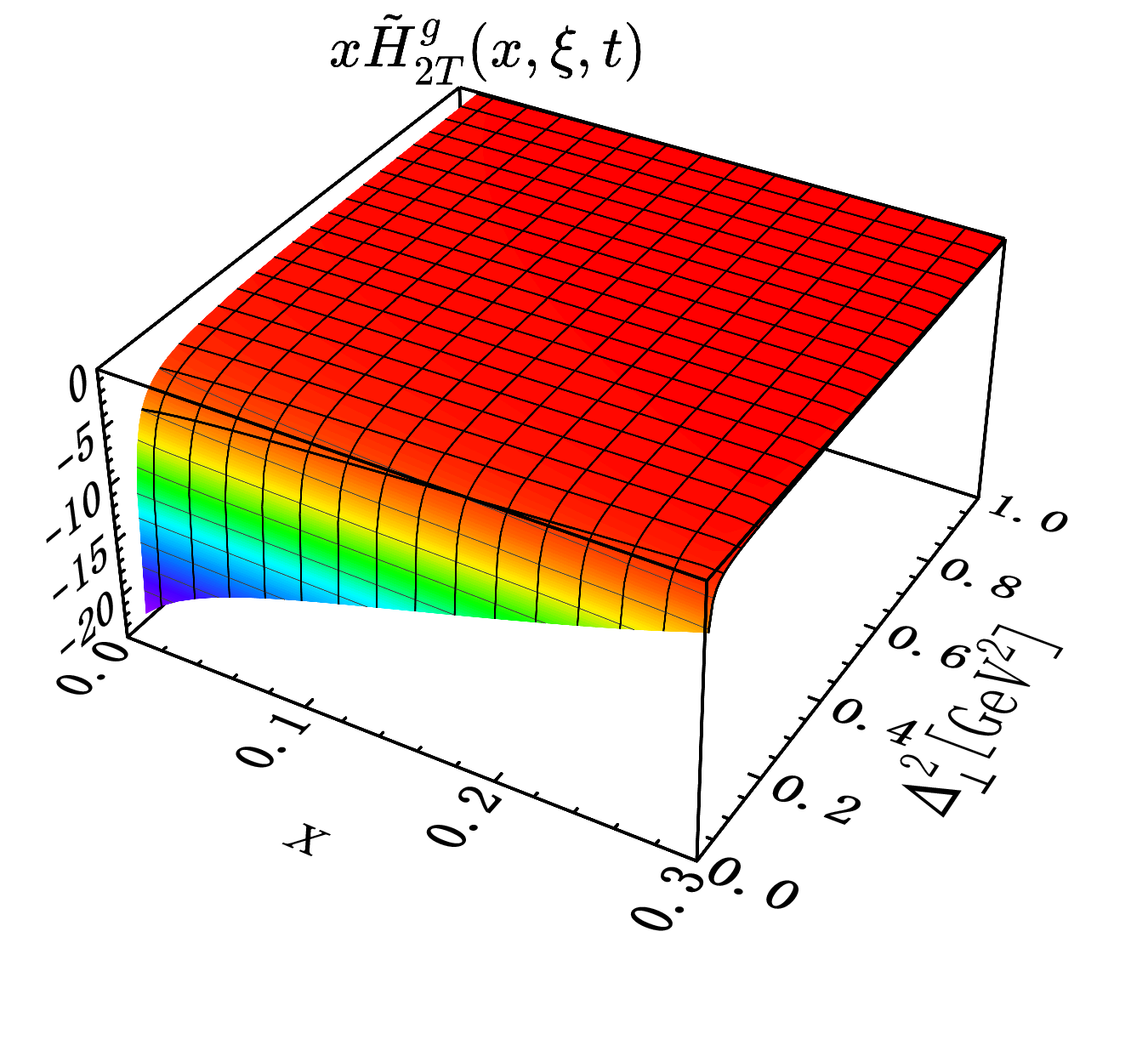}
	\includegraphics[width=0.24\columnwidth]{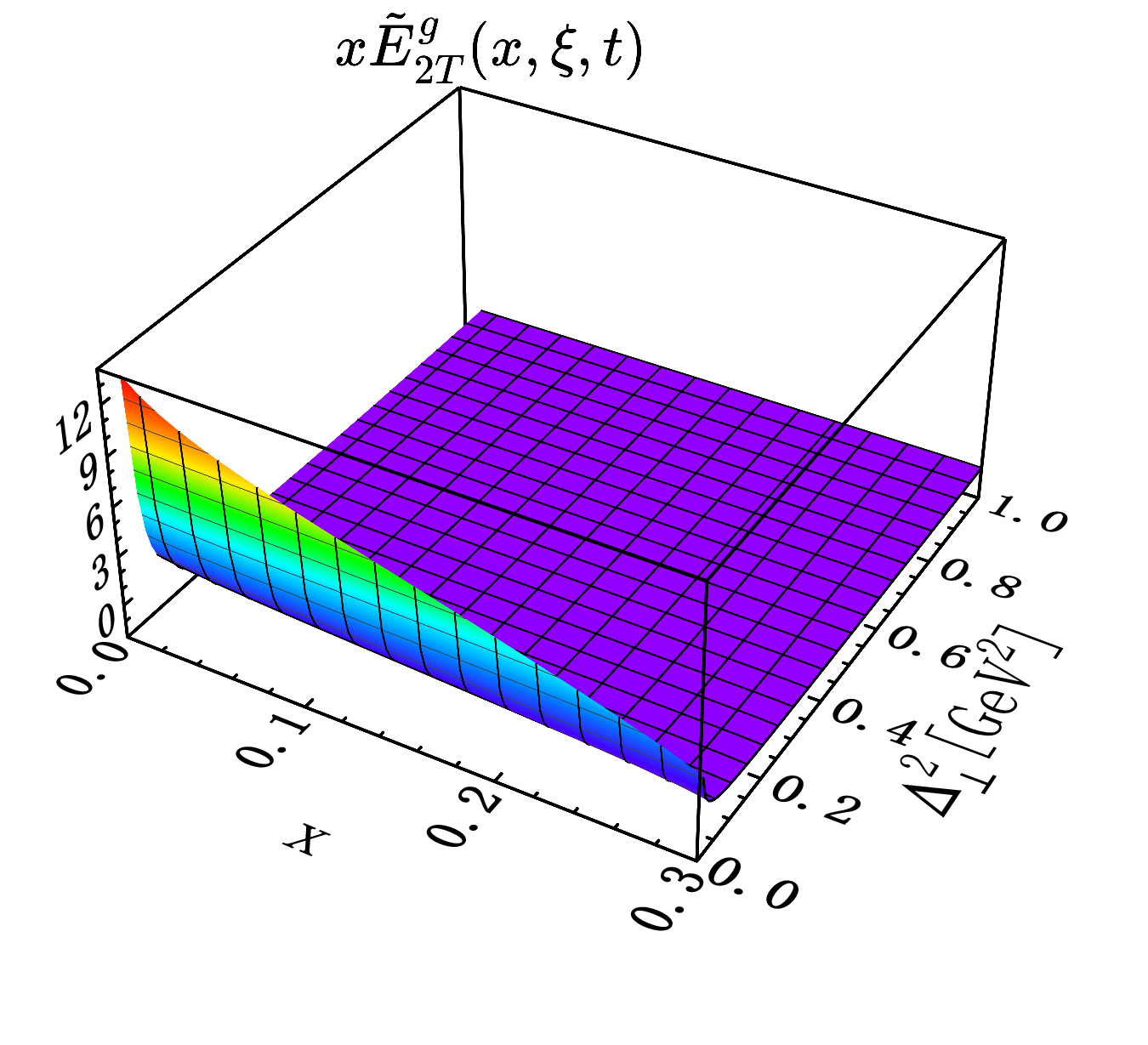}\\
	\includegraphics[width=0.24\columnwidth]{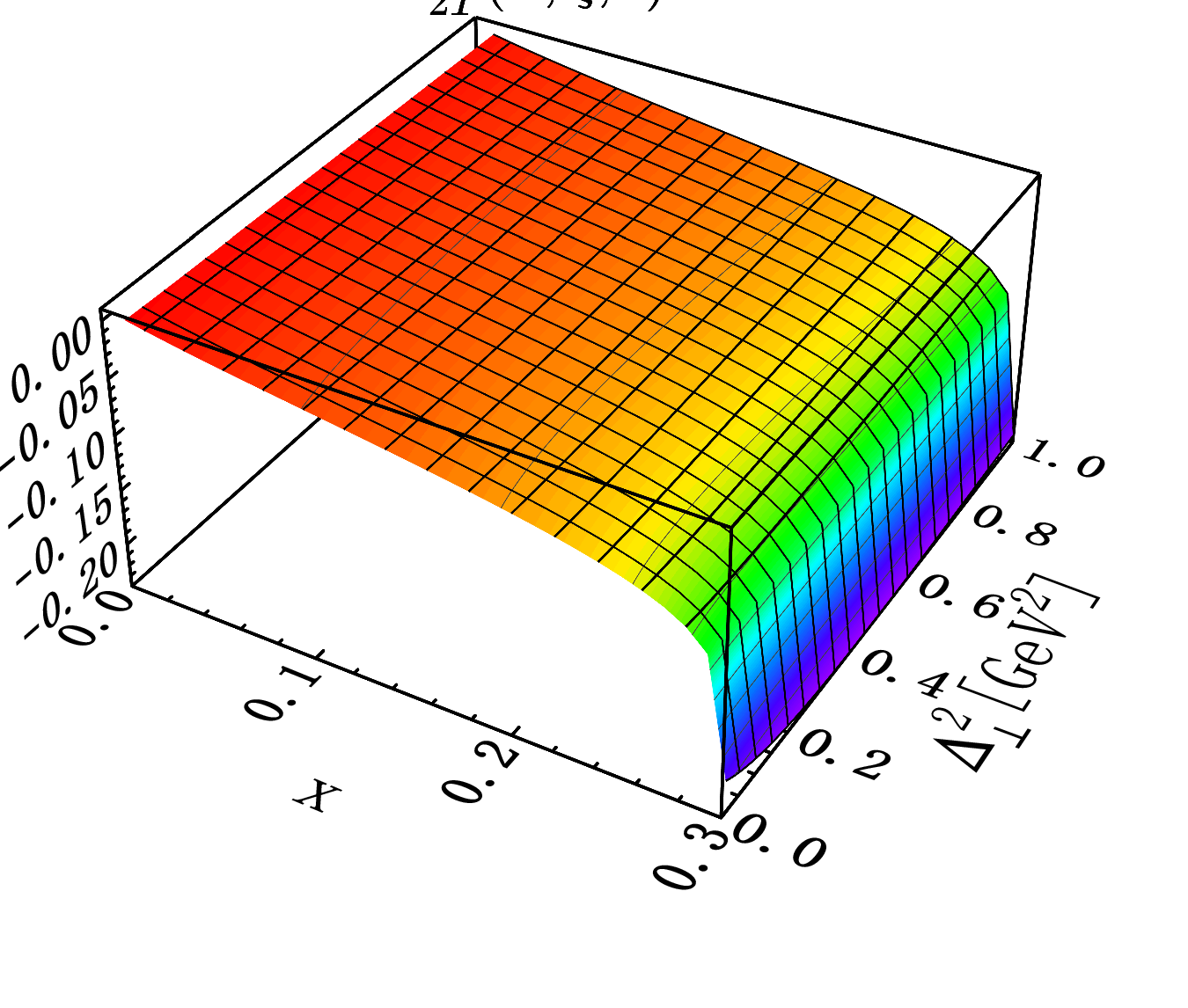}
	\includegraphics[width=0.24\columnwidth]{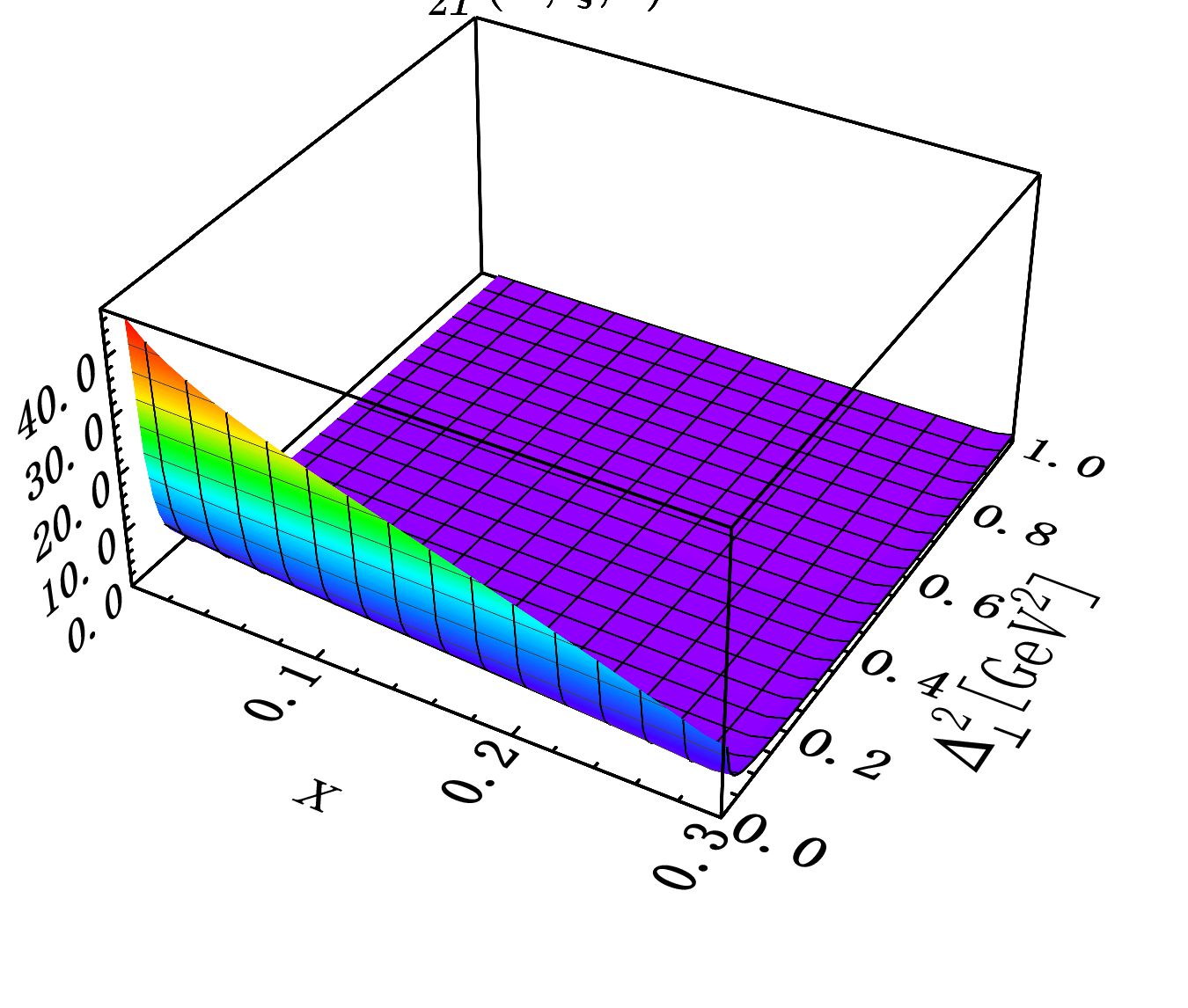}
	\includegraphics[width=0.24\columnwidth]{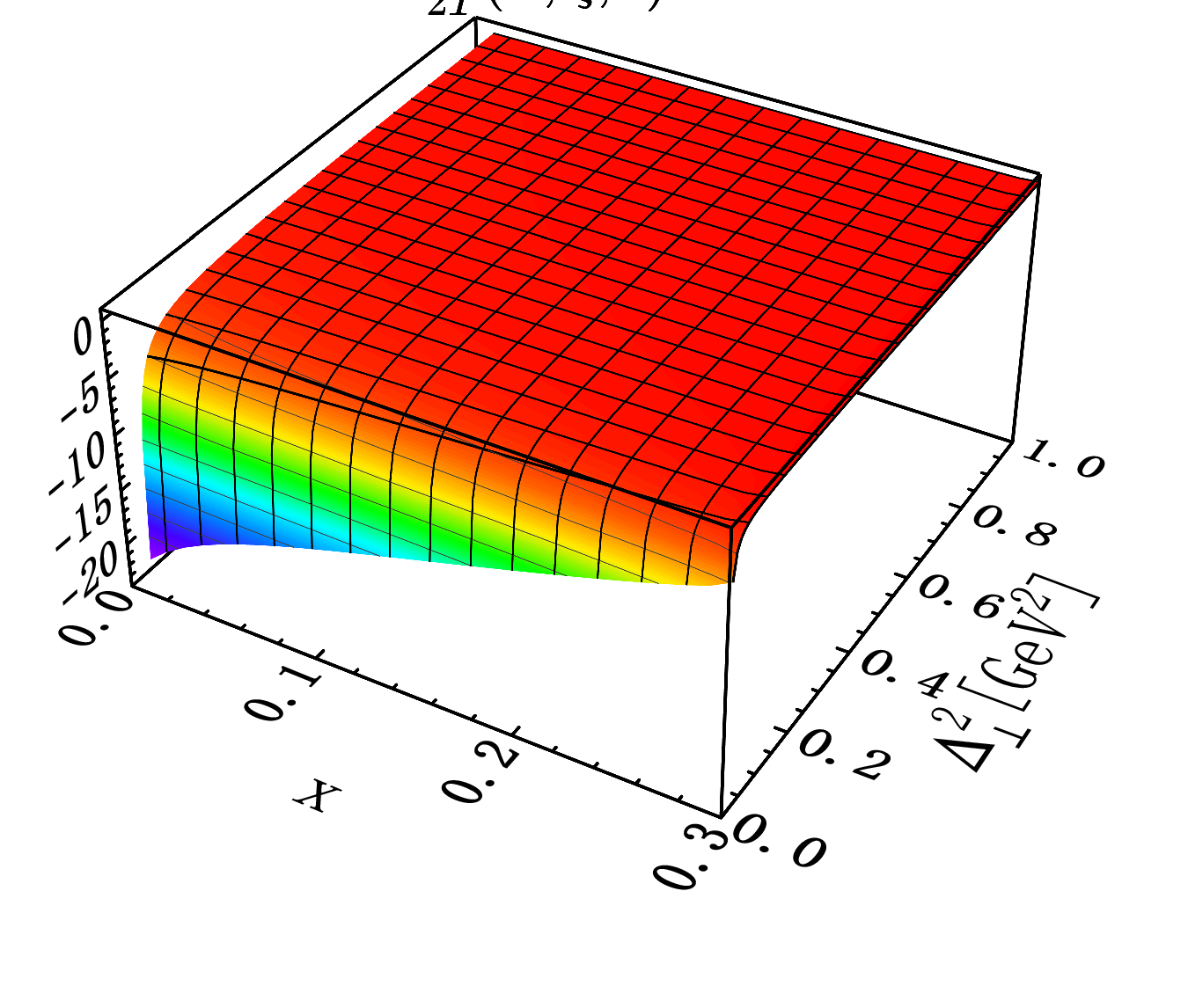}
	\includegraphics[width=0.24\columnwidth]{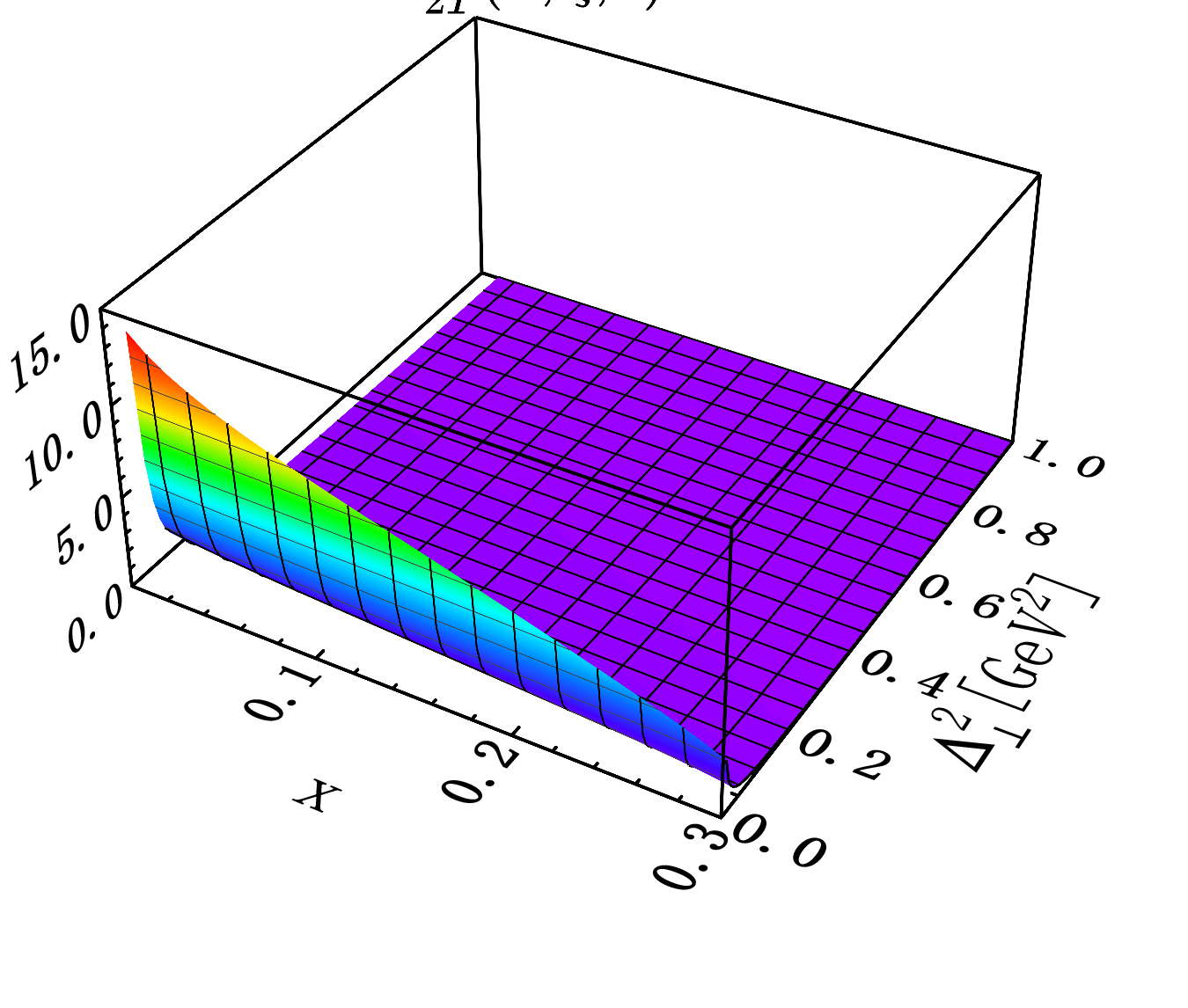}\\
	\includegraphics[width=0.24\columnwidth]{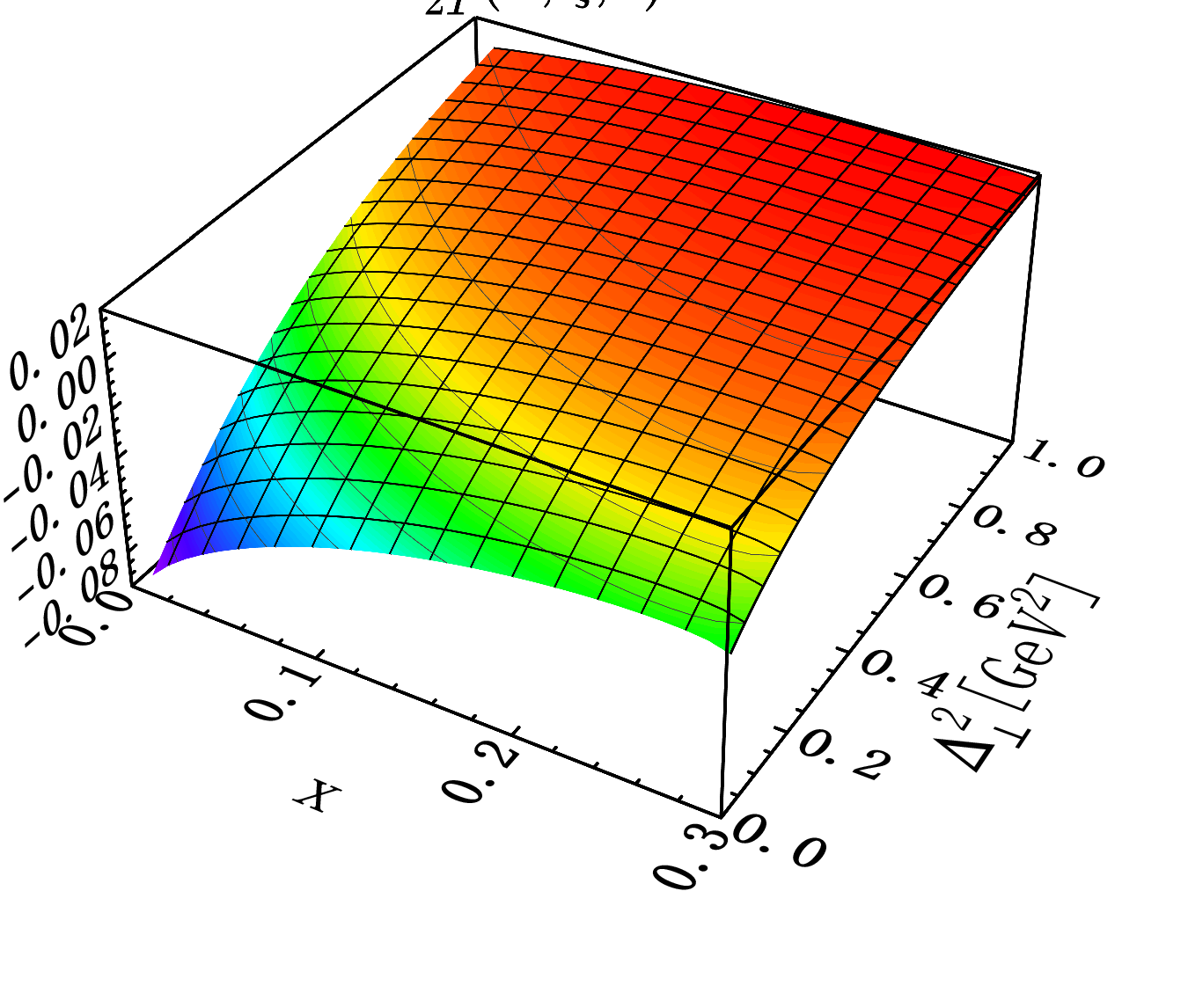}
	\includegraphics[width=0.24\columnwidth]{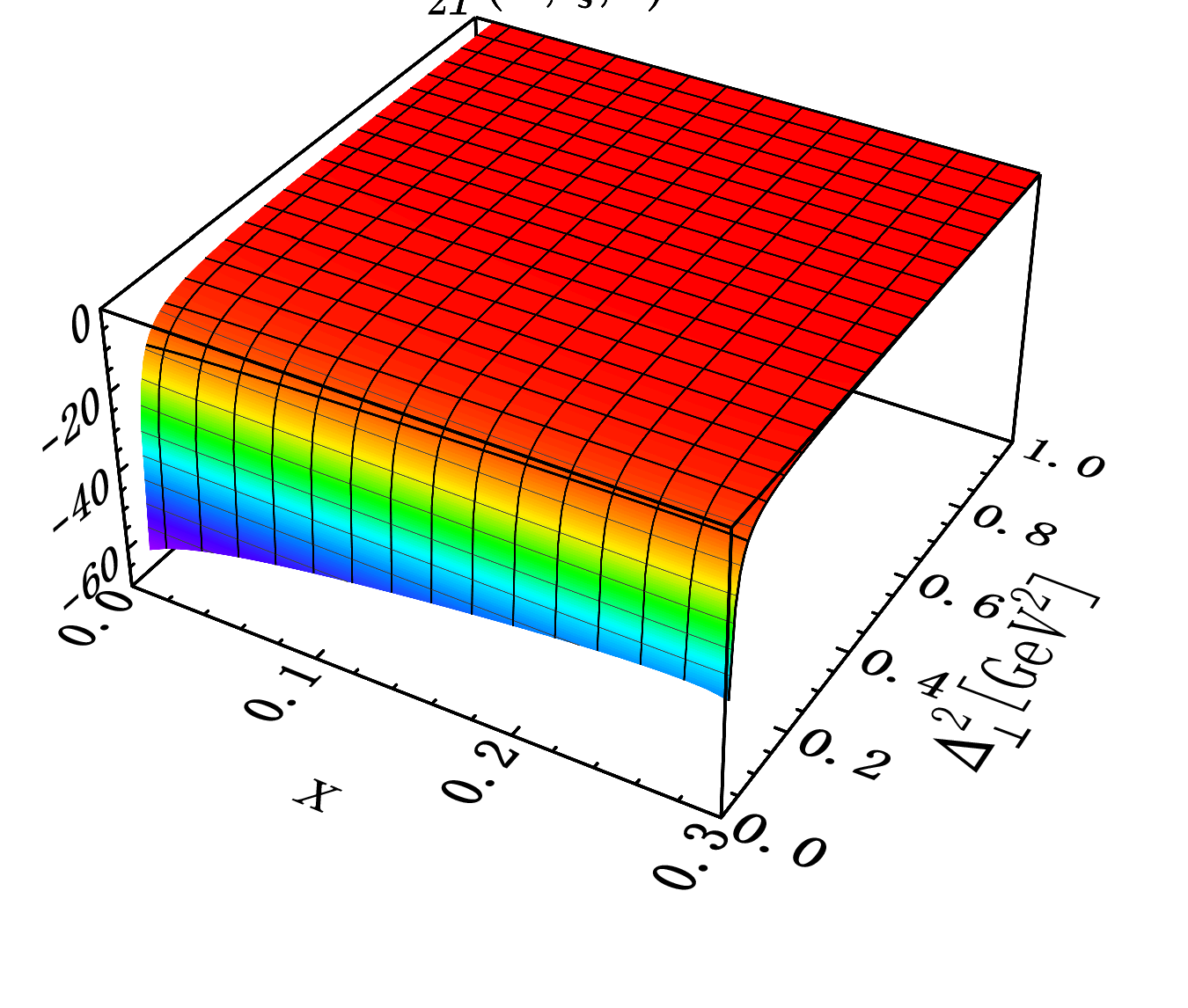}
	\includegraphics[width=0.24\columnwidth]{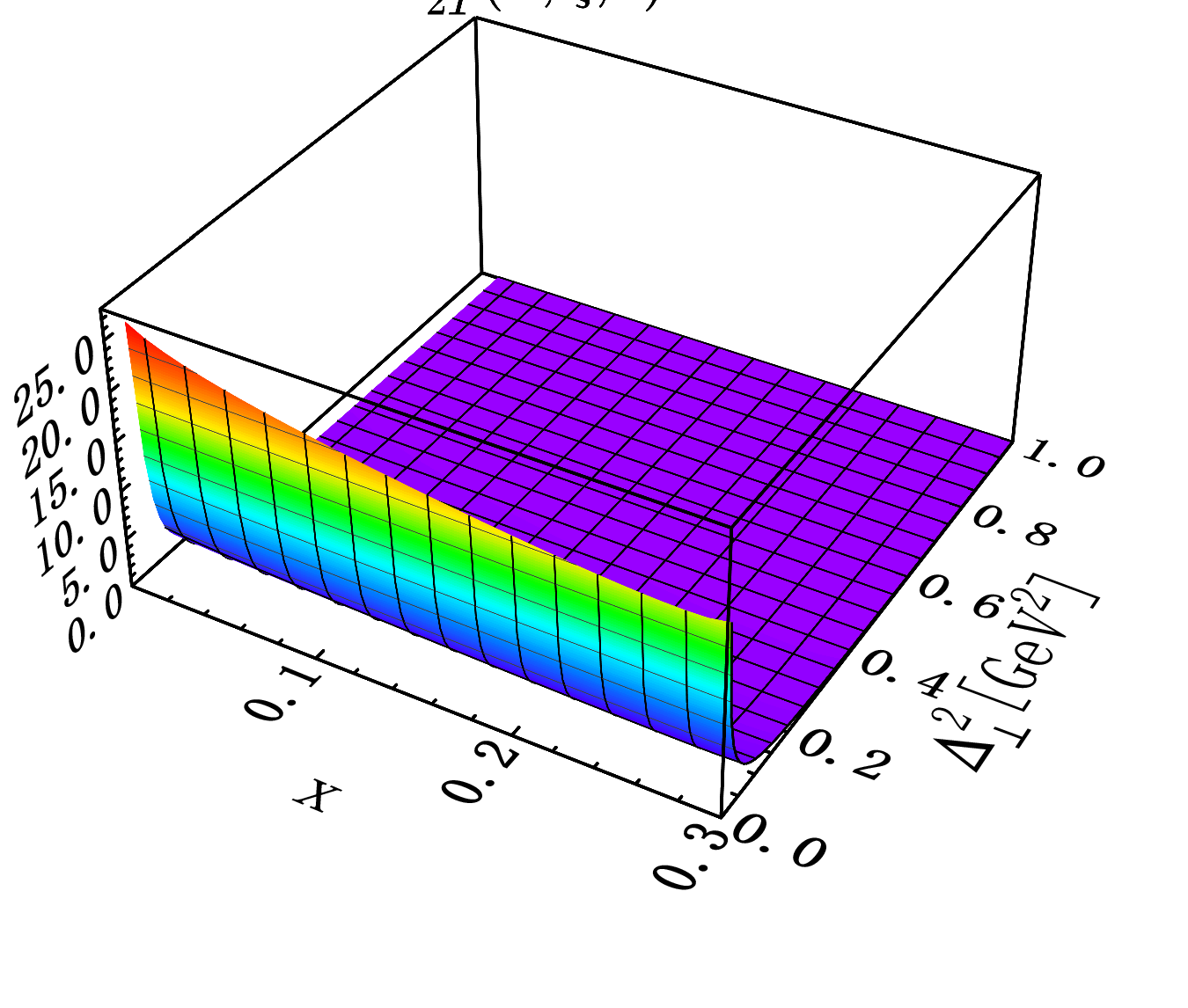}
	\includegraphics[width=0.24\columnwidth]{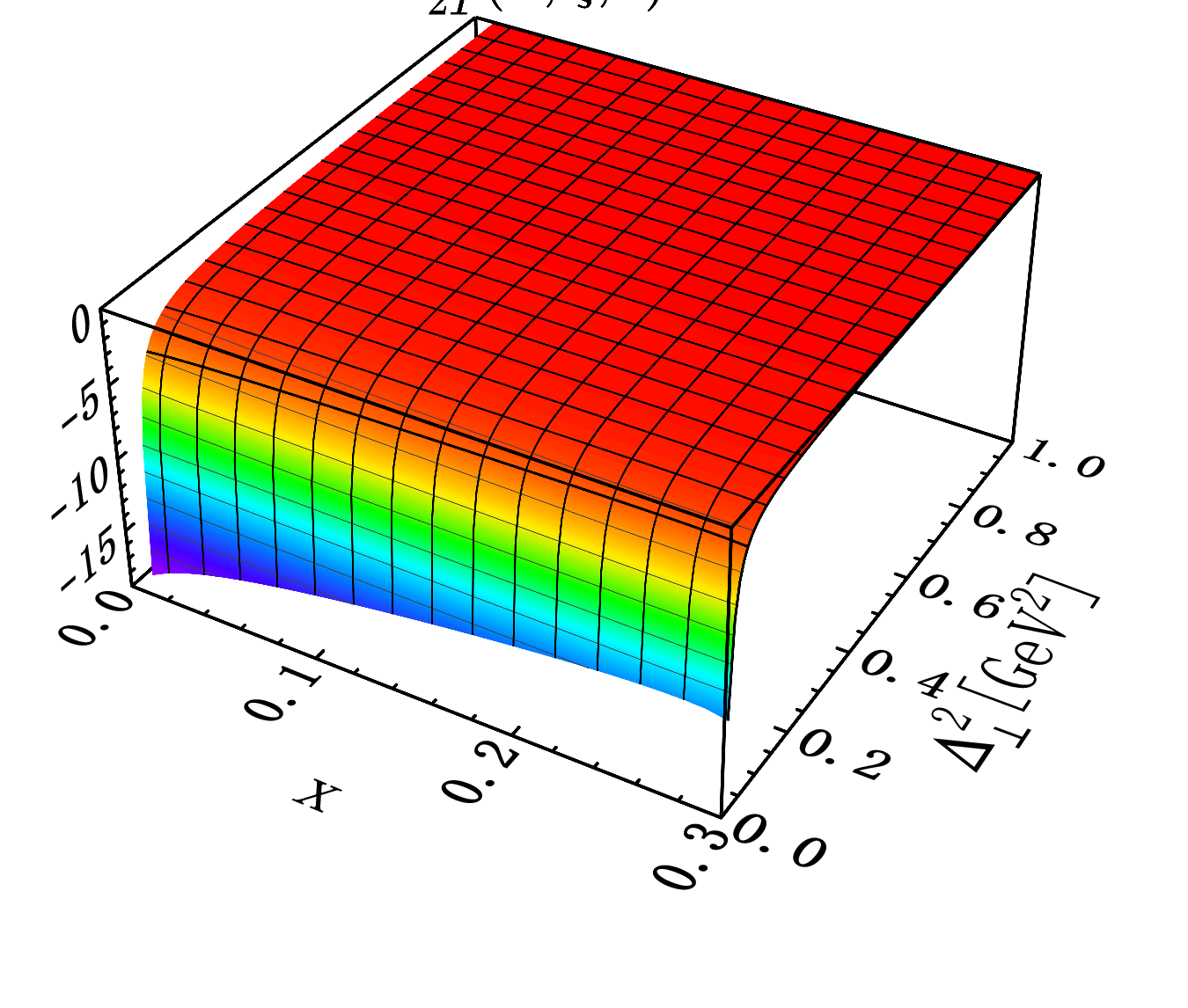}\\
	\includegraphics[width=0.24\columnwidth]{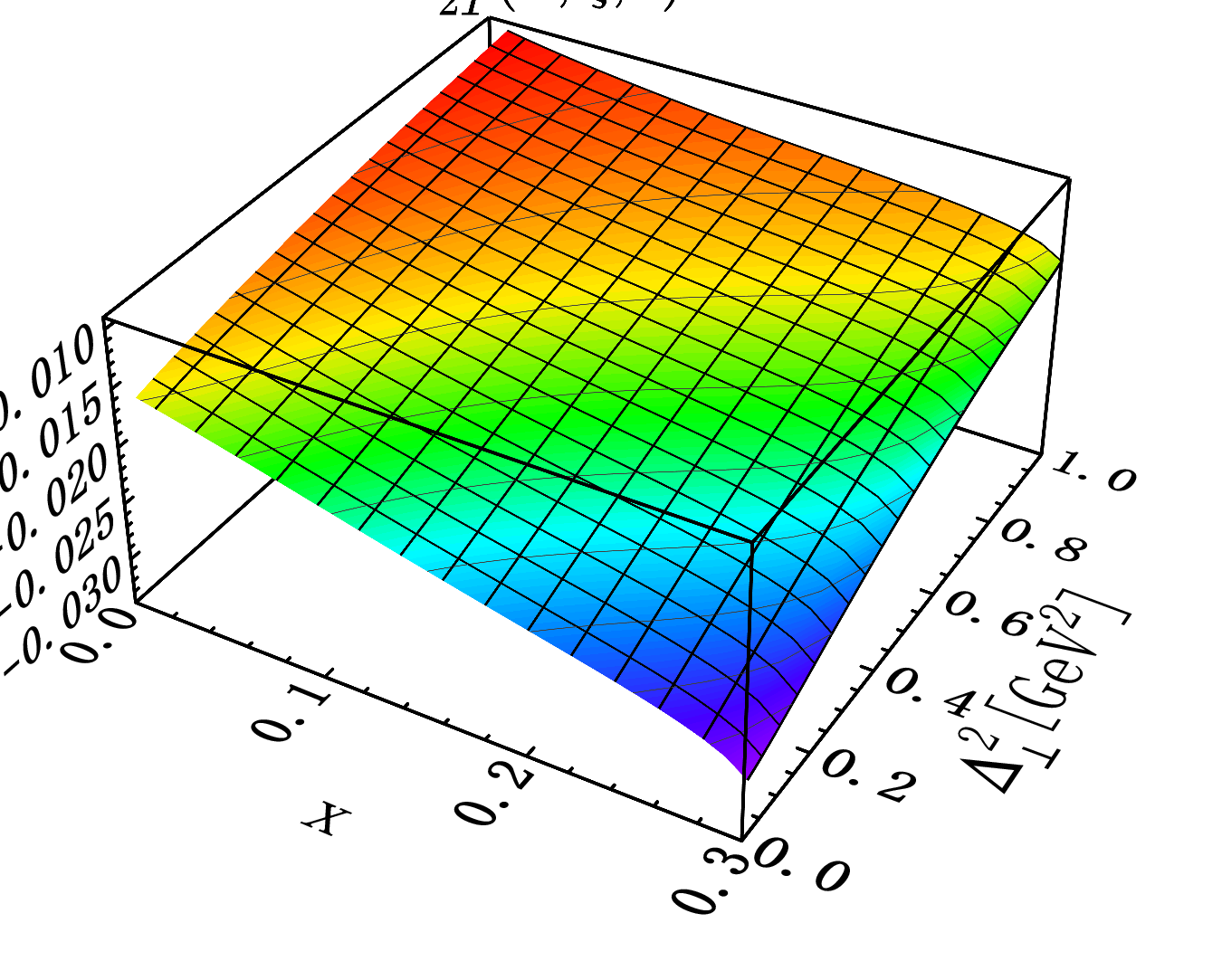}
	\includegraphics[width=0.24\columnwidth]{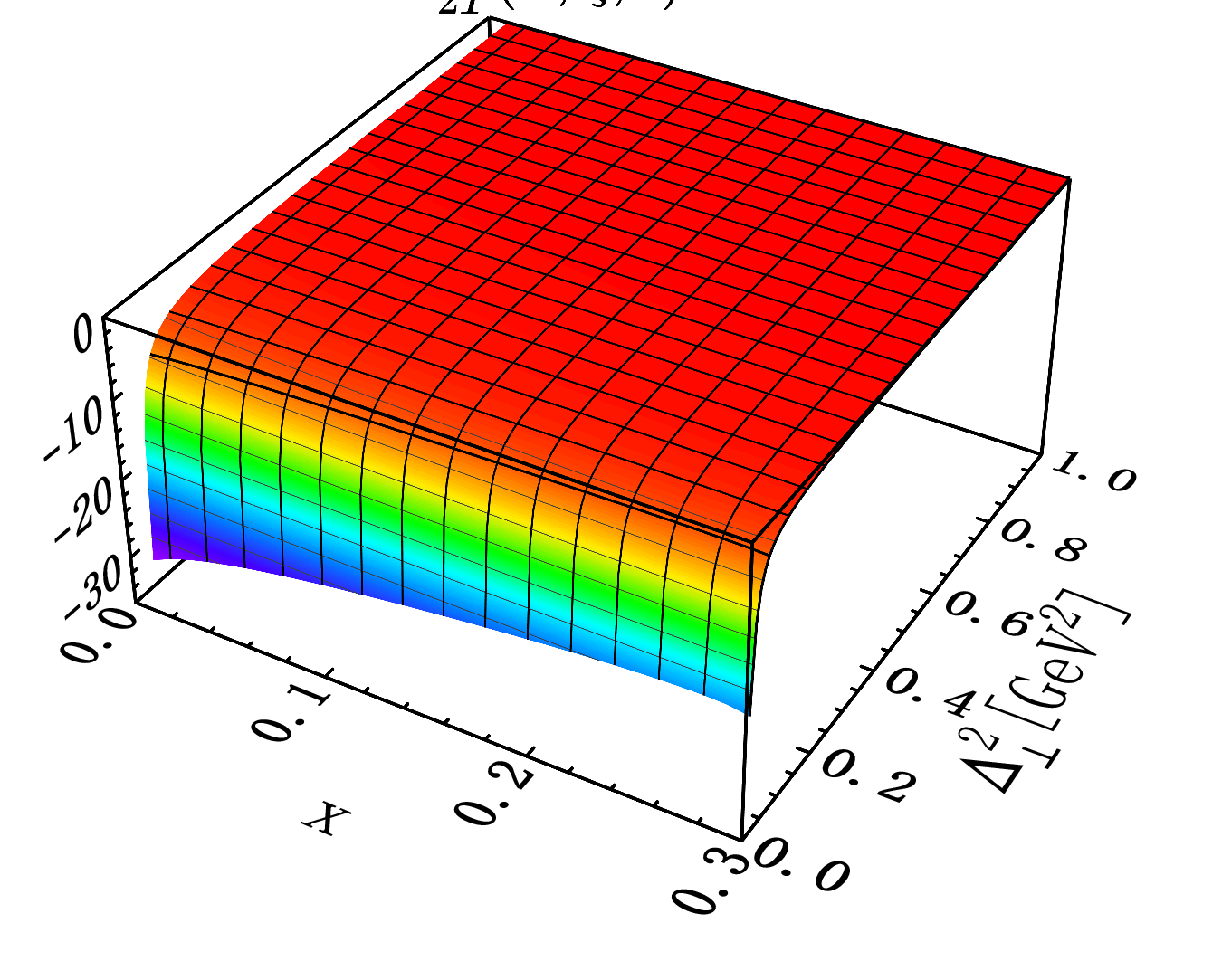}
	\includegraphics[width=0.24\columnwidth]{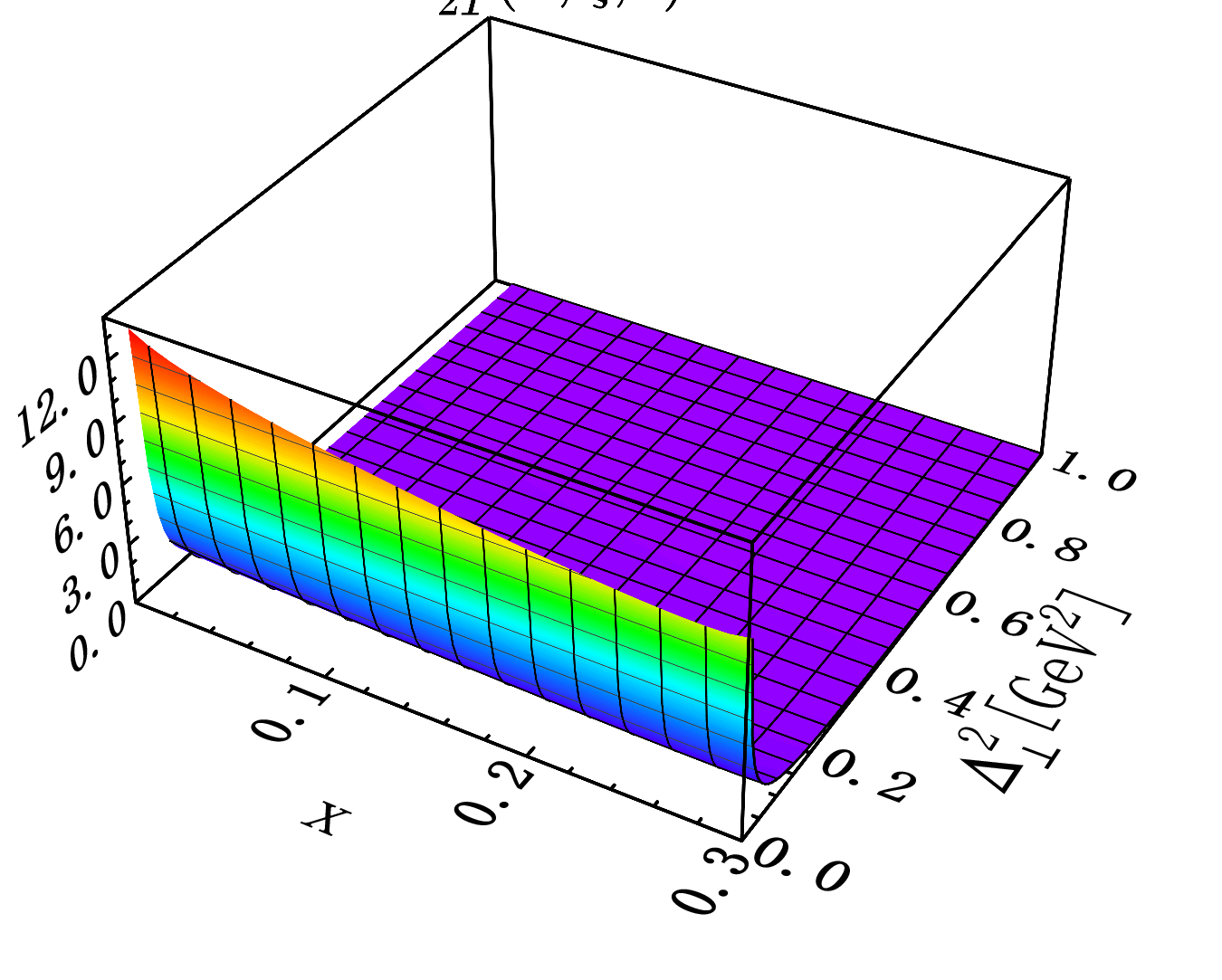}
	\includegraphics[width=0.24\columnwidth]{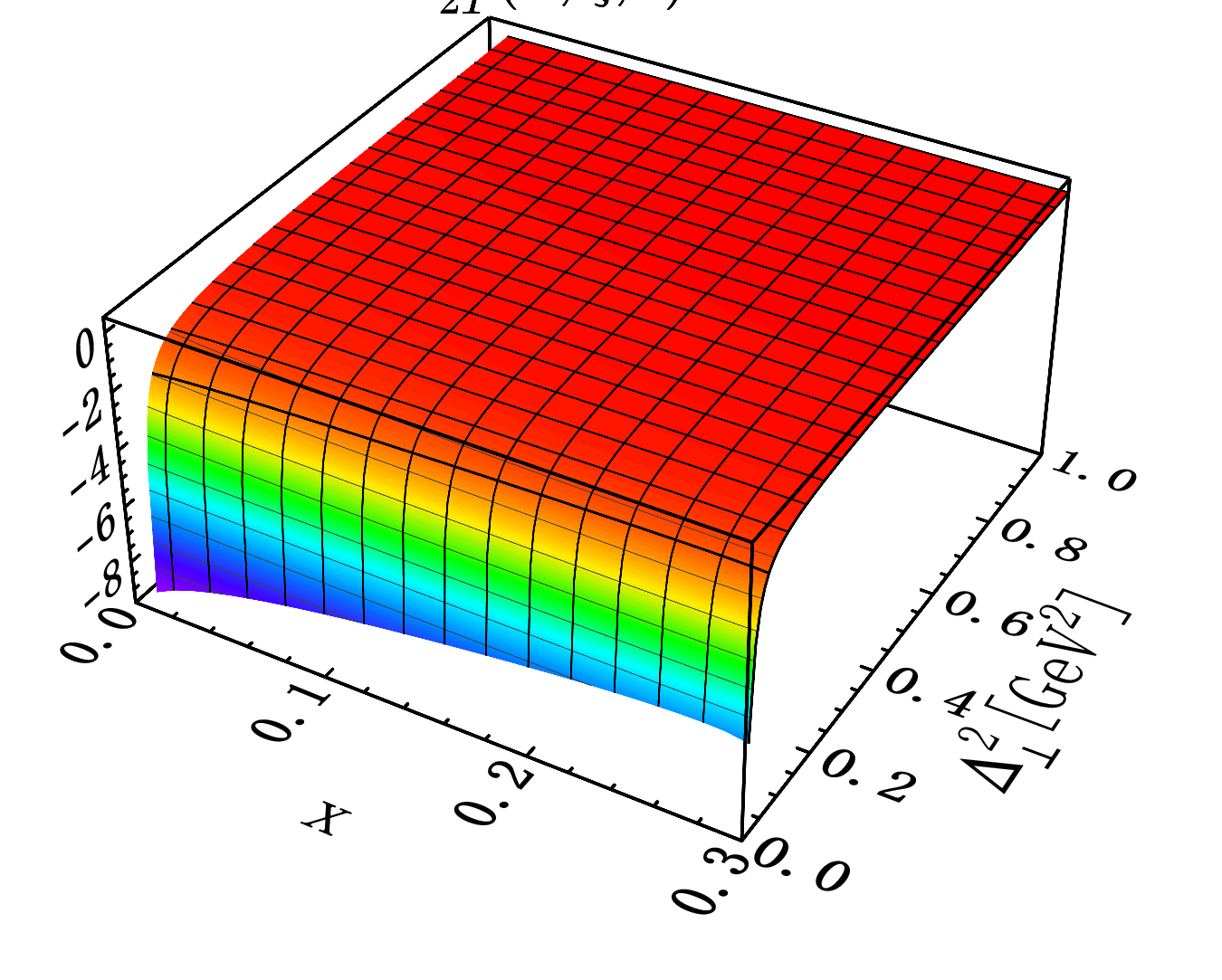}
	\caption{The twist-3 GPDs are plotted with respect to $x$ and $\bm{\Delta}_\perp^2$ in the kinematic range $x\in [0,0.3]$ and $\bm{\Delta}_\perp^2\in [0,1]\,\text{GeV}^2$ at fixed $\xi=0.3$.}
	\label{fig:GPD4}
\end{figure}

\begin{figure}
	\centering
	\includegraphics[width=0.24\columnwidth]{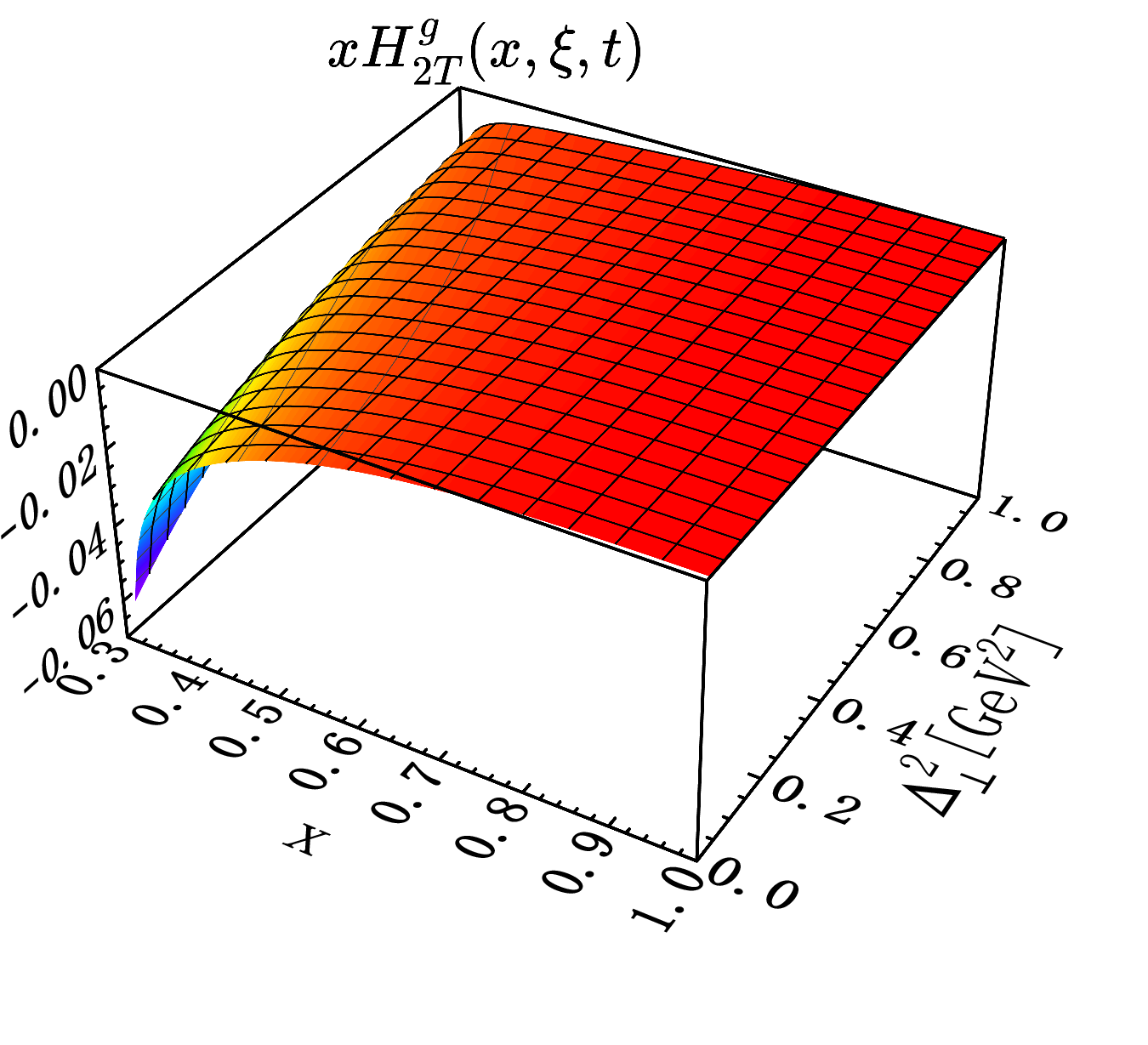}
	\includegraphics[width=0.24\columnwidth]{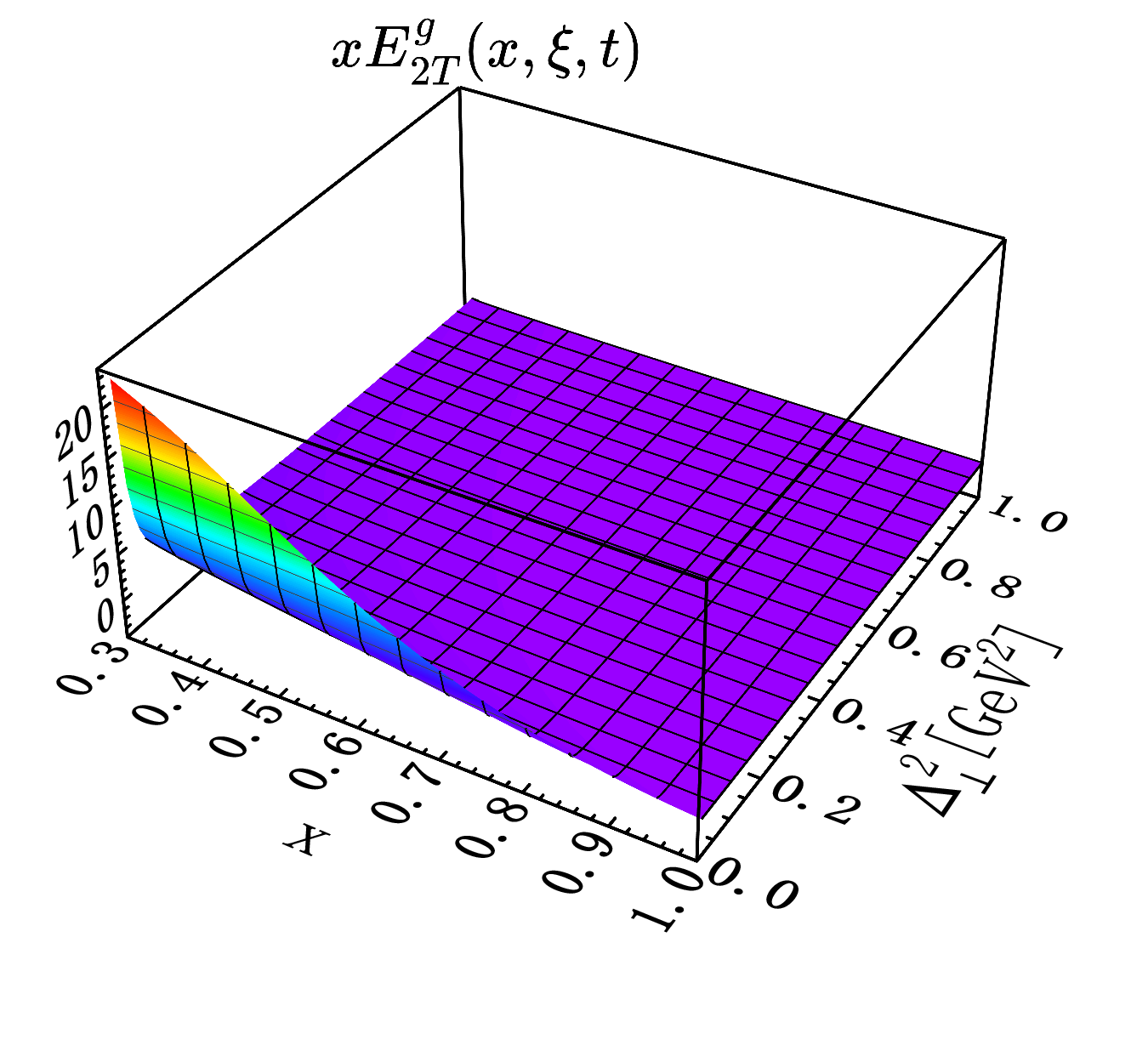}
	\includegraphics[width=0.24\columnwidth]{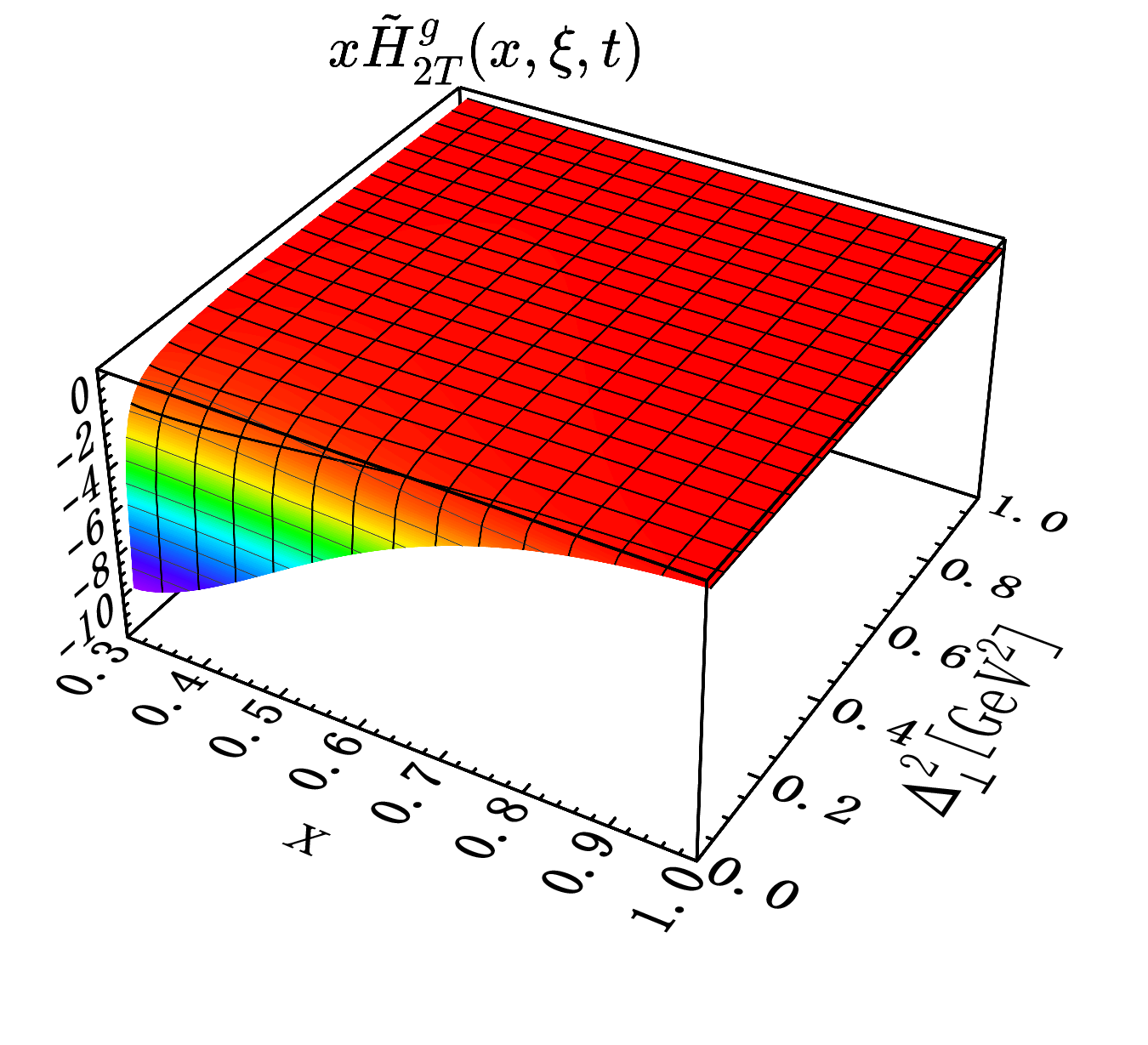}
	\includegraphics[width=0.24\columnwidth]{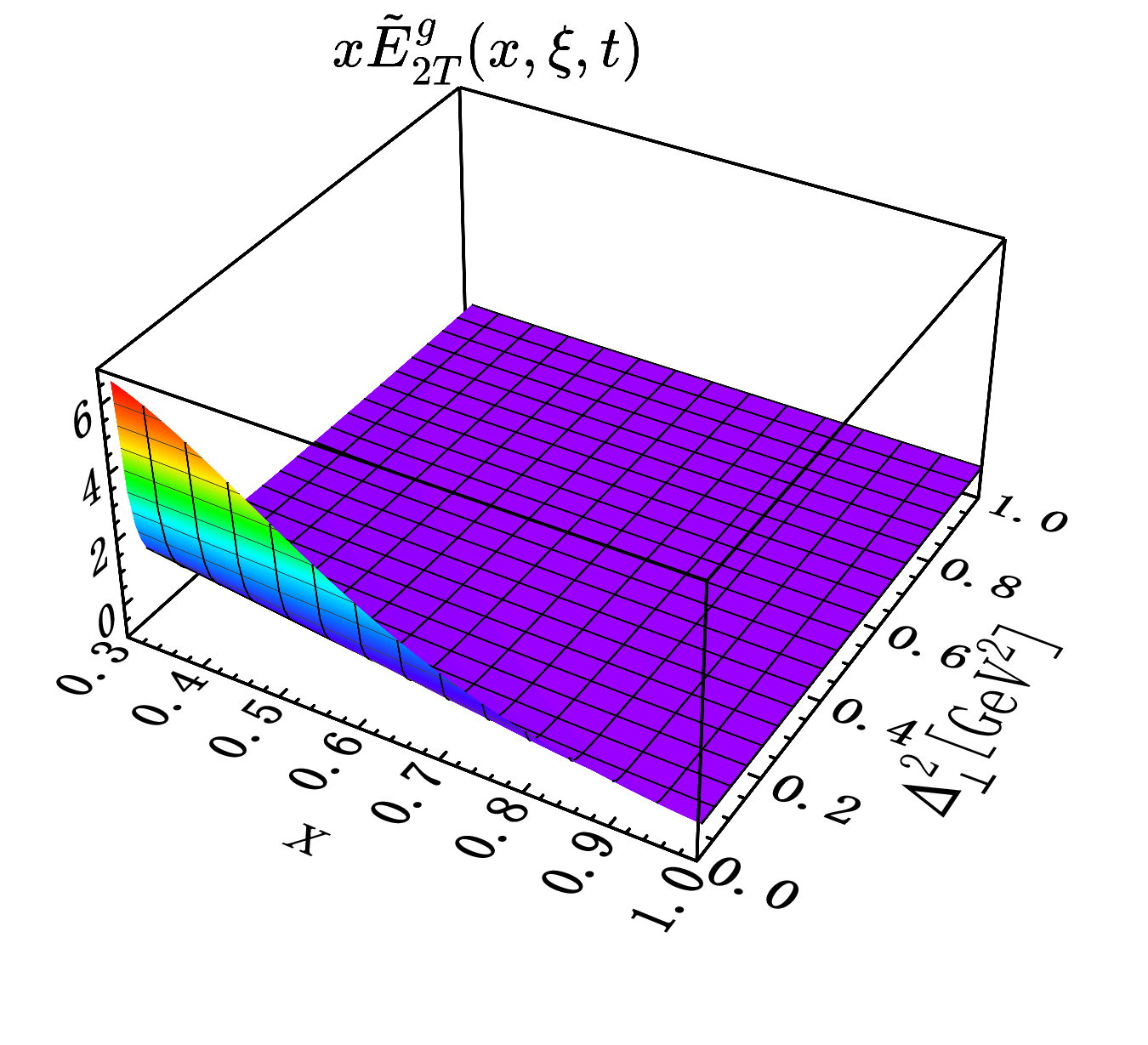}\\
	\includegraphics[width=0.24\columnwidth]{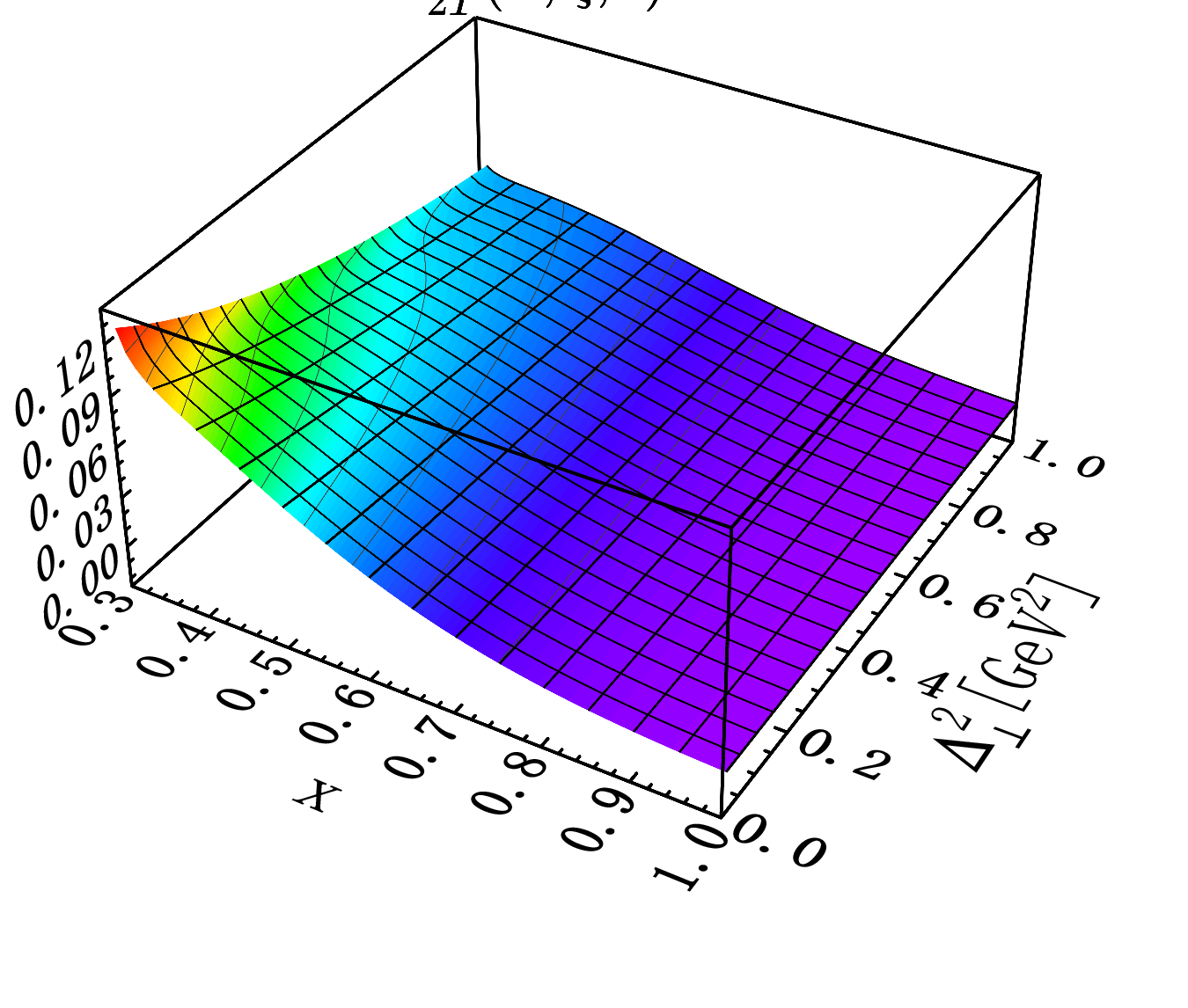}
	\includegraphics[width=0.24\columnwidth]{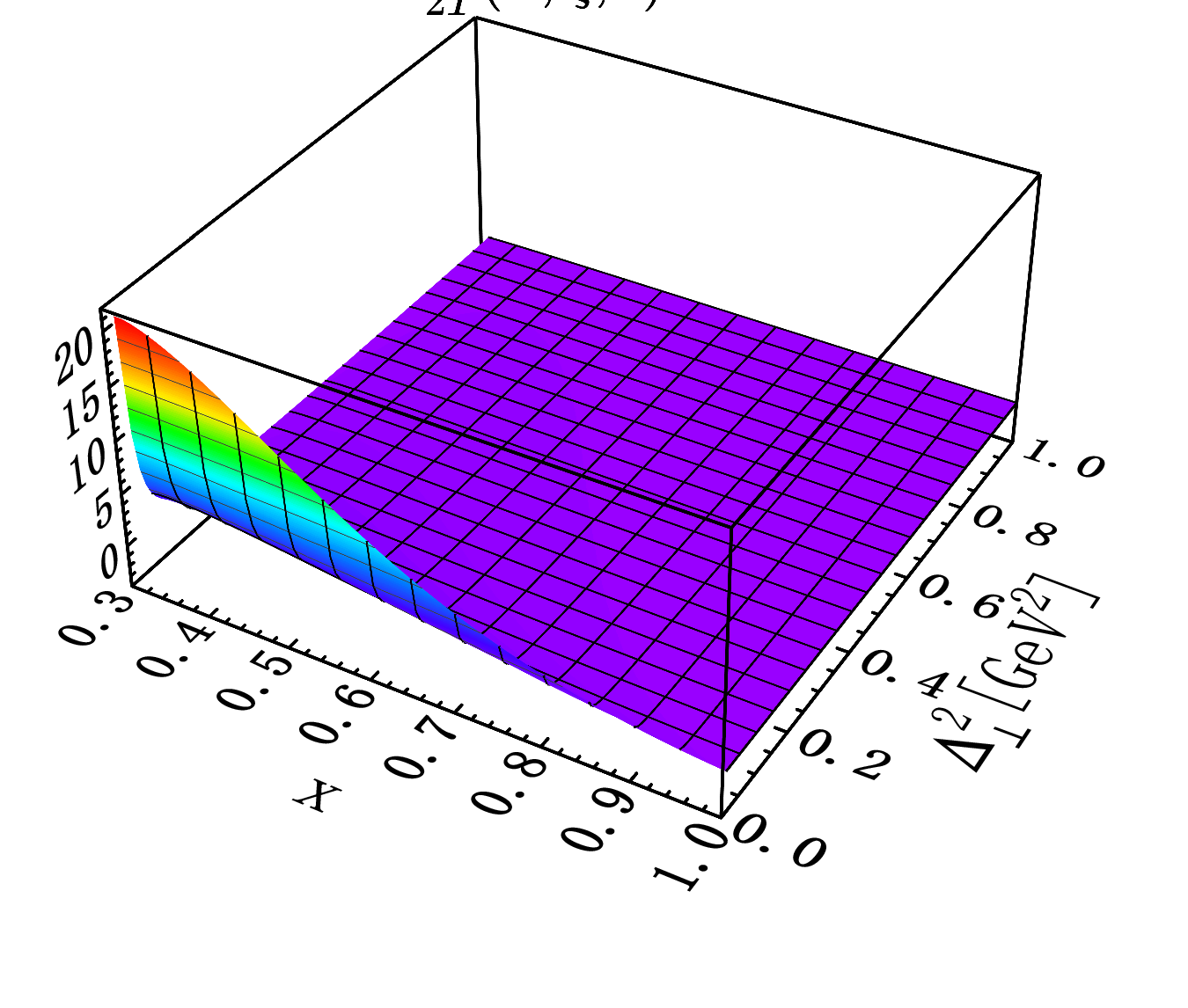}
	\includegraphics[width=0.24\columnwidth]{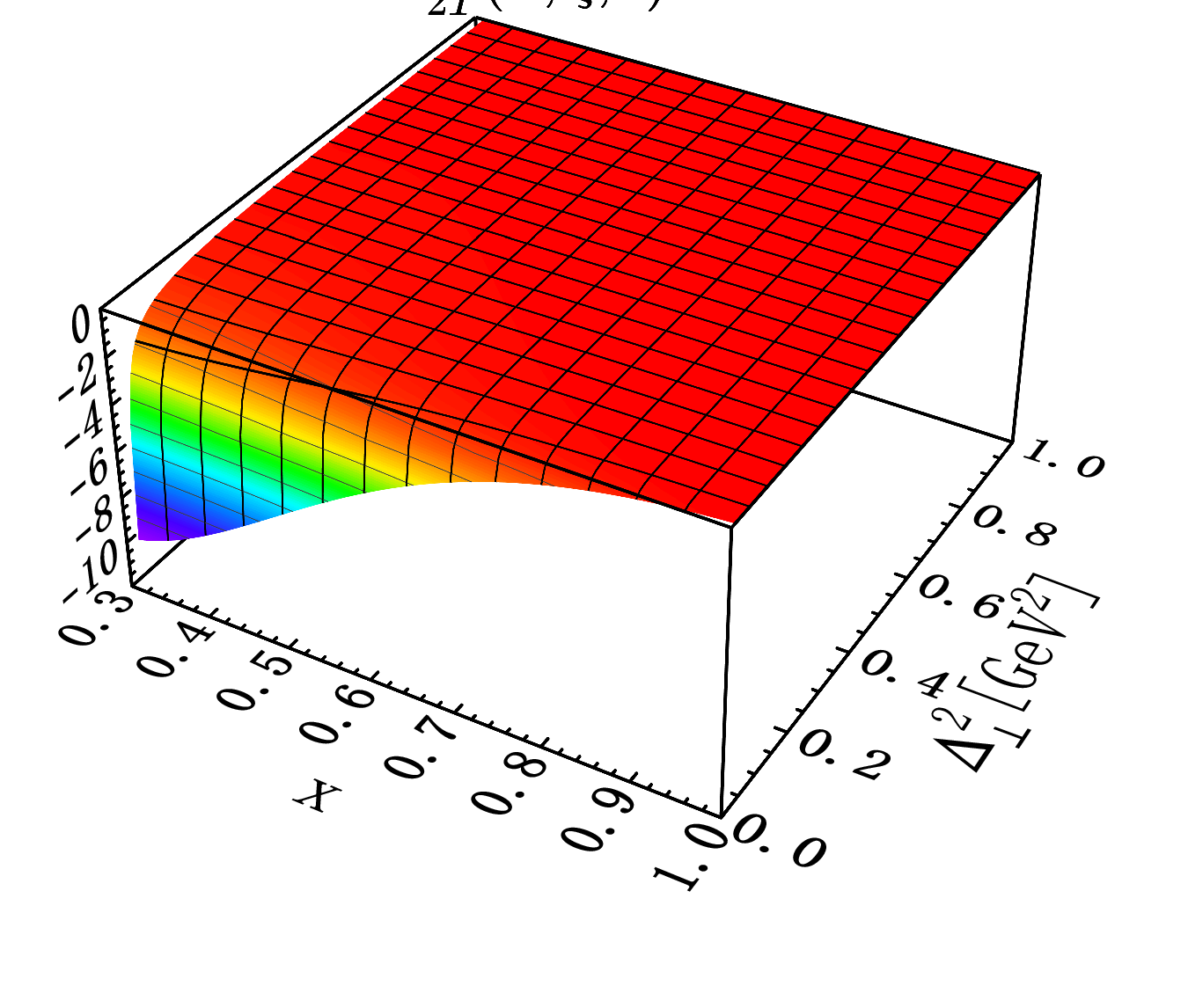}
	\includegraphics[width=0.24\columnwidth]{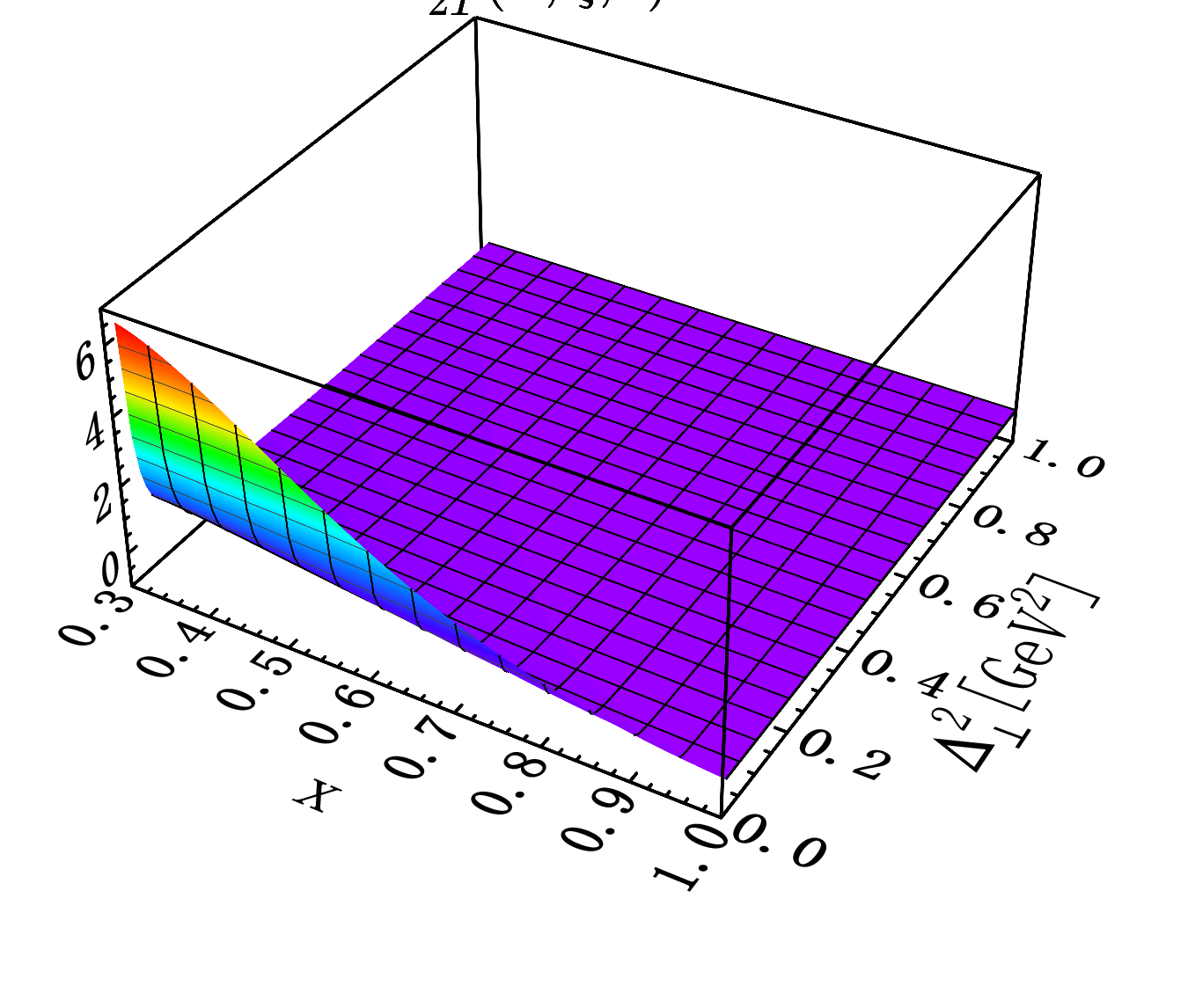}\\
	\includegraphics[width=0.24\columnwidth]{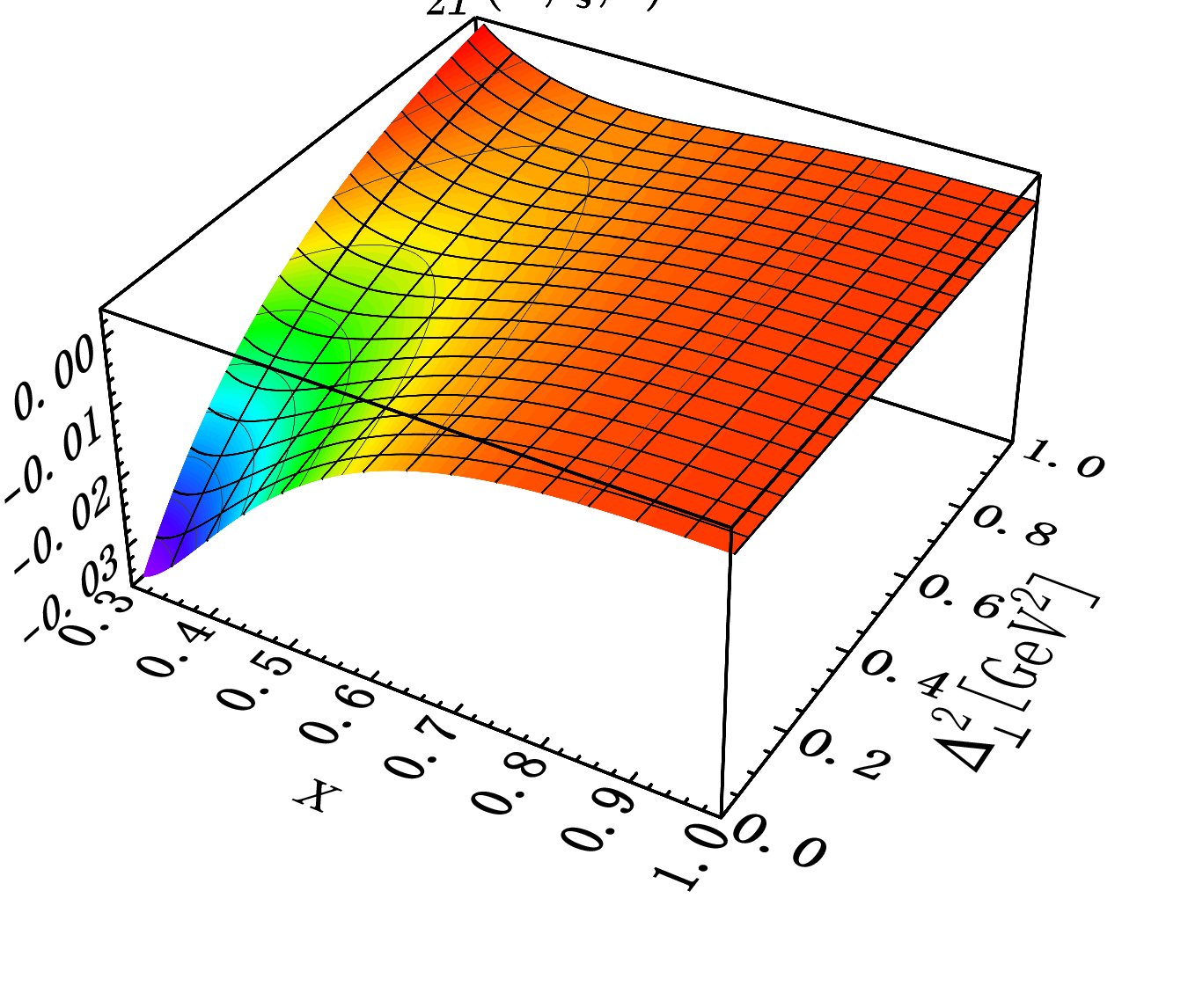}
	\includegraphics[width=0.24\columnwidth]{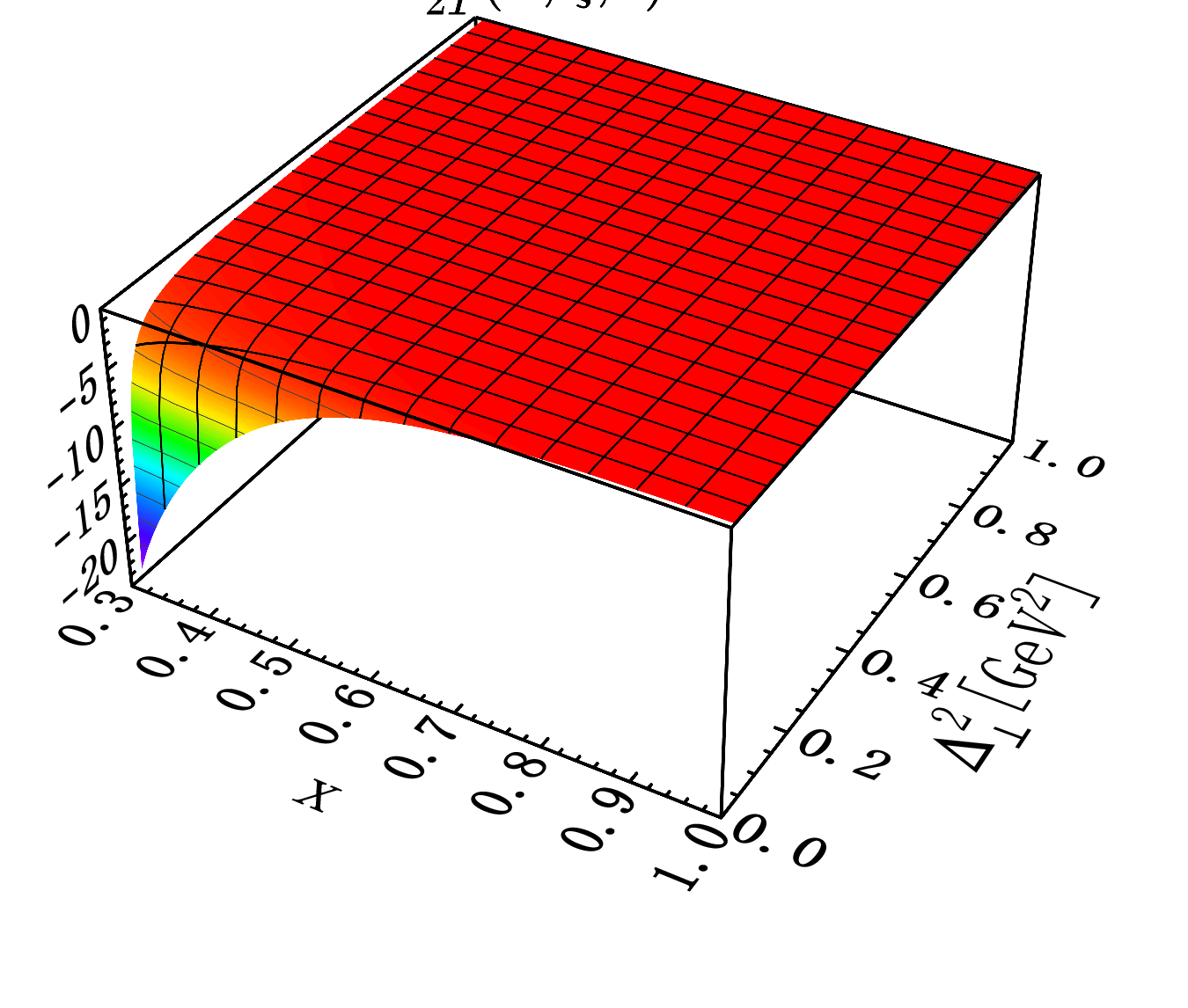}
	\includegraphics[width=0.24\columnwidth]{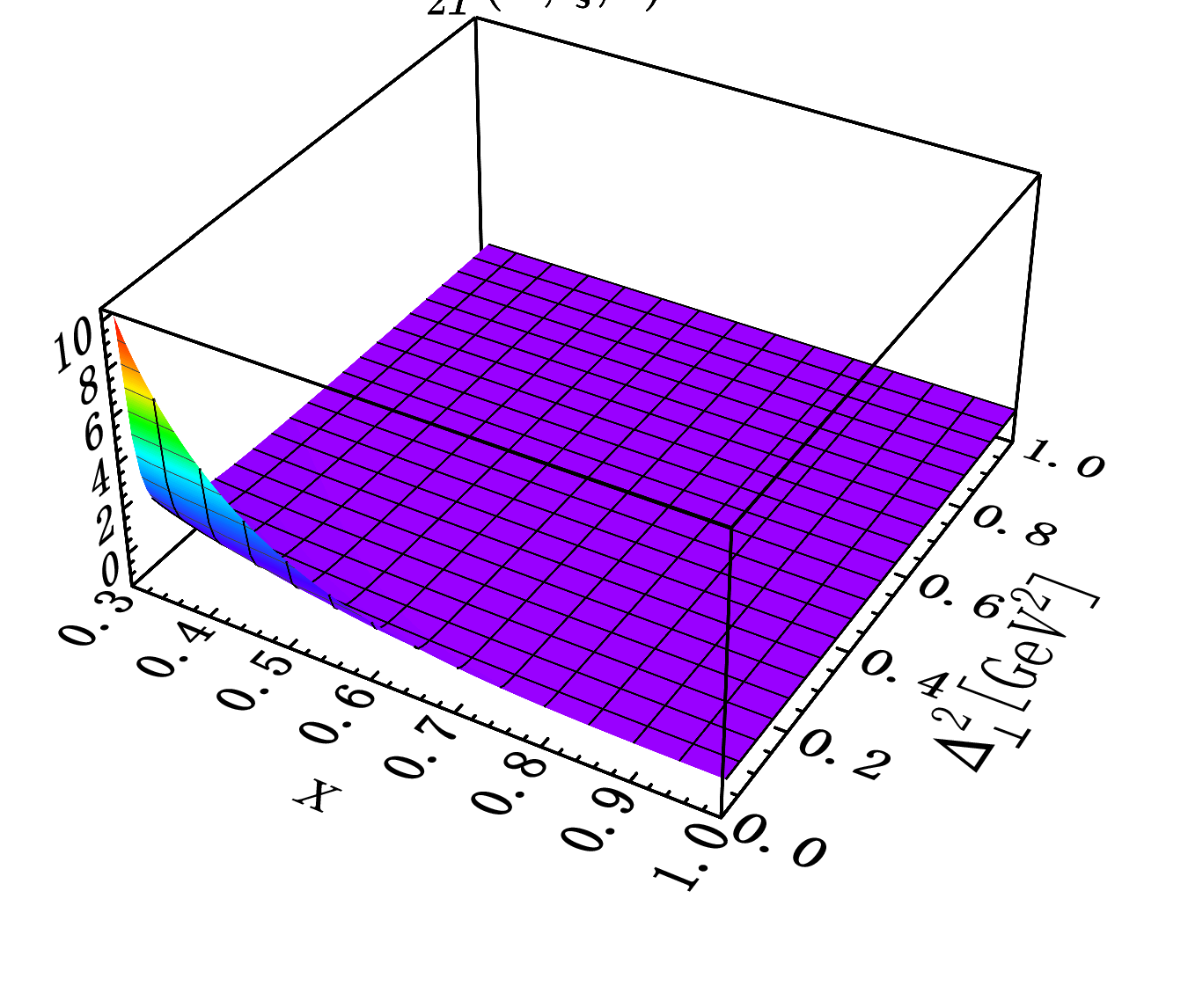}
	\includegraphics[width=0.24\columnwidth]{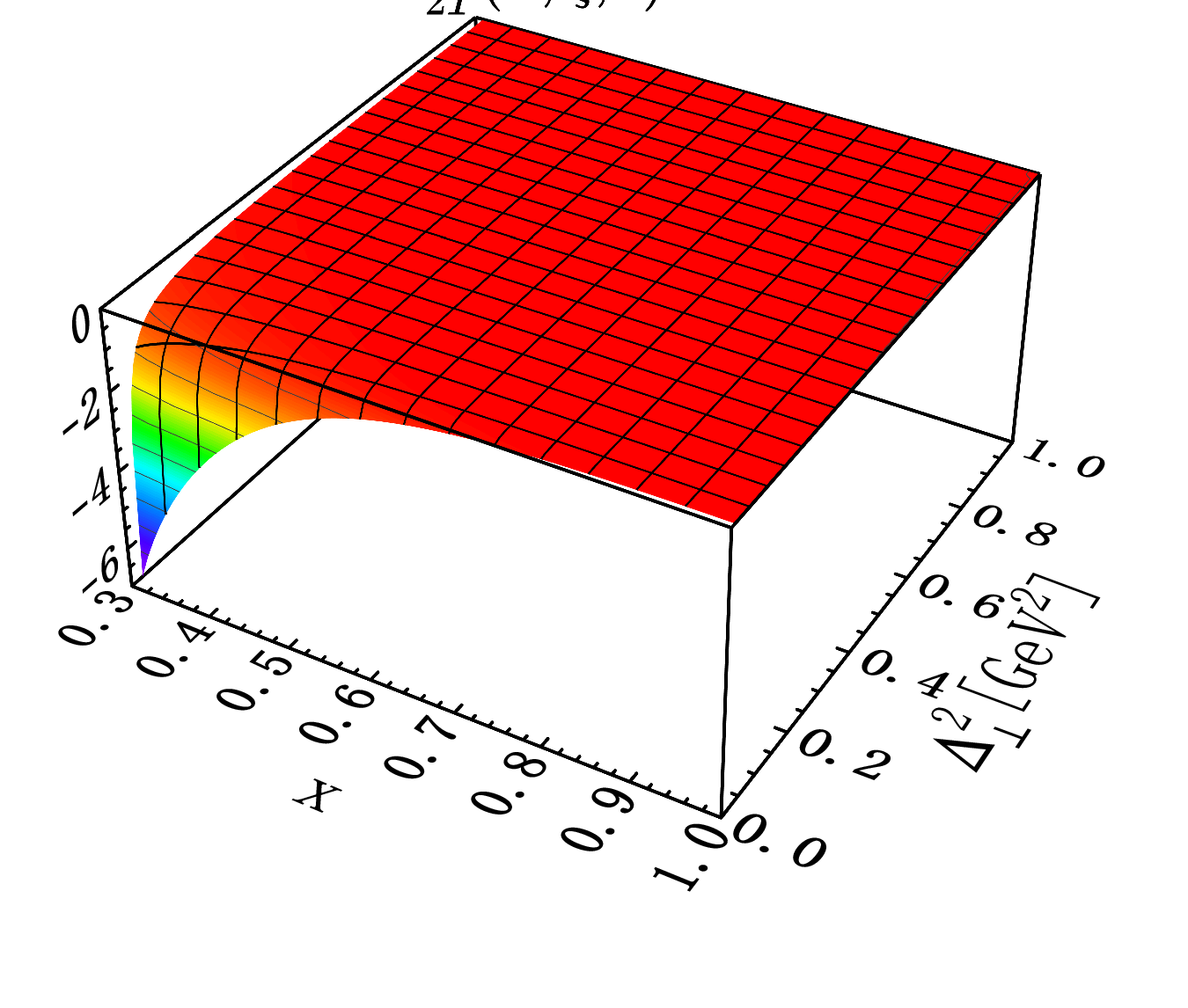}\\
	\includegraphics[width=0.24\columnwidth]{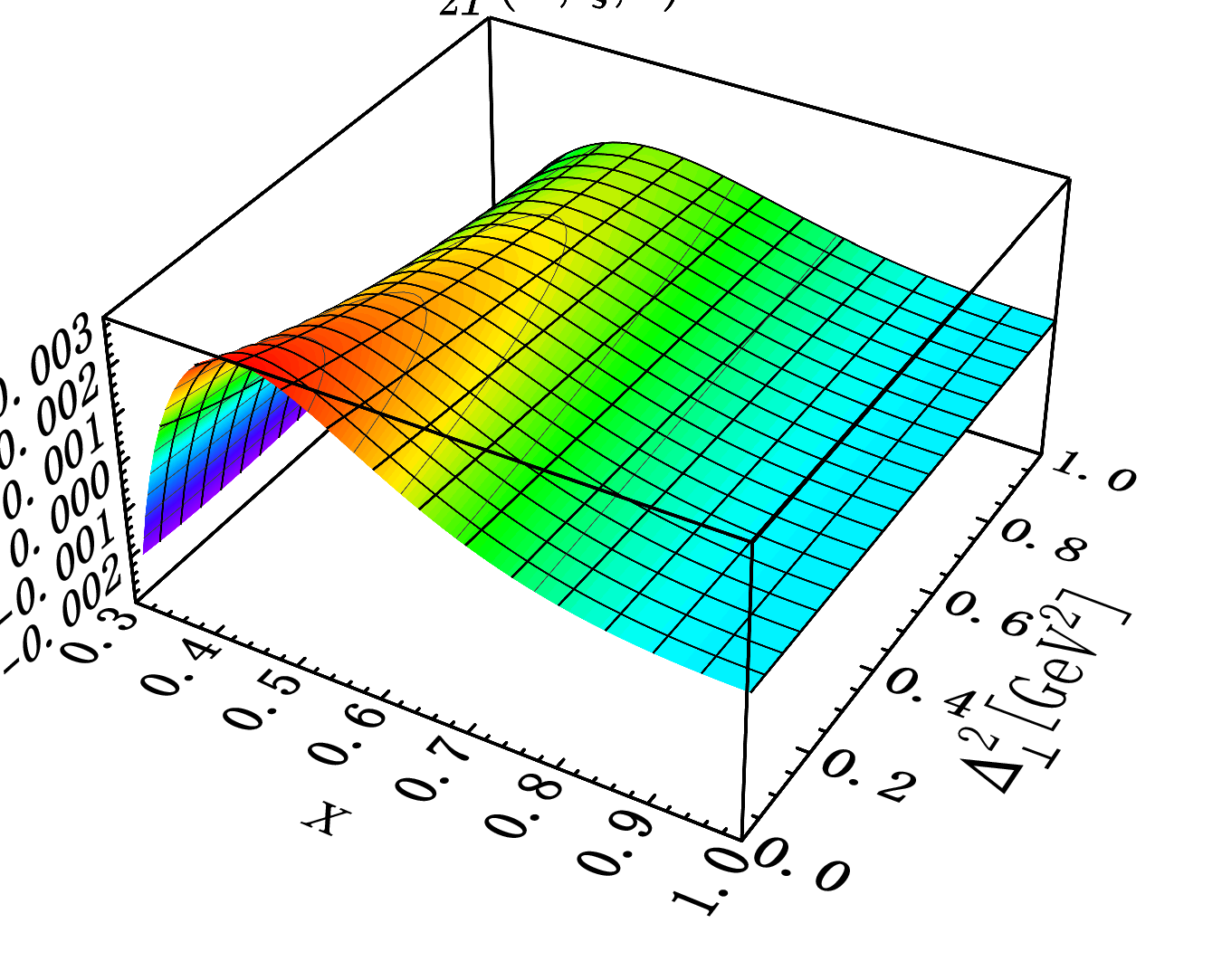}
	\includegraphics[width=0.24\columnwidth]{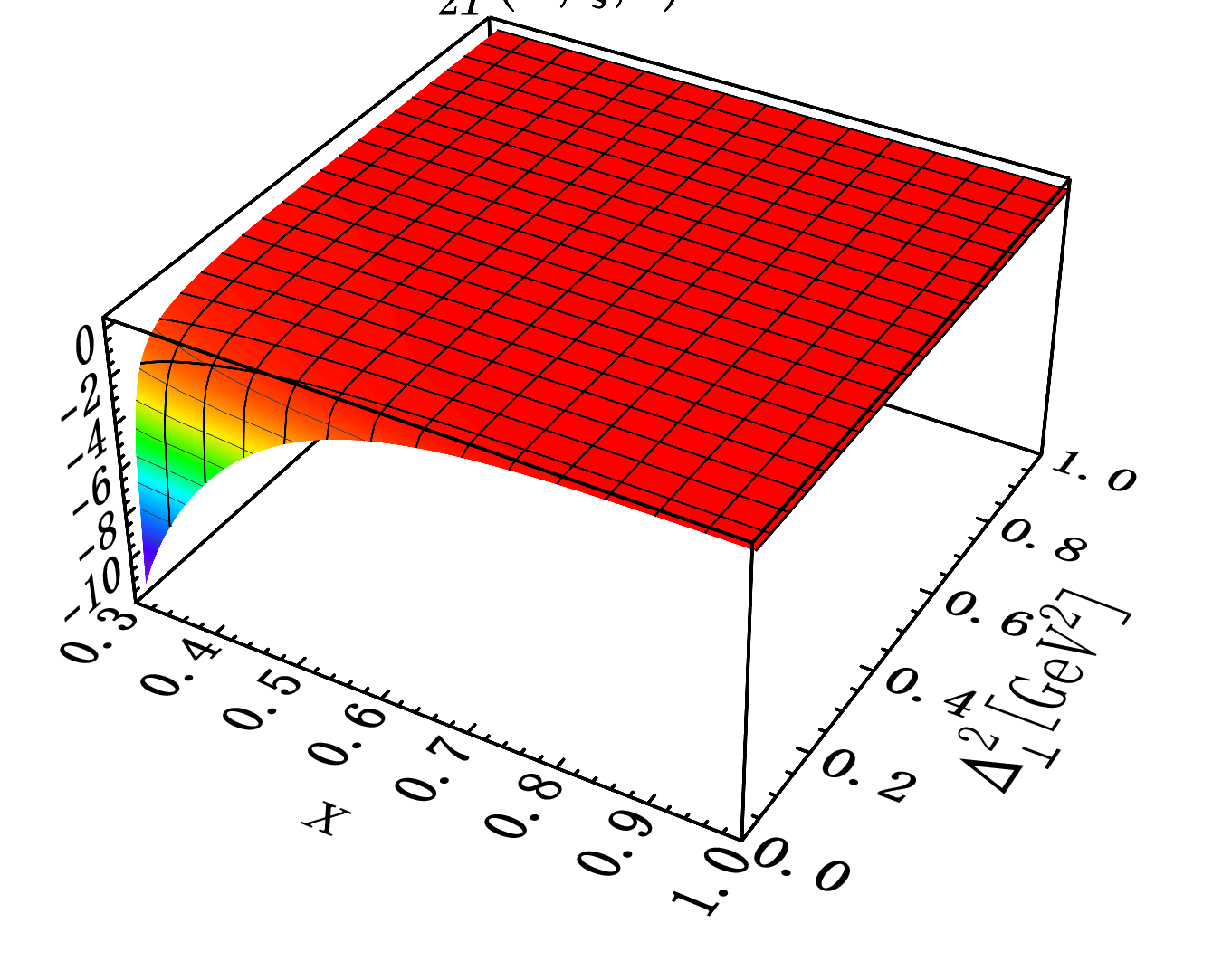}
	\includegraphics[width=0.24\columnwidth]{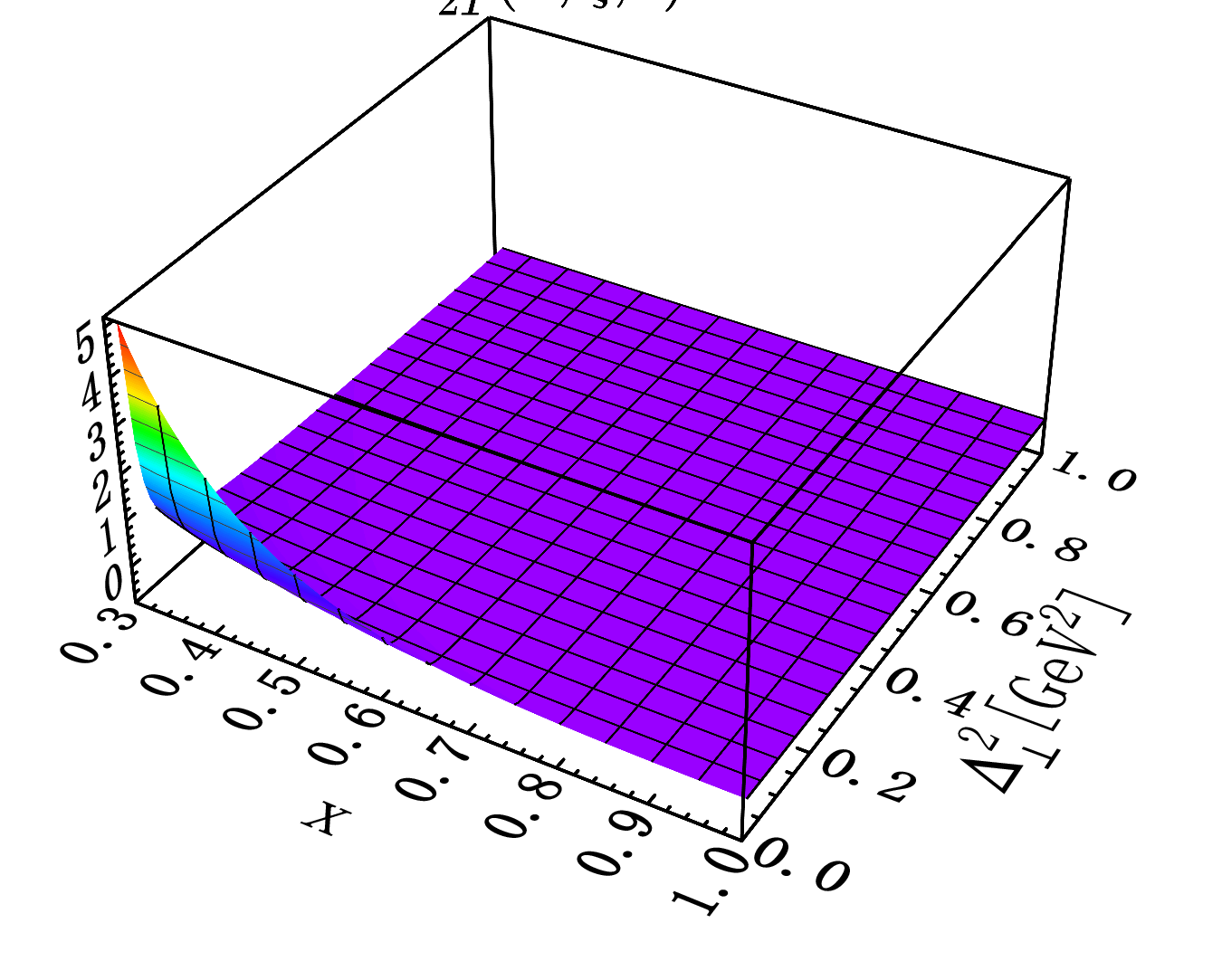}
	\includegraphics[width=0.24\columnwidth]{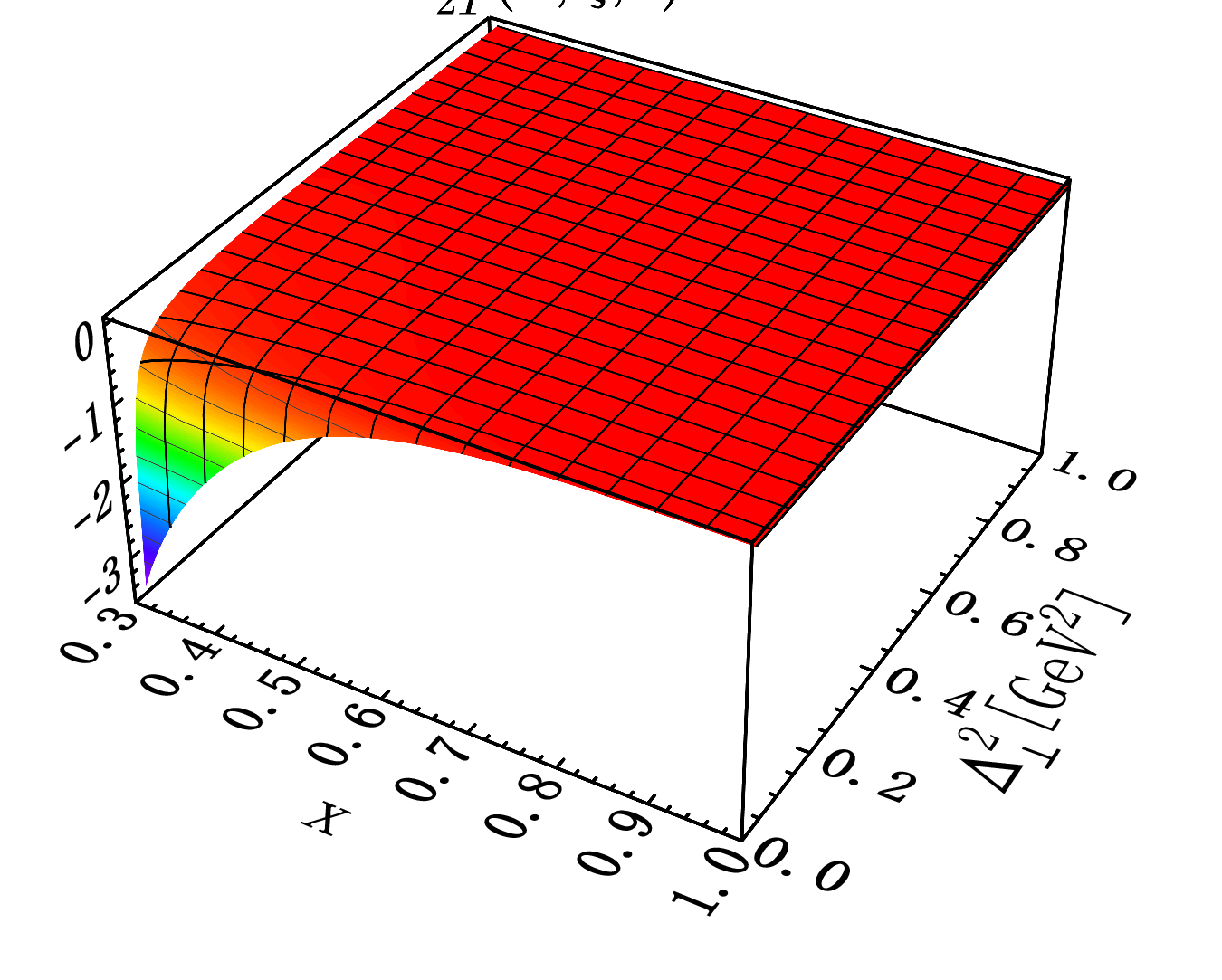}\\
	\caption{The twist-3 GPDs are plotted with respect to $x$ and $\bm{\Delta}_\perp^2$ in the kinematic range $x\in (0.3,1]$ and $\bm{\Delta}_\perp^2\in [0,1]\,\text{GeV}^2$ at fixed $\xi=0.3 $.}
	\label{fig:GPD5}
\end{figure}

At fixed $\xi=0.3$, Figs.~\ref{fig:GPD4} and~\ref{fig:GPD5} show the $\bm{\Delta}_\perp^2$ dependence in the ERBL and DGLAP regions, respectively. In the ERBL region, most GPDs exhibit a pronounced rise toward $\bm{\Delta}_\perp^2\to0$ and a subsequent monotonic decrease, whereas the distributions in the first column decrease much more slowly. This behavior reflects the transverse localization of gluons in the nucleon. The dominant distributions peak either toward $x\to0$ or near the ERBL boundary $x=\xi$, again indicating an enhanced gluon density at small $x$. In the DGLAP region, most GPDs show qualitatively similar behavior to their ERBL counterparts, while smoothly approaching zero as $x\to1$. In particular, $H_{2T}^g$ and $\bar{H}_{2T}^g$ peak near $x=\xi$ and $\bm{\Delta}_\perp^2=0$, whereas $\bar{H}_{2T}^{\prime g}$ peaks at intermediate $x$. Compared with the DGLAP region, the corresponding ERBL distributions generally have larger magnitudes, indicating that the dominant contributions are concentrated toward small $x$.

These results illustrate the general structure generated by the twist-3 gluon correlator. The detailed magnitudes and shapes are model-dependent, while the qualitative features found here---including the concentration toward small $x$, the suppression at large $x$, and the decrease with increasing transverse momentum transfer---provide useful benchmarks for future studies of twist-3 gluon GPDs.

\subsection{twist-3 IPDs}

The transverse spatial structure of partons can be investigated through impact-parameter-dependent distributions. At zero skewness, the IPDs are defined as the two-dimensional Fourier transforms of GPDs with respect to the transverse momentum transfer:
\begin{align} \mathcal{H}(x,\bm{b}_\perp)=\int\frac{d^2\bm{\Delta}_\perp}{(2\pi)^2}e^{-i\bm{\Delta}_\perp\cdot\bm{b}_\perp}H(x,0,\bm{\Delta}_\perp^2),
\end{align}
where $\bm{b}_\perp$ is the transverse coordinate conjugate to $\bm{\Delta}_\perp$ and represents the impact parameter of the parton relative to the nucleon's transverse center of momentum. This framework provides simultaneous information on the longitudinal momentum fraction and transverse spatial dependence of partons~\cite{Burkardt:2000za,Burkardt:2002hr,Burkardt:2002ks,Diehl:2002he}. At $\xi=0$, twelve of the twist-3 gluon GPDs considered here survive, and therefore twelve corresponding IPDs can in principle be constructed. 
To illustrate their dependence on $x$ and $b_\perp$, we focus on the following eight combinations:
\begin{align}
	&\frac{1}{2}\mathcal{E}_{2T}^g+\tilde{\mathcal{H}}_{2T}^g,\qquad \tilde{\mathcal{E}}_{2T}^g,\qquad \bar{\mathcal{H}}_{2T}^g,\qquad \frac{1}{2}\bar{\mathcal{E}}_{2T}^g+\tilde{\bar{\mathcal{H}}}_{2T}^g,  \nonumber\\
	&\mathcal{H}_{2T}^{\prime g},\qquad -\left(\frac{1}{2}\mathcal{E}_{2T}^{\prime g}+\tilde{\mathcal{H}}_{2T}^{\prime g}\right),\qquad \frac{1}{2}\bar{\mathcal{E}}_{2T}^{\prime g}+\tilde{\bar{\mathcal{H}}}_{2T}^{\prime g},\qquad\tilde{\bar{\mathcal{E}}}_{2T}^{\prime g},\nonumber
\end{align}
where $\frac{1}{2}\bar{\mathcal{E}}_{2T}^g+\tilde{\bar{\mathcal{H}}}_{2T}^g$ and $-\left(\frac{1}{2}\mathcal{E}_{2T}^{\prime g}+\tilde{\mathcal{H}}_{2T}^{\prime g}\right)$ vanish in our calculations. We do not include the uncertainties induced by the model-parameter uncertainties in the following discussion.

\begin{figure}
	\centering
	\includegraphics[width=0.32\columnwidth]{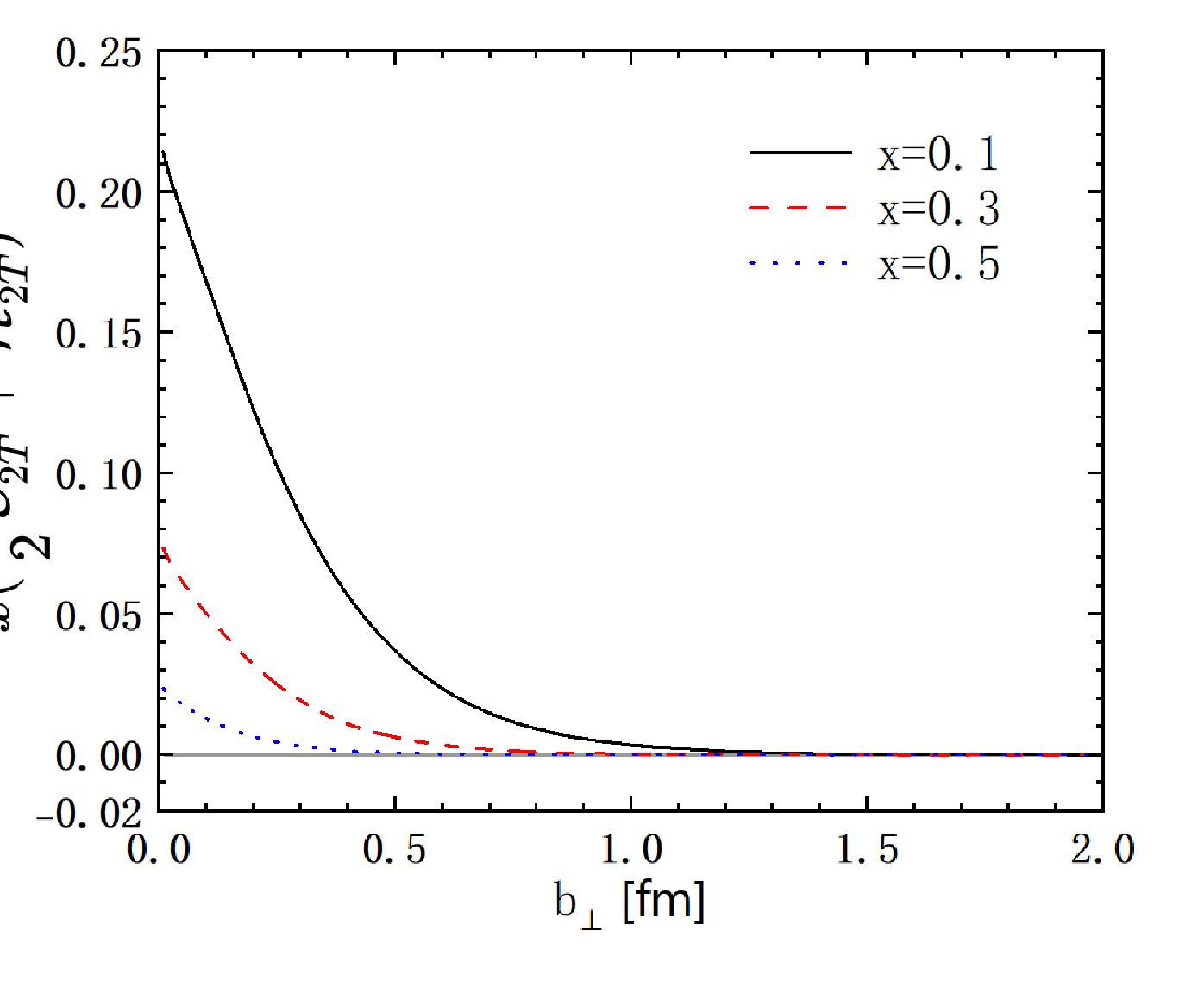}
	\includegraphics[width=0.32\columnwidth]{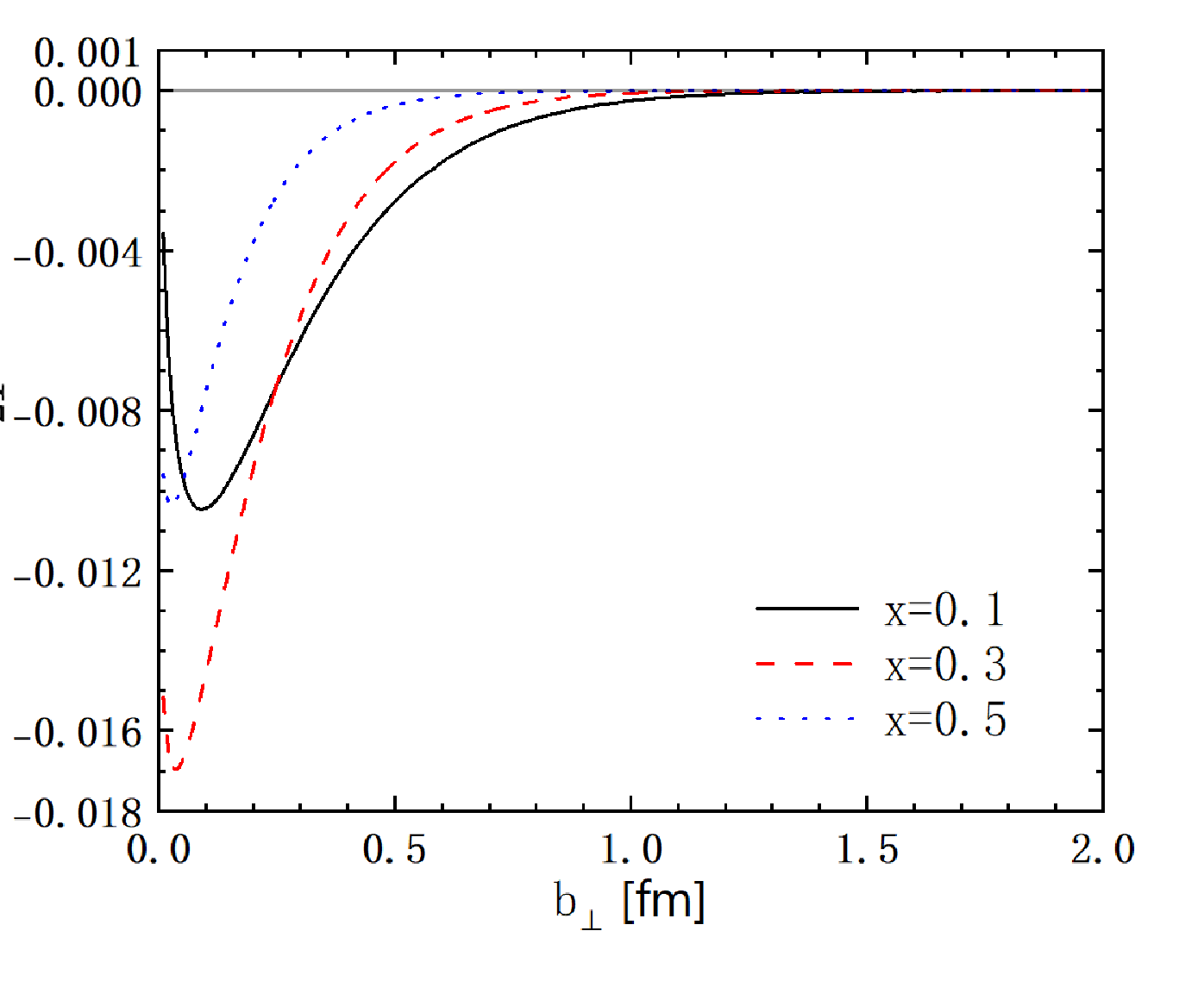}
	\includegraphics[width=0.32\columnwidth]{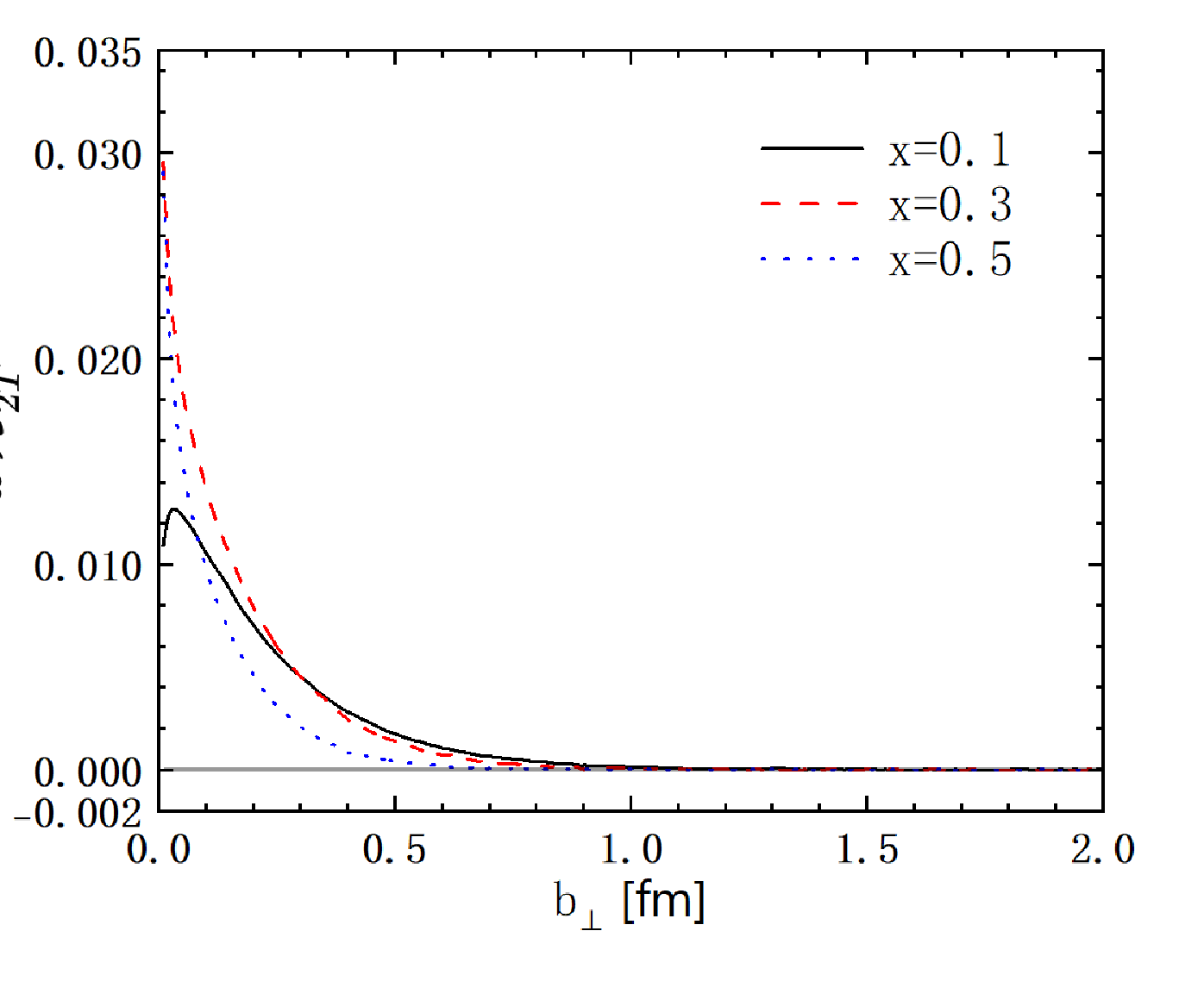}\\
	\includegraphics[width=0.32\columnwidth]{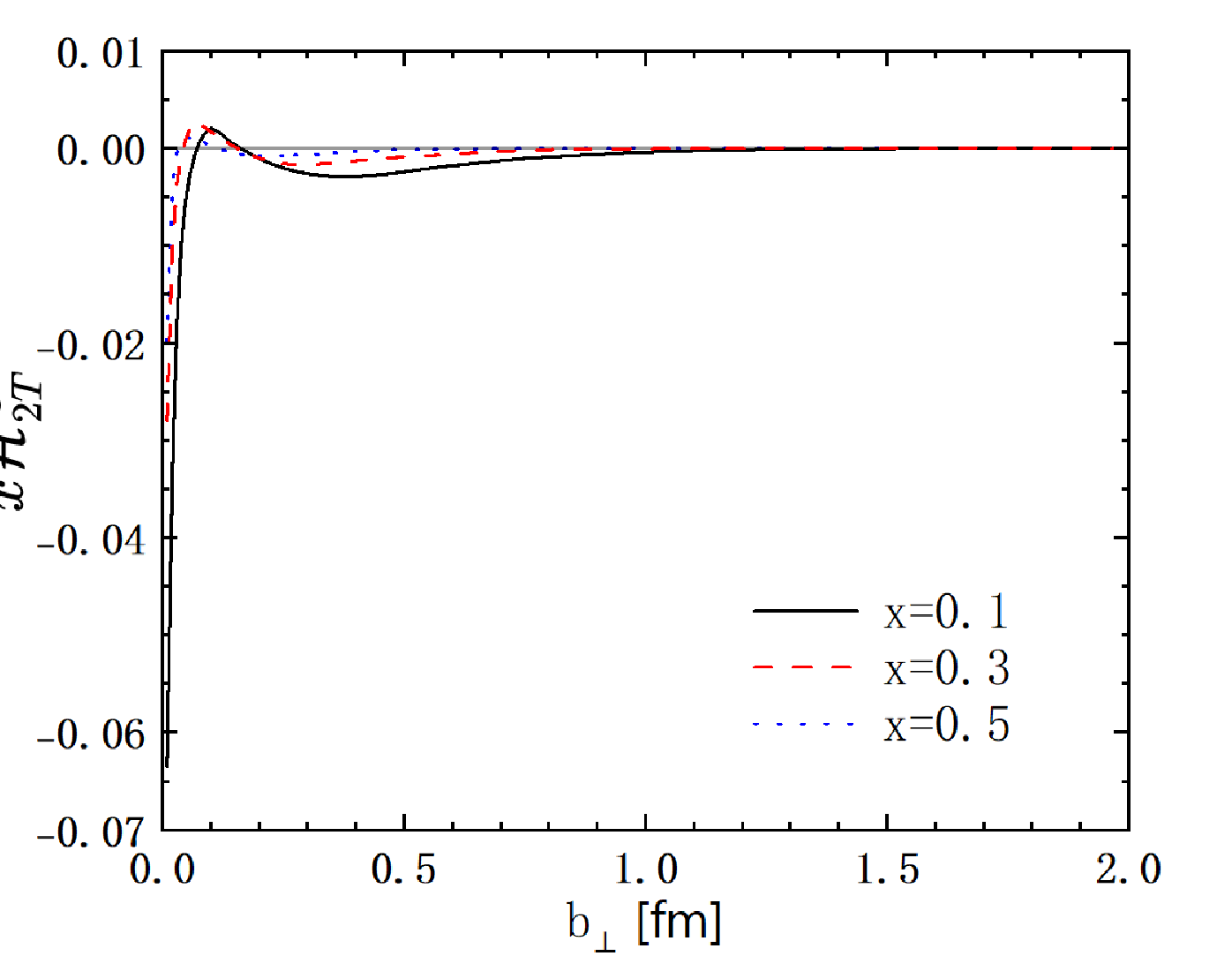}
	\includegraphics[width=0.32\columnwidth]{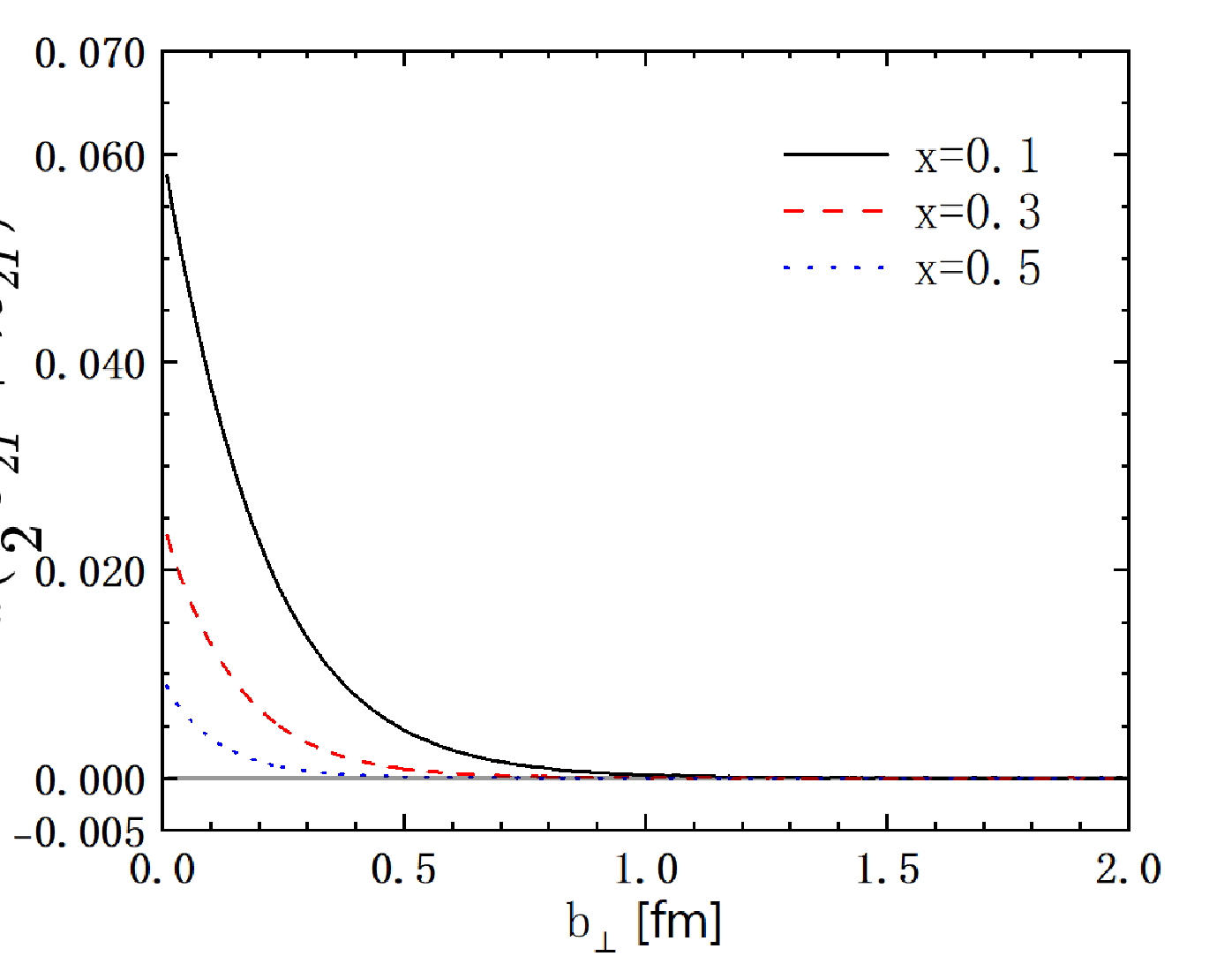}
	\includegraphics[width=0.32\columnwidth]{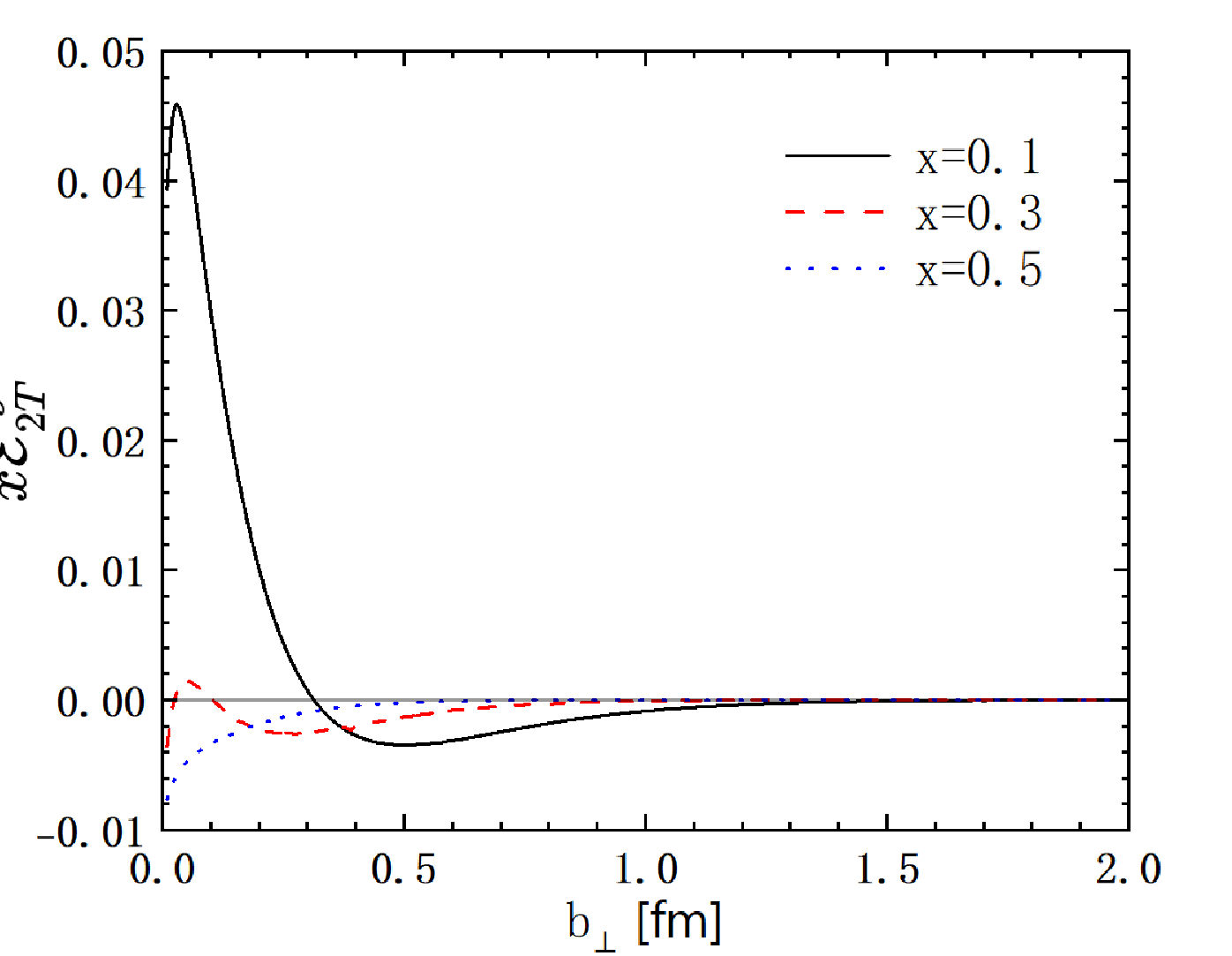}
	\caption{The twist-3 IPDs are plotted with respect to $b_\perp$ in the kinematic range $b_\perp\in [0,2]$ fm at fixed $x=0.1,0.3,$ and 0.5.}
	\label{fig:IPD1}
\end{figure}

Fig.~\ref{fig:IPD1} shows the model predictions for the twist-3 gluon IPDs as functions of $b_\perp$ for $x=0.1$ (solid line), 0.3 (dashed line), and 0.5 (dotted line). All distributions exhibit similar overall shapes and comparable magnitudes. They are localized at small $b_\perp$ and rapidly decrease toward zero as $b_\perp$ increases. The largest amplitudes occur for $b_\perp<0.1$ fm, with $\mathcal{H}_{2T}^{\prime g}$ particularly localized in this region.

The IPD amplitudes generally decrease with increasing $x$, although the detailed $b_\perp$-dependence varies among the distributions. The IPDs $\frac{1}{2}\mathcal{E}_{2T}^g+\tilde{\mathcal{H}}_{2T}^g$, $\bar{\mathcal{H}}_{2T}^g$, and $\frac{1}{2}\bar{\mathcal{E}}_{2T}^{\prime g}+\tilde{\bar{\mathcal{H}}}_{2T}^{\prime g}$ remain positive, whereas $\tilde{\mathcal{E}}_{2T}^g$ is negative throughout the $b_\perp$ range. $\mathcal{H}_{2T}^{\prime g}$ changes sign near $b_\perp=0.1$ fm, with a node that remains approximately stable at $x\approx0.2$, while $\tilde{\bar{\mathcal{E}}}_{2T}^{\prime g}$ changes sign at larger $b_\perp$ and $x$. For $\tilde{\mathcal{E}}_{2T}^g$ and $\mathcal{H}_{2T}^{\prime g}$, the peak positions shift toward smaller $b_\perp$ as $x$ increases.

\begin{figure}
	\centering
	\includegraphics[width=0.32\columnwidth]{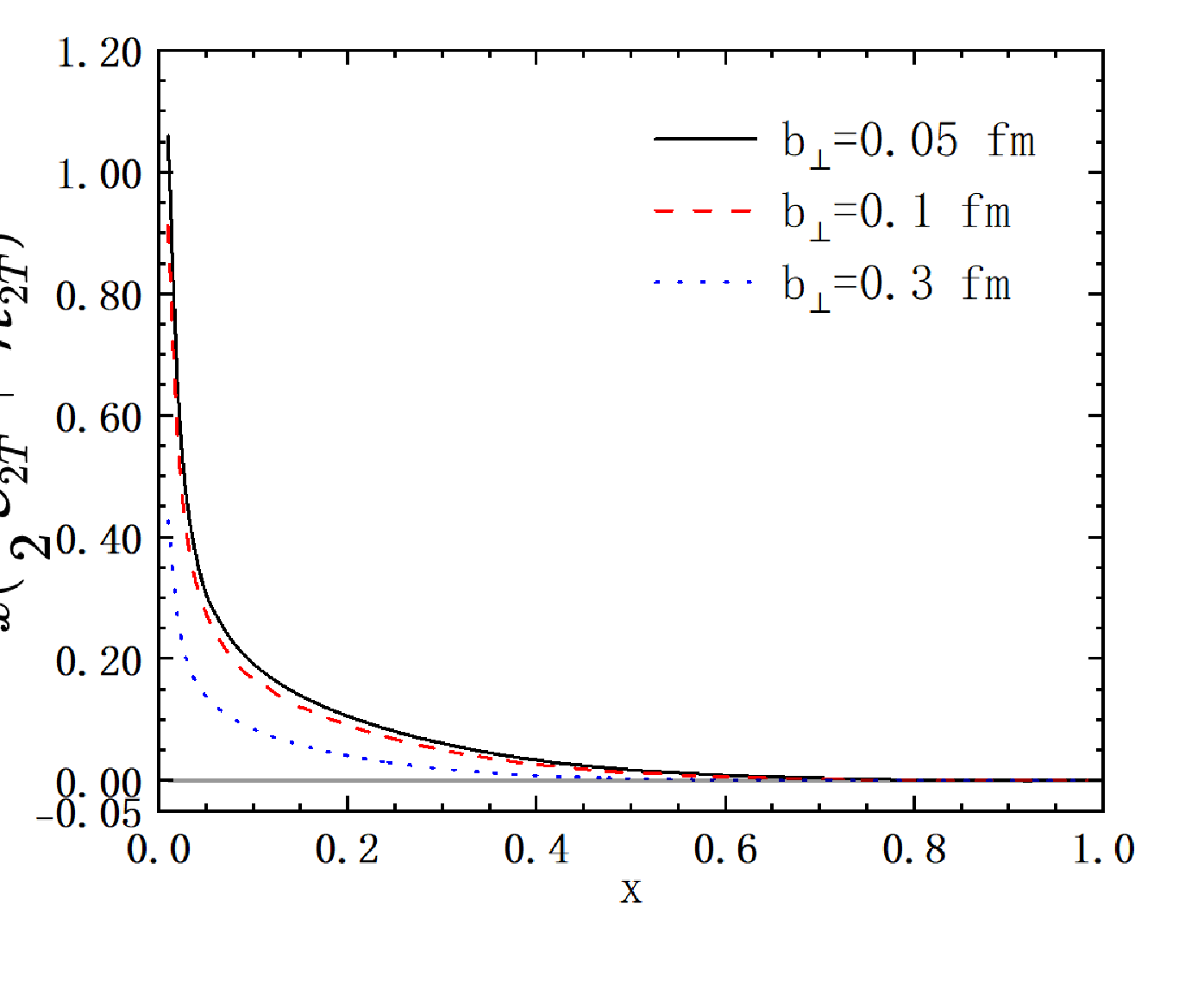}
	\includegraphics[width=0.32\columnwidth]{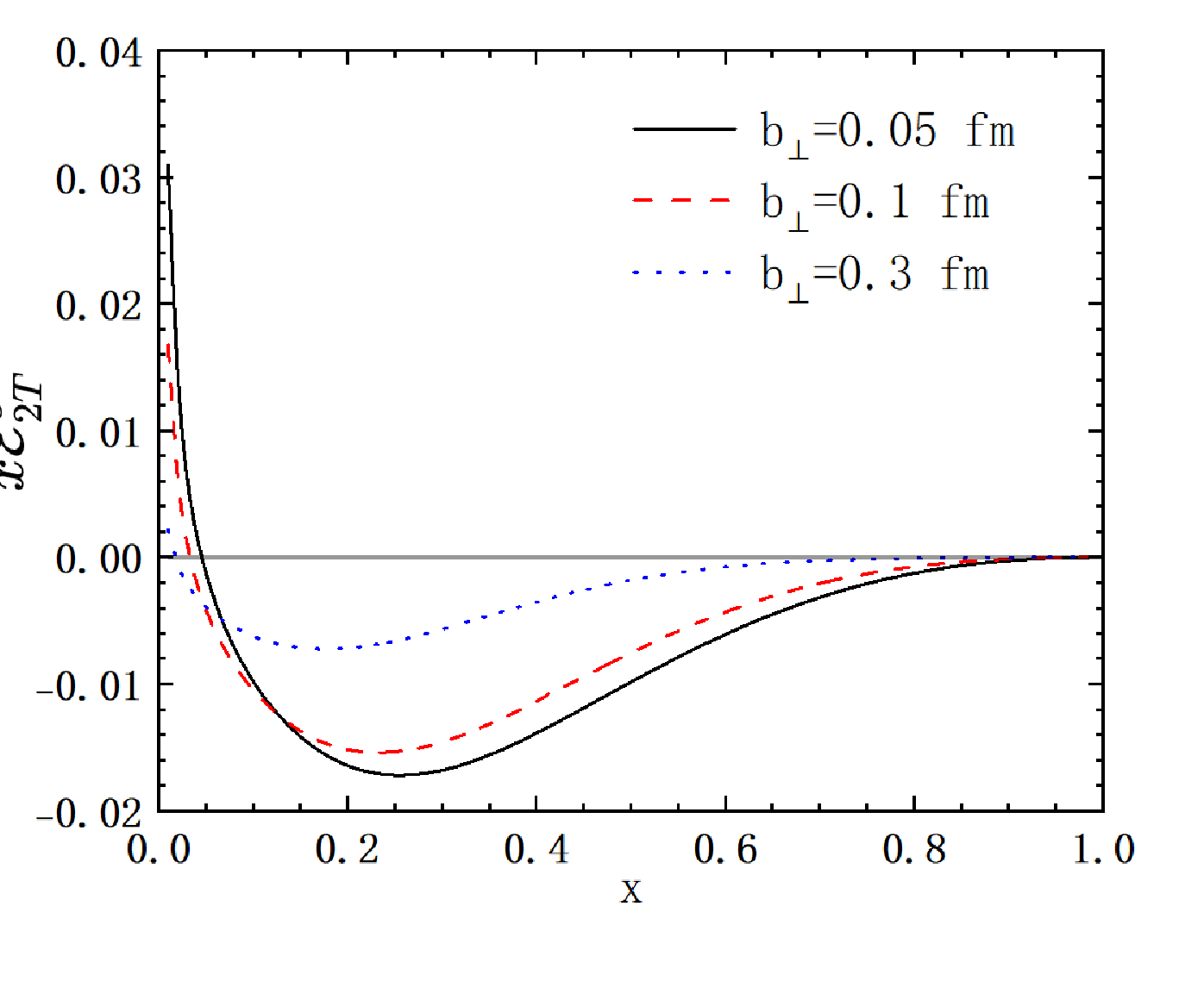}
	\includegraphics[width=0.32\columnwidth]{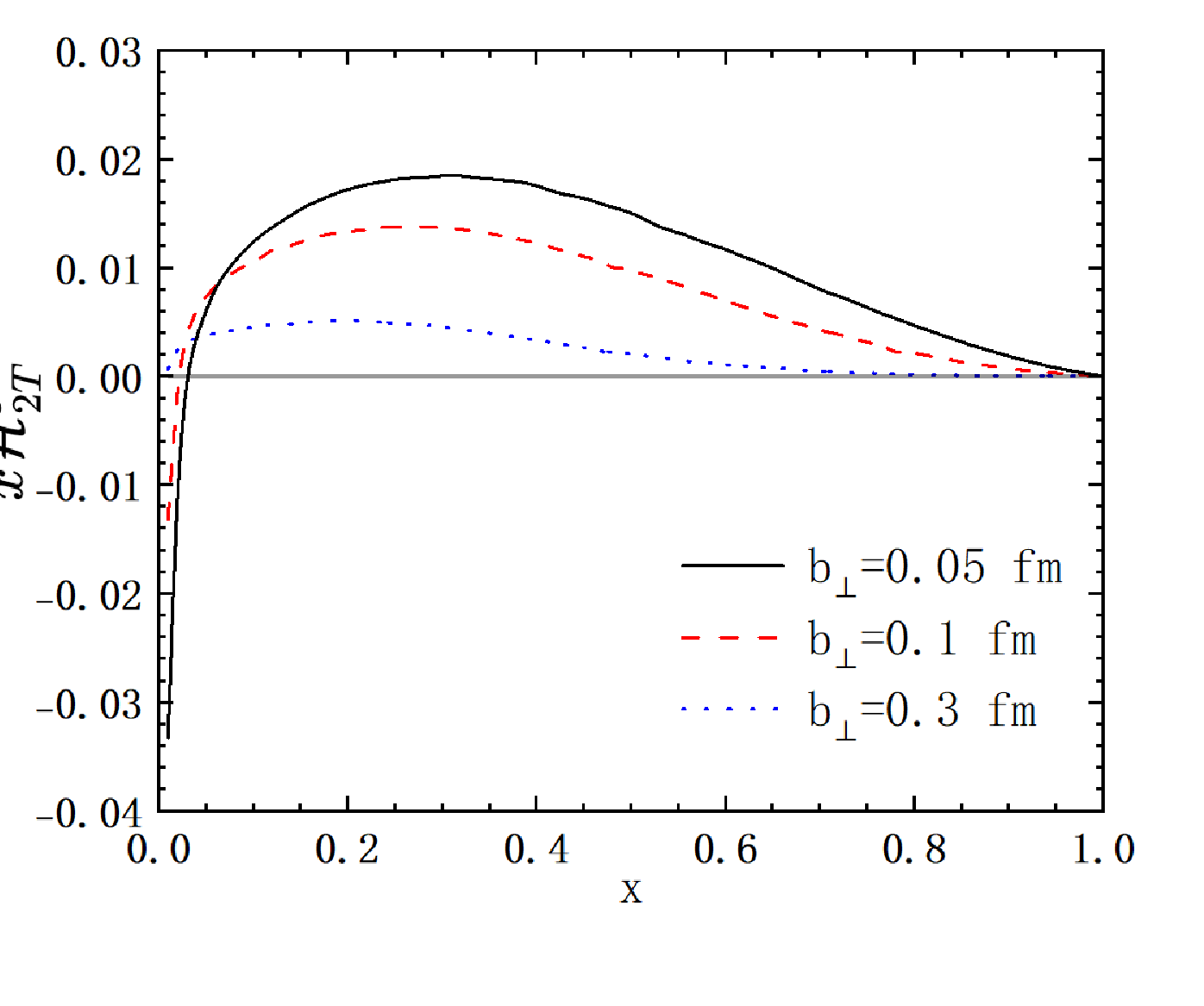}\\
	\includegraphics[width=0.32\columnwidth]{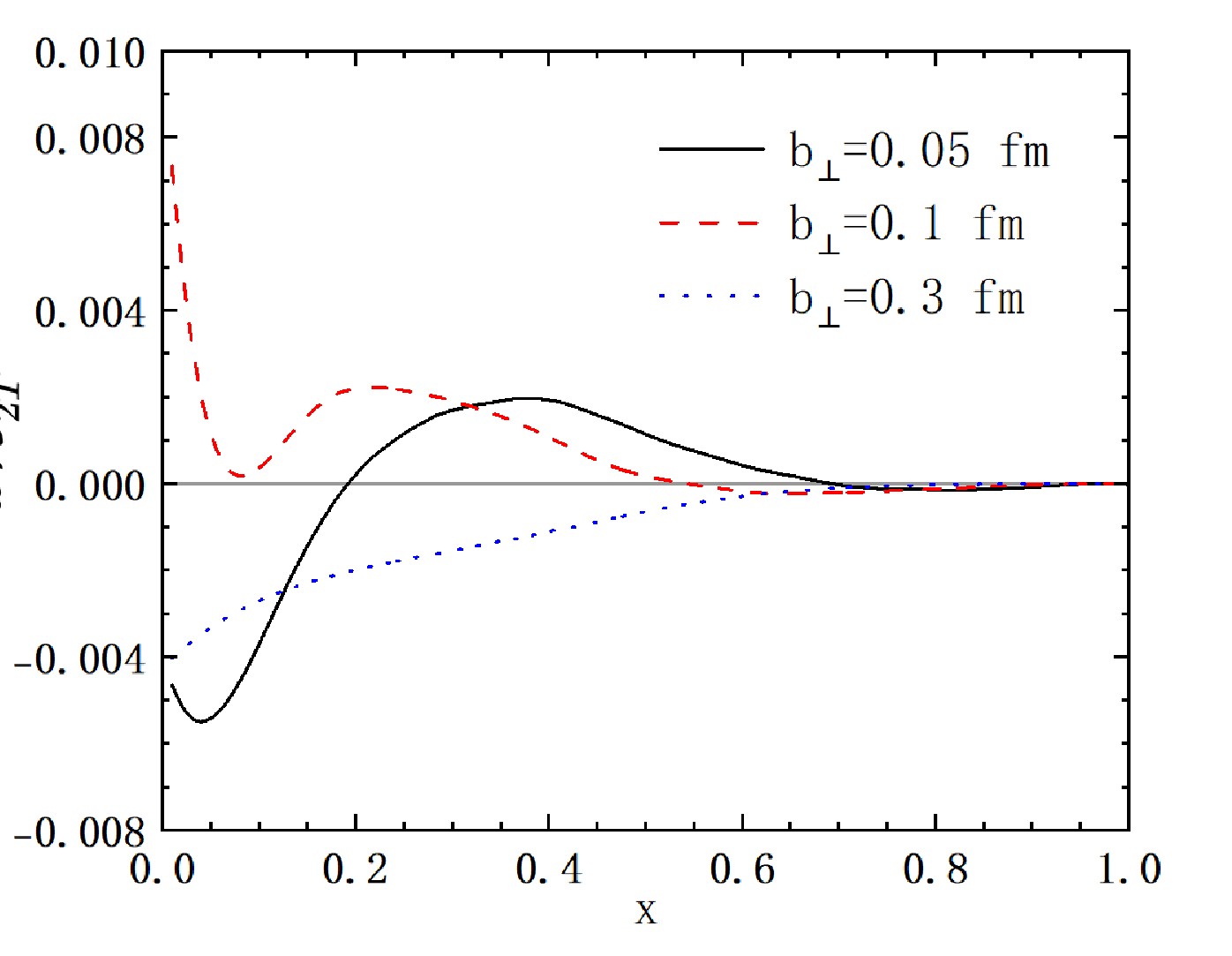}
	\includegraphics[width=0.32\columnwidth]{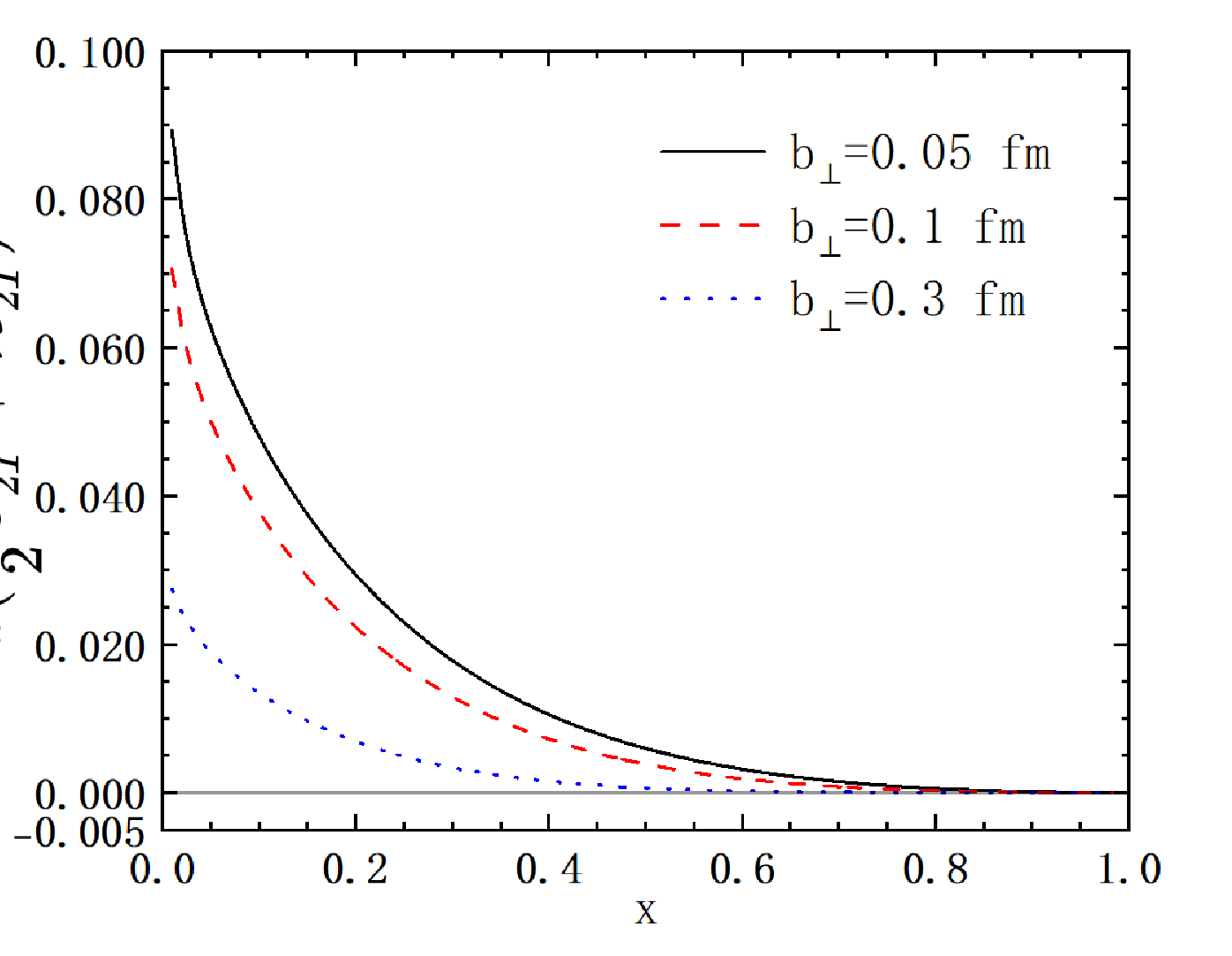}
	\includegraphics[width=0.32\columnwidth]{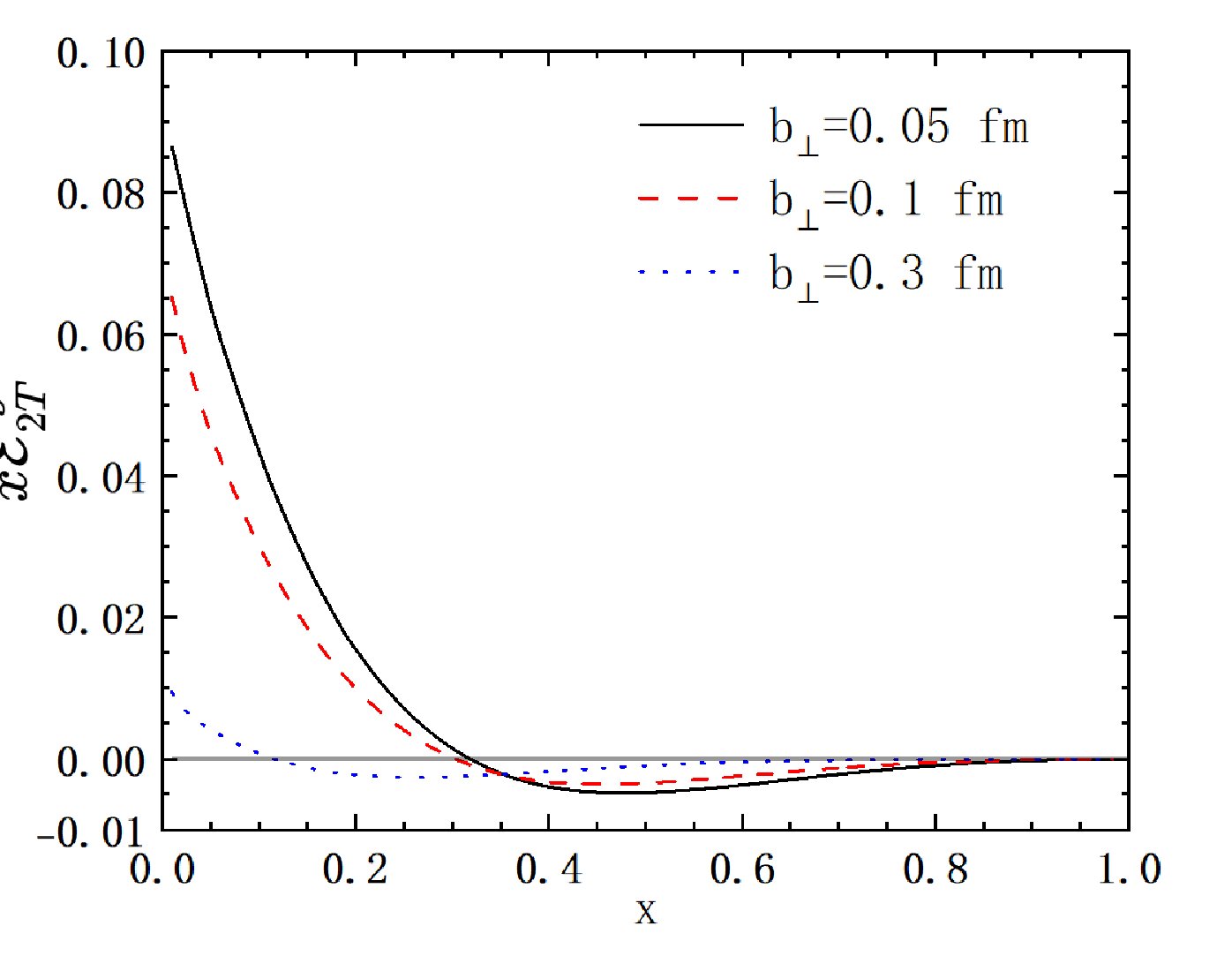}
	\caption{The twist-3 IPDs are plotted with respect to $x$ in the kinematic range $x\in [0,1]$ at fixed $b_\perp=0.05,0.1,$ and 0.3 fm.}
	\label{fig:IPD2}
\end{figure}

Fig.~\ref{fig:IPD2} presents the model predictions for the IPDs as functions of $x$ at different $b_\perp$ values. The distributions display similar overall behavior: they decrease toward zero as $x\to1$ and increase as $x\to0$. Their amplitudes peak in the region $x<0.3$, with $\frac{1}{2}\mathcal{E}_{2T}^g+\tilde{\mathcal{H}}_{2T}^g$ providing the largest contribution. The amplitudes generally decrease with increasing $b_\perp$.

The combinations $\frac{1}{2}\mathcal{E}_{2T}^g+\tilde{\mathcal{H}}_{2T}^g$ and $\frac{1}{2}\bar{\mathcal{E}}_{2T}^{\prime g}+\tilde{\bar{\mathcal{H}}}_{2T}^{\prime g}$ remain positive throughout the displayed $x$ range. The distribution $\tilde{\mathcal{E}}_{2T}^g$ is positive at very small $x$ and changes sign at larger $x$, while $\bar{\mathcal{H}}_{2T}^g$ exhibits an analogous shape with the opposite sign. The distribution $\tilde{\bar{\mathcal{E}}}_{2T}^{\prime g}$ is positive at small $x$ and changes sign at intermediate $x$, with a node near $x=0.35$. 
For $\tilde{\mathcal{E}}_{2T}^g$, $\bar{\mathcal{H}}_{2T}^g$, and $\tilde{\bar{\mathcal{E}}}_{2T}^{\prime g}$, the peak positions shift toward lower $x$ as $b_\perp$ increases.

\subsection{twist-3 PDFs}

PDFs were originally introduced in the context of deep-inelastic scattering and characterize the longitudinal momentum structure of hadrons. 
At leading twist, PDFs admit a probabilistic interpretation as parton densities. 
Here we consider the twist-3 gluon PDFs obtained from the corresponding GPDs in the forward limit, $\Delta\to0$. In particular, two pairs of GPDs and PDFs are related to each other~\cite{Lorce:2013pza}:
\begin{align}
	\hat{\bar{H}}_{2T}^g(x,0,0;M_X)=&\hat{\bar{f}}_T^g(x;M_X)\nonumber\\
	=&\frac{1}{384 M^3 \pi^2 [ M^2 (x-1) x + \Lambda_X^2 + x (M_X - \Lambda_X) (M_X + \Lambda_X) ]^3}(1 - x) [2 M \kappa_1 - (M + M_X) \kappa_2]\nonumber\\
	&\times \{ M^4 (1 - x)^2 [2 (2 - x (2 + x)) \kappa_1 - (2 - 5 x) x \kappa_2] + M^3 M_X (1 - x)^2 [ x (5 + x) \kappa_2-2 (2 + 3 x) \kappa_1 ] \nonumber\\
	&- M^2 (1 - x) [ M_X^2 (2 (2 - x (4 + x)) \kappa_1 + x (1 + 8 x) \kappa_2) - (1 - x) x (2 \kappa_1 - 5 \kappa_2) \Lambda_X^2 ] \nonumber\\
	&- M M_X (1 - x) [ M_X^2 ( x (5 + x) \kappa_2-2 (2 + x) \kappa_1 ) + (1 - x) ( \kappa_2 + x \kappa_2-2 \kappa_1 ) \Lambda_X^2 ]\nonumber\\
	& + \kappa_2 [ 3 M_X^4 x (1 + x) + M_X^2 (1 + (4 - 5 x) x) \Lambda_X^2 + 2 (1 - x)^2 \Lambda_X^4 ] \},
	\label{eq:fTbarg}\\
	\hat{H}_{2T}^{\prime g}(x,0,0;M_X)=&\hat{g}_T^g(x;M_X)\nonumber\\
	=&\frac{1}{384 M^3 \pi^2 x [M^2 (x-1 ) x + \Lambda_X^2 + x (M_X - \Lambda_X) (M_X + \Lambda_X)]^2}(1 - x)^2 [ (M + M_X) \kappa_2-2 M \kappa_1 ]\nonumber\\
	&\times \{3 M_X^2 x \kappa_2 - 2 M^2 ( 1-x ) [(1 - x) \kappa_1 + x \kappa_2] - M M_X (1 - x) (x \kappa_2-2 \kappa_1) + 2 (1 - x) \kappa_2 \Lambda_X^2\}.
	\label{eq:gTg}
\end{align}	
These two twist-3 PDFs have no partonic probability interpretation. 

\begin{figure}
	\centering
	\includegraphics[width=0.4\columnwidth]{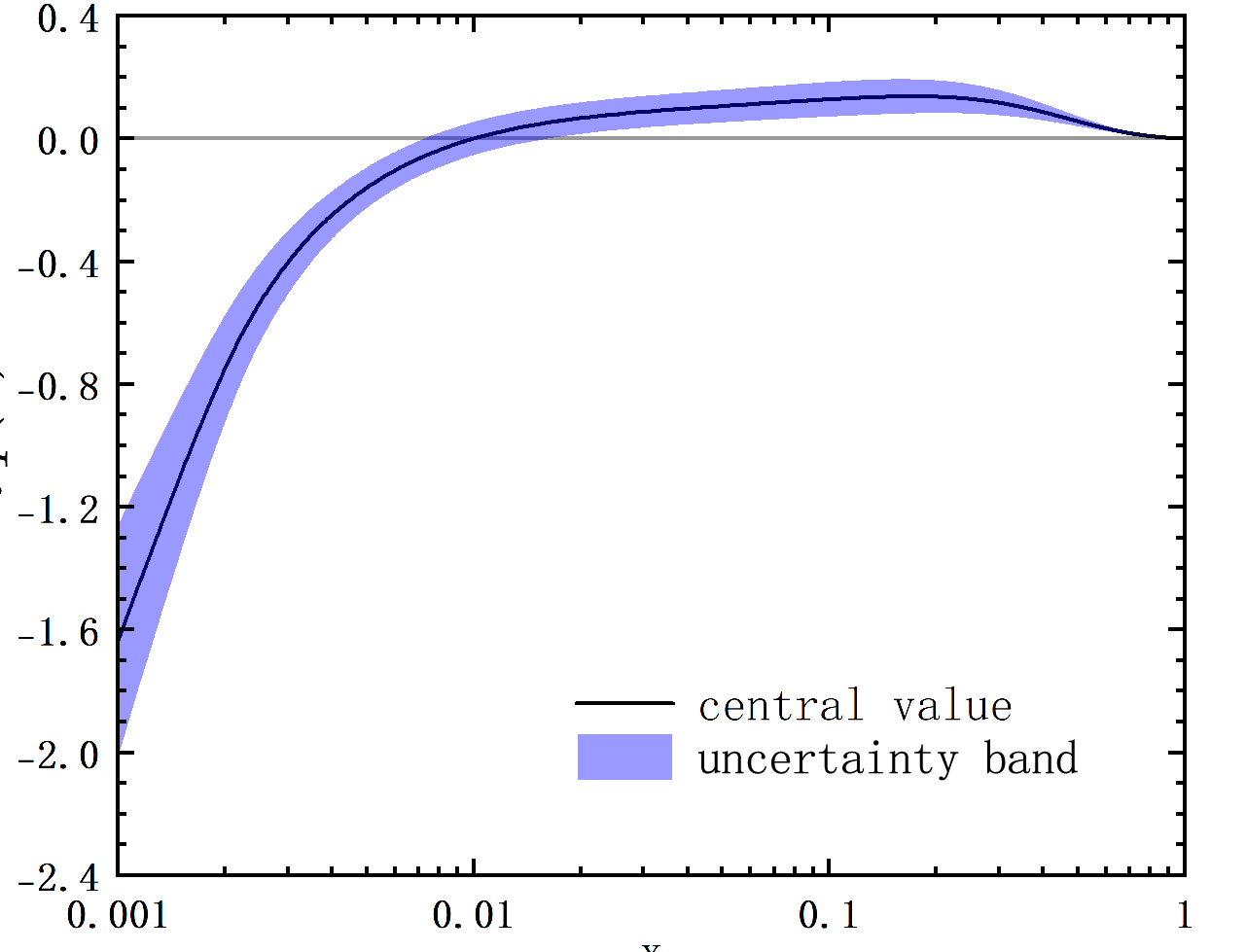}
	\includegraphics[width=0.4\columnwidth]{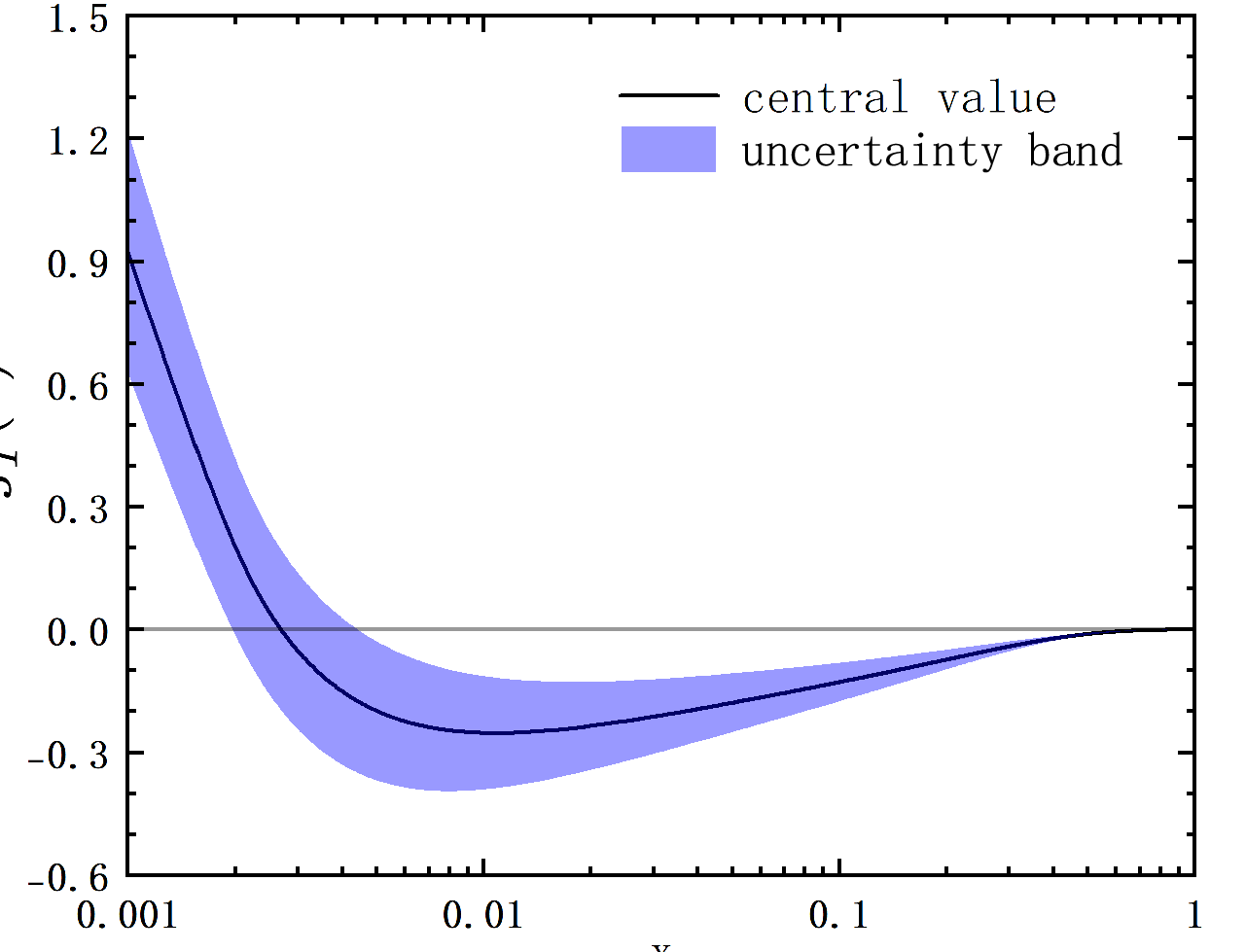}
	\caption{The twist-3 PDFs are plotted with respect to $x$ in the kinematic range $x\in [0.001,1]$.}
	\label{fig:pdf}
\end{figure}

Fig.~\ref{fig:pdf} shows the model predictions for the twist-3 gluon PDFs as functions of $x$. The bands represent the uncertainties induced by the model parameters, while the solid lines correspond to the central parameter values. The two distributions exhibit similar overall shapes and comparable magnitudes, but have opposite signs. 
Their dominant contributions are concentrated in the region $x<0.1$, consistent with the general small-$x$ behavior of gluon distributions.
More specifically, $x\bar{f}_T^g(x)$ is negative at very small $x$ and changes sign at approximately $x=0.01$, reaching a positive peak around $x=0.2$. 
In contrast, $xg_T^g(x)$ is positive at very small $x$ and changes sign at approximately $x=0.003$, followed by a negative peak around $x=0.01$.

\section{Summary}\label{Sec:5}

In this work, we investigated the complete set of twist-3 gluon GPDs of the nucleon at nonzero skewness within a spectator model. 
The spectator mass was allowed to vary continuously according to a spectral function, while the nucleon-gluon-spectator vertex was modeled using two dipolar form factors. 
We derived expressions for the twist-3 gluon GPDs in the two kinematic regions $0\leq x\leq\xi$ and $\xi<x\leq1$ and evaluated them numerically over a broad kinematic range. 
We presented the dependence of the GPDs on $x$ and $\xi$ at fixed $\bm{\Delta}_\perp^2$, as well as their dependence on $x$ and $\bm{\Delta}_\perp^2$ at fixed $\xi$. 
Results at $\xi=0$ were also examined in order to clarify the transverse-momentum-transfer dependence of the distributions.
The model results reveal variations in the magnitude and sign of the different twist-3 gluon GPDs. 
The GPDs in the ERBL and DGLAP regions exhibit a pronounced change across their common boundary at $x=\xi$.
We also investigated the corresponding twist-3 IPDs, which provided information on the transverse spatial dependence of the gluon distributions at zero skewness. 
The results indicate that the IPDs considered here were predominantly concentrated at small impact parameters, with substantial suppression at $b_\perp\gtrsim1$ fm. 
Finally, we obtained two twist-3 gluon PDFs in the forward limit. 
Although these distributions do not admit a direct probability interpretation, they exhibit substantial magnitudes in the small-$x$ region and provided supplementary information on the longitudinal structure of the nucleon.
Future measurements at the EIC and EicC may help constrain gluonic contributions to the multidimensional structure of the nucleon and provide opportunities to test the model predictions presented here.

\section*{Acknowledgements}
This work is partially supported by the National Natural Science Foundation of China under grant numbers 12447136 and 12150013.

\appendix
\section{The numerators and the denominator of twist-3 gluon GPDs at $\xi=0$}\label{Appendix}

The numerators are listed as follows:
\allowdisplaybreaks
\begin{align}
	N_{E_{2T}^g}=&(1-x)\{16[k_\perp^2 + (M_X^2 - M^2(1-x))x]^2 + 8[M_X^2 - M^2(1-x)]x\Delta_\perp^2 + \Delta_\perp^4 + 8k_\perp^2\Delta_\perp^2\cos(2\theta)\}\{(1 - x)\Delta_\perp^2\nonumber\\
	&\times [8(2k_\perp^4 - M_X^4 x(1 - 2x) + k_\perp^2 M_X^2(2 + x^2)) - 2(1 - x)^2(4k_\perp^2 + M_X^2(2 - 3x))\Delta_\perp^2+ (1 - x)^4\Delta_\perp^4]\kappa_2^2 \nonumber\\
	&+ 32M^6 (1 - x)^3 x(2\kappa_1 - \kappa_2)(2\kappa_1 - x\kappa_2) - 2M M_X x\kappa_2[8(1 - x)^2\Delta_\perp^2( (1 - x)\Delta_\perp^2-2k_\perp^2 - 4M_X^2)\kappa_1 \nonumber\\
	& + (16(k_\perp^2 + M_X^2)(k_\perp^2 + M_X^2 x) + 4(1 - x)^2(2k_\perp^2 + M_X^2(3 + x))\Delta_\perp^2 - (1 - x)^3(3 + x)\Delta_\perp^4)\kappa_2]- 32M^5 M_X \nonumber\\
	&\times(1 - x)^2 x[4(1 + x)\kappa_1^2 - 8x\kappa_1\kappa_2 + x(1 + x)\kappa_2^2] - 32M^4(1 - x)[ (x - 1)^3\Delta_\perp^2(\kappa_1 - \kappa_2)( x\kappa_2-2\kappa_1 ) \nonumber\\
	&-k_\perp^2 x^2(2\kappa_1 - \kappa_2)^2+ M_X^2 x((4x-2 )\kappa_1 + (2 - 3x)\kappa_2)( x\kappa_2-2\kappa_1)] + 8M^3 M_X(1 - x)x[4k_\perp^2(2\kappa_1 - \kappa_2)\nonumber\\
	& \times(2\kappa_1- (1 + 2x)\kappa_2)+ 4M_X^2 (4\kappa_1^2 - 8x\kappa_1\kappa_2 + x(2 + x)\kappa_2^2) - (1 - x)\Delta_\perp^2(4(1 + x)\kappa_1^2 + 8(1 - 2x)\kappa_1\kappa_2 \nonumber\\
	&+ (x(3 + 2x)-3 )\kappa_2^2)] + 2M^2[16k_\perp^4 x(2\kappa_1 - \kappa_2)\kappa_2 + 16M_X^4 x(1 - 2x)\kappa_2( x\kappa_2-2\kappa_1 ) - (1 - x)^3\Delta_\perp^4 \nonumber\\
	& \times(4(2 - x)\kappa_1^2+ 2( x + x^2-4 )\kappa_1\kappa_2+ ( (8 - 5x)x-2 )\kappa_2^2) - 4M_X^2(1 - x)\Delta_\perp^2(4(2 + x( x + x^2-3 ))\kappa_1\kappa_2\nonumber\\
	&-4(2 - x(3 - 2x))\kappa_1^2+ x( 2(5 - 3x)x-5 )\kappa_2^2) - 4k_\perp^2(4M_X^2 x^2\kappa_2( \kappa_2 + x\kappa_2-4\kappa_1 ) - (1 - x)\Delta_\perp^2\nonumber\\
	&\times(4(2 - (2 - x)x)\kappa_1^2 - 4(2 - (2 - x)x)\kappa_1\kappa_2- (2 - 3 (2 - x)x)\kappa_2^2))]-8 k_\perp^2 (1 - x)^2 x (4 M^2 + \Delta_\perp^2) \nonumber\\
	&\times[2 M \kappa_1 - (M + M_X) \kappa_2] [M (2 \kappa_1 + \kappa_2)-M_X \kappa_2 ] \cos(2\theta)\}/(2 M^2 \pi^3 x \Delta_\perp^2),\\
	N_{\tilde{H}_{2T}^g}=&(-1 + x) \{16 [k_\perp^2 + (M_X^2 + M^2 (-1 + x)) x]^2 + 8 [M_X^2 + M^2 (-1 + x)] x \Delta_\perp^2 + \Delta_\perp^4 + 8 k_\perp^2 \Delta_\perp^2 \cos(2\theta)\}\nonumber\\
	&\times\{M_X x [-16 (k_\perp^2 + M_X^2) (k_\perp^2 + M_X^2 x) - 4 M_X^2 (3 - x) (1 - x)^2 \Delta_\perp^2 + (1 - x)^4 \Delta_\perp^4] \kappa_2^2 - 16 M^5 (1 - x)^3 x\nonumber\\
	&\times (2 \kappa_1 - \kappa_2) ( x \kappa_2-2 \kappa_1 ) - 16 M^4 M_X (1 - x)^2 x [4 (1 + x) \kappa_1^2 - 8 x \kappa_1 \kappa_2 + x (1 + x) \kappa_2^2] + 4 M^2 M_X (1 - x) x \nonumber\\
	&\times[4 k_\perp^2 (2 \kappa_1 - \kappa_2) (2 \kappa_1 - (1 + 2 x) \kappa_2) - (-1 + x) \Delta_\perp^2 (2 \kappa_1 - \kappa_2) (2 (-3 + x) \kappa_1 + (-3 + 5 x) \kappa_2) + 4 M_X^2\nonumber\\
	&\times (4 \kappa_1^2 - 8 x \kappa_1 \kappa_2 + x (2 + x) \kappa_2^2)] + 4 M^3 (1 - x) [4 k_\perp^2 x^2 ( \kappa_2-2 \kappa_1 )^2 - 4 M_X^2 x ((4 x-2 ) \kappa_1 + (2 - 3 x) \kappa_2)\nonumber\\
	&\times ( x \kappa_2-2 \kappa_1 ) + (1 - x)^2 \Delta_\perp^2 (4 x \kappa_1^2 + 4 (2 - 3 x) \kappa_1 \kappa_2 + x ( 4 x-3) \kappa_2^2)] - M \kappa_2 [16 k_\perp^4 x ( \kappa_2-2 \kappa_1) + 16 M_X^4 x\nonumber\\
	&\times (1 - 2 x) (2 \kappa_1 - x \kappa_2) + (1 - x)^4 \Delta_\perp^4 (2 (-4 + x) \kappa_1 + 3 x \kappa_2) - 4 M_X^2 (1 - x)^3 \Delta_\perp^2 (4 (-2 + x) \kappa_1 + 3 x \kappa_2)\nonumber\\
	& + 16 k_\perp^2 ((1 - x)^2 \Delta_\perp^2 (2 \kappa_1 - x \kappa_2) + M_X^2 x^2 ( \kappa_2 + x \kappa_2-4 \kappa_1 ))]-16 k_\perp^2 M (1 - x)^2 x [2 M \kappa_1 - (M + M_X) \kappa_2]\nonumber\\
	&\times [-M_X \kappa_2 + M (2 \kappa_1 + \kappa_2)] \cos(2\theta)\}/(2 M \pi^3 x \Delta_\perp^2),\\
	N_{\tilde{E}_{2T}^g}=&(-1 + x)^2 \{16 [k_\perp^2 + (M_X^2 + M^2 (-1 + x)) x]^2 + 8 [M_X^2 + M^2 (-1 + x)] x \Delta_\perp^2 + \Delta_\perp^4 + 8 k_\perp^2 \Delta_\perp^2 \cos(2\theta)\}\nonumber\\
	&\times\{8 k_\perp^4 \kappa_2^2 + 2 k_\perp^2 [8 M^2 (-2 + x) x \kappa_1^2 + 8 M x (2 M + M_X - (M + M_X) x) \kappa_1 \kappa_2 - (4 M M_X (1 - x) x + 2 M^2\nonumber\\
	&\times (2 + (-2 + x) x) - 2 M_X^2 (2 + x^2) + (-1 + x)^2 \Delta_\perp^2) \kappa_2^2] + x [-4 M_X^2 + 4 M^2 (-1 + x)^2 + (-1 + x)^2 \Delta_\perp^2]\nonumber\\
	&\times [-M_X^2 \kappa_2^2 + M^2 (\kappa_2-2 \kappa_1 )^2]+2 k_\perp^2 [8 M^2 (-1 + x) x \kappa_1^2 + 8 M (M + M_X) (1 - x) x \kappa_1 \kappa_2 - (-4 k_\perp^2 - 4 M_X^2 \nonumber\\
	&+ 2 (-1 + x) (M^2 (-2 + x) - 2 M M_X x - M_X^2 x) + (-1 + x)^2 \Delta_\perp^2) \kappa_2^2] \cos(2\theta)\}/(M^2 \pi^3 x),\\
	N_{\bar{H}_{2T}^g}=& [2 M \kappa_1 - (M + M_X) \kappa_2] \{16 [k_\perp^2 + (M_X^2 - M^2 (1 - x)) x]^2 + 8 [M_X^2 - M^2 (1 - x)] x \Delta_\perp^2 + \Delta_\perp^4 + 8 k_\perp^2 \Delta_\perp^2 \nonumber\\
	&\times\cos(2\theta)\}( x-1 )\{[16(k_\perp^2 + M_X^2)(k_\perp^2 + M_X^2 x) - 4M_X^2(1 - x)^3\Delta_\perp^2 + (1 - x)^4\Delta_\perp^4]\kappa_2 - (\kappa_1 + x\kappa_1 - x\kappa_2)\nonumber\\
	&\times 32M^3 M_X(1 - x)^2 - 16M^4(1 - x)^3( x\kappa_2-2\kappa_1 ) - 8M M_X(1 - x)[(\kappa_1 + x\kappa_1 - x\kappa_2)-4M_X^2(\kappa_1 - x\kappa_2) \nonumber\\
	& \times(1 - x)\Delta_\perp^2 + 2k_\perp^2( \kappa_2 + x\kappa_2-2\kappa_1)] - 4M^2(1 - x)[4k_\perp^2 x( \kappa_2-2\kappa_1 ) - (1 - x)^2\Delta_\perp^2(2\kappa_1 + \kappa_2 - 2x\kappa_2)\nonumber\\
	& + 4M_X^2((2 - 4x)\kappa_1 + x^2\kappa_2)]+8 k_\perp^2 (1 - x)^2 [ \Delta_\perp^2 \kappa_2 + 2 M^2 (2 \kappa_1 + \kappa_2)-2 M M_X \kappa_2 ] \cos(2\theta)\}/(4 M^3 \pi^3),\\
	N_{\bar{E}_{2T}^g}=&(1 - x) \{16 [k_\perp^2 + (M_X^2 + M^2 (-1 + x)) x]^2 + 8 [M_X^2 + M^2 (-1 + x)] x \Delta_\perp^2 + \Delta_\perp^4 + 8 k_\perp^2 \Delta_\perp^2 \cos(2\theta)\}\nonumber\\
	&\times\{-M x [2 M \kappa_1 - (M + M_X) \kappa_2] [ - (16 (k_\perp^2 + M_X^2) (k_\perp^2 + M_X^2 x) + 4 M_X^2 (-1 + x)^3 \Delta_\perp^2 + (1 - x)^4 \Delta_\perp^4) \kappa_2 \nonumber\\
	&+ 32 M^3 M_X (1 - x)^2 (\kappa_1 + x \kappa_1 - x \kappa_2) + 16 M^4 (1 - x)^3 (  x \kappa_2-2 \kappa_1) + 8 M M_X (1 - x) ( -4 M_X^2 (\kappa_1 - x \kappa_2)\nonumber\\
	& + (1 - x) \Delta_\perp^2 (\kappa_1 + x \kappa_1 - x \kappa_2) + 2 k_\perp^2 (-2 \kappa_1 + \kappa_2 + x \kappa_2) ) - 4 M^2 (1 - x) ( k_\perp^2 x (8 \kappa_1 - 4 \kappa_2) + (1 - x)^2 \Delta_\perp^2\nonumber\\
	&\times (2 \kappa_1 + \kappa_2 - 2 x \kappa_2) - 4 M_X^2 ((2 - 4 x) \kappa_1 + x^2 \kappa_2) ) ] + 8 k_\perp^2 M (1 - x)^2 x [2 M \kappa_1 - (M + M_X) \kappa_2] [ \Delta_\perp^2 \kappa_2 \nonumber\\
	& -2 M M_X \kappa_2+ 2 M^2 (2 \kappa_1 + \kappa_2) ] \cos(2\theta)+k_\perp \Delta_\perp [16 k_\perp^4 \kappa_2^2 + (8 M_X^4 x (1 + x) - 2 M_X^2 (2 - x) (1 - x)^3 \Delta_\perp^2 \nonumber\\
	&+ (1 - x)^4 \Delta_\perp^4) \kappa_2^2 + 32 M^3 M_X (1 - x)^2 x \kappa_2 ((-3 + x) \kappa_1 + \kappa_2) - 8 M^4 (-1 + x)^3 (4 (-2 + x) \kappa_1^2 + 8 \kappa_1 \kappa_2\nonumber\\
	& - 3 x \kappa_2^2) - 8 M M_X (1 - x) x \kappa_2 (4 M_X^2 (\kappa_2-\kappa_1 ) - (1 - x) \Delta_\perp^2 ((-3 + x) \kappa_1 + \kappa_2)) + 8 k_\perp^2 (4 M M_X (1 - x) x\nonumber\\
	&\times (\kappa_1 - \kappa_2) \kappa_2 + (M_X^2 (2 + x + x^2) - (-1 + x)^2 \Delta_\perp^2) \kappa_2^2 + M^2 (-1 + x) (4 (-2 + x) \kappa_1^2 - 8 (-1 + x) \kappa_1 \kappa_2 \nonumber\\
	&+ (2 - x) \kappa_2^2)) + 2 M^2 (1 - x) ((1 - x)^2 \Delta_\perp^2 (4 (-2 + x) \kappa_1^2 + 8 \kappa_1 \kappa_2 + (2 - 5 x) \kappa_2^2) + 4 M_X^2 (4 (2 - x) \kappa_1^2 \nonumber\\
	&+ 8 (-1 + x) \kappa_1 \kappa_2 + x (2 + (-4 + x) x) \kappa_2^2))] \sin(\theta)\}/(M^2 \pi^3 x \Delta_\perp^2),\\
	N_{\tilde{\bar{H}}_{2T}^g}=&( x-1) \{16 [k_\perp^2 + (M_X^2 - M^2 (1 - x)) x]^2 + 8 [M_X^2 - M^2 (1 - x)] x \Delta_\perp^2 + \Delta_\perp^4 + 8 k_\perp^2 \Delta_\perp^2 \cos(2\theta)\}\{-x [2 M \kappa_1\nonumber\\
	& - (M + M_X) \kappa_2] [ (4 M_X^2 (1 - x)^3 \Delta_\perp^2 -16 (k_\perp^2 + M_X^2) (k_\perp^2 + M_X^2 x) - (1 - x)^4 \Delta_\perp^4) \kappa_2+ 32 M^3 M_X (1 - x)^2\nonumber\\
	& \times (\kappa_1 + x \kappa_1 - x \kappa_2) + 16 M^4 (1 - x)^3 (x \kappa_2-2 \kappa_1 ) + 8 M M_X (1 - x) ( 4 M_X^2 (x \kappa_2-\kappa_1)+(\kappa_1 + x \kappa_1 - x \kappa_2) \nonumber\\
	& \times(1 - x) \Delta_\perp^2  + 2 k_\perp^2 ( \kappa_2 + x \kappa_2-2 \kappa_1) ) - 4 M^2 (1 - x) ( k_\perp^2 x (8 \kappa_1 - 4 \kappa_2) + (1 - x)^2 \Delta_\perp^2(2 \kappa_1 + \kappa_2 - 2 x \kappa_2)\nonumber\\
	& - 4 M_X^2 ((2 - 4 x) \kappa_1 + x^2 \kappa_2) ) ]+8 k_\perp^2 (1 - x)^2 x [2 M \kappa_1 - (M + M_X) \kappa_2] [\Delta_\perp^2 \kappa_2 -2 M M_X \kappa_2+ (2 \kappa_1 + \kappa_2) \nonumber\\
	& \times2 M^2 ]\cos(2\theta)-4 k_\perp (1 - x) \Delta_\perp [ 16 M^2 (M_X - M (1 - x)) (1 - x) x \kappa_1^2+ 2 M( 4 k_\perp^2 (2 - x) + 4 (M ( x-1 ) \nonumber\\
	& +M_X ) (2 (M + M_X) - 3 (M + M_X) x + (M + 2 M_X) x^2)- (2 - x) (1 - x)^2 \Delta_\perp^2 )\kappa_1 \kappa_2  + (M - M_X)x \nonumber\\
	& \times( 4 (M_X - M (1 - x))-4 k_\perp^2  (M_X (x-2 ) + 2 M (x-1))+ (1 - x)^2 \Delta_\perp^2 ) \kappa_2^2 ] \sin(\theta)\}/(2 M \pi^3 x \Delta_\perp^2),\\
	N_{H_{2T}^{\prime g}}=&(1 - x)^2 [2 M \kappa_1 - (M + M_X) \kappa_2] \{8 M [M_X - M (1 - x)] (x-1 ) (2 k_\perp^2 - \Delta_\perp^2) \kappa_1 + [8 k_\perp^2 ( M_X (M (1-x )\nonumber\\
	& -M_X ) x-k_\perp^2)- 2 (1 - x) (k_\perp^2 (x-1 ) + 2 M (M_X - M (1 - x)) x) \Delta_\perp^2 - (1 - x)^3 \Delta_\perp^4] \kappa_2+ 2 k_\perp^2 \nonumber\\
	& \times [8 M (1 -x)(M_X - M(1 - x)) \kappa_1 + (4 k_\perp^2 + 4 M_X (M_X - M (1 - x)) x - (3 - x) (1 - x) \Delta_\perp^2) \kappa_2] \cos(2\theta)\} \nonumber\\
	& \times\{4 k_\perp^2+ 4 [M_X^2 - M^2 (1 - x)] x + \Delta_\perp^2 - 4 k_\perp \Delta_\perp \sin(\theta)\} \{4 k_\perp^2 + 4 [M_X^2 - M^2 (1 - x)] x + \Delta_\perp^2 \nonumber\\
	&+ 4 k_\perp \Delta_\perp \sin(\theta)\}/(2 M^3 \pi^3 x),\\
	N_{E_{2T}^{\prime g}}=&-2 (1 - x)^2 [2 M \kappa_1 - (M + M_X) \kappa_2] \{16 [k_\perp^2 + (M_X^2 - M^2 (1 - x)) x]^2 + 8 [M_X^2 - M^2 (1 - x)] x \Delta_\perp^2 + \Delta_\perp^4\nonumber\\
	& + 8 k_\perp^2 \Delta_\perp^2 \cos(2\theta)\} \{M [8 M (M_X - M (1 - x)) (1-x ) (2 k_\perp^2 - \Delta_\perp^2) \kappa_1 + (8 k_\perp^2 (k_\perp^2 + M_X (M_X - M (1 - x)) x)\nonumber\\
	& + 2 (1 - x) (k_\perp^2 ( x-1) + 2 M (M_X - M (1 - x)) x) \Delta_\perp^2 + (1 - x)^3 \Delta_\perp^4) \kappa_2] - 2 k_\perp^2 M [8 M ( M (1 - x)-M_X )\nonumber\\
	&\times ( x-1 ) \kappa_1 + (4 k_\perp^2 + 4 M_X (M_X - M (1 - x)) x - (3 - x) (1 - x) \Delta_\perp^2) \kappa_2] \cos(2\theta) + k_\perp \Delta_\perp[x (4 M^2 (1 - x)^2 \nonumber\\
	& -4 M_X^2 + (-1 + x)^2 \Delta_\perp^2) (-2 M \kappa_1 + (M + M_X) \kappa_2) + k_\perp^2 (8 M x \kappa_1 - 4 (M + M_X x) \kappa_2)] \sin(\theta)\nonumber\\
	& + 4 k_\perp^3 M (-1 + x) \Delta_\perp \kappa_2 \sin(3\theta)\}/(M^2 \pi^3 x \Delta_\perp^2),\\
	N_{\tilde{H}_{2T}^{\prime g}}=&(-1 + x)^2 [2 M \kappa_1 - (M + M_X) \kappa_2] \{[4 k_\perp^2 + 4 (M_X^2 + M^2 (-1 + x)) x + \Delta_\perp^2]^2 - 16 k_\perp^2 \Delta_\perp^2 \sin^2(\theta)\}\nonumber\\
	&\times \{ -8 M [M_X + M (-1 + x)] (-1 + x) (2 k_\perp^2 - \Delta_\perp^2) \kappa_1 + [8 k_\perp^2 (k_\perp^2 + M_X (M_X - M (1 - x)) x) + 2 (1 - x)\nonumber\\
	&\times (k_\perp^2 (-1 + x) + 2 M (M_X + M (-1 + x)) x) \Delta_\perp^2 + (1 - x)^3 \Delta_\perp^4] \kappa_2 - 8 k_\perp (1 0- x) \Delta_\perp [k_\perp^2 \kappa_2 + (M_X + M\nonumber\\
	&\times ( x-1 )) x ((M + M_X) \kappa_2-2 M \kappa_1 )] \sin(\theta) + 2 k_\perp^2 \cos(2\theta) [8 M (M_X - M (1 - x)) (x-1 ) \kappa_1+ ((3 - x)\Delta_\perp^2\nonumber\\
	& \times (1 - x) -4 (k_\perp^2 + M_X (M_X - M (1 - x)) x) ) \kappa_2 - 4 k_\perp (1 - x) \Delta_\perp \kappa_2 \sin(\theta) ] \}/(M \pi^3 x \Delta_\perp^2),\\
	N_{\bar{E}_{2T}^{\prime g}}=&-(1 - x)^2 \{16 [k_\perp^2 + (M_X^2 - M^2 (1 - x)) x]^2 + 8 [M_X^2 - M^2 (1 - x)] x \Delta_\perp^2 + \Delta_\perp^4 + 8 k_\perp^2 \Delta_\perp^2 \cos(2\theta)\}\nonumber\\
	&\times \{x [-4 M_X^2 + 4 M^2 (-1 + x)^2 + (-1 + x)^2 \Delta_\perp^2] [-2 M \Delta_\perp \kappa_1 + (M + M_X) \Delta_\perp \kappa_2]^2 - 8 k_\perp^4 \kappa_2 [-4 M^2 \kappa_1 \nonumber\\
	&+ (2 M (M + M_X) + \Delta_\perp^2) \kappa_2] + 2 k_\perp^2 [-8 M^2 (4 M (M_X + M (-1 + x)) (-1 + x) + (2 + (-2 + x) x) \Delta_\perp^2) \kappa_1^2 \nonumber\\
	&+ 4 M (4 M (M_X - M (1 - x)) (M (x-1 ) + M_X ( 2 x-1)) + (M (3 - (2 - x) x) + 2 M_X (1 - (1 - x) x)) \Delta_\perp^2) \nonumber\\
	&\times\kappa_1 \kappa_2 + (-8 M M_X (M + M_X) (M_X + M (-1 + x)) x - 2 (M^2 (-1 - 2 (-2 + x) x) + M M_X (1 + x^2) \nonumber\\
	&+ M_X^2 (2 + x^2)) \Delta_\perp^2 + (1 - x)^2 \Delta_\perp^4) \kappa_2^2] + 2 k_\perp^2 [8 M^2 (1 - x) (4 M (M_X - M (1 - x)) - (2 - x) \Delta_\perp^2) \kappa_1^2+ 4 M  \nonumber\\	
	&\times(4 M (k_\perp^2 + (M_X - M (1 - x)) (M ( x-1 ) - M_X (1 - 2 x))) + (M ( x-3) - 2 M_X (1 - x)) ( x-1) \Delta_\perp^2)\kappa_1 \kappa_2  \nonumber\\
	&- (8 M (M + M_X) (k_\perp^2 + M_X (M_X - M (1 - x)) x) + 2 (2 k_\perp^2 + M M_X (1 - x)^2 + M^2 ((3 - 2 x) x-1 )+ M_X^2 \nonumber\\
	&\times (2 - (1 - x) x)) \Delta_\perp^2 - (1 - x)^2 \Delta_\perp^4) \kappa_2^2] \cos(2\theta) + 4 k_\perp M \Delta_\perp \kappa_2 [ (M + M_X) \kappa_2-2 M \kappa_1] [k_\perp^2 (8 x-4 )+ 4  \nonumber\\
	&\times (M_X - M (1 - x)) x (M + M_X - M x) - (1 - x)^2 x \Delta_\perp^2- 4 k_\perp^2 (1 - x) \cos(2\theta)] \sin(\theta)\}/(2 M^2 \pi^3 x \Delta_\perp^2),\\
	N_{\tilde{\bar{H}}_{2T}^{\prime g}}=&2 (1 - x)^2 \{ -4 k_\perp^4 \kappa_2 [(M + M_X) \kappa_2-2 M \kappa_1 ] + 2 [M_X - M (1 - x)] (-1 + x) x \Delta_\perp^2 [ (M + M_X) \kappa_2-2 M \kappa_1]^2\nonumber\\
	& + k_\perp^2 [ 16 M^2 (M_X - M (1 - x)) (1 - x) \kappa_1^2 + 2 M (4 (M_X + M (-1 + x)) (M (-1 + x) + M_X (-1 + 2 x)) \nonumber\\
	&+ (5 - x) (1 - x) \Delta_\perp^2) \kappa_1 \kappa_2 - (4 M_X (M + M_X) (M_X - M (1 - x)) x + (1 - x) (M + M_X + 3 M x - M_X x)\nonumber\\
	&\times \Delta_\perp^2) \kappa_2^2 ] + k_\perp^2 [ 16 M^2 (M_X - M (1 - x)) (1 - x) \kappa_1^2 + 2 M (4 k_\perp^2 + 4 (M_X + M (-1 + x)) (M (-1 + x) \nonumber\\
	&+ M_X (-1 + 2 x)) + (5 - x) (1 - x) \Delta_\perp^2) \kappa_1 \kappa_2 - (4 (M + M_X) (k_\perp^2 + M_X (M_X - M (1 - x)) x) + (1 - x)\nonumber\\
	&\times (M + M_X + 3 M x - M_X x) \Delta_\perp^2) \kappa_2^2 ] \cos(2\theta) + k_\perp \Delta_\perp \kappa_2 [-2 M \kappa_1 + (M + M_X) \kappa_2] [ k_\perp^2 (-4 + 8 x) \nonumber\\
	& + 4 (M_X- M (1 - x)) x (M + M_X - M x) - (1 - x)^2 x \Delta_\perp^2 - 4 k_\perp^2 (1 - x) \cos(2\theta) ] \sin(\theta) \}  \nonumber\\
	&\times\{ [4 k_\perp^2 + 4 (M_X^2- M^2 (1 - x)) x + \Delta_\perp^2]^2 - 16 k_\perp^2 \Delta_\perp^2 \sin^2(\theta) \}/(2 M \pi^3 x \Delta_\perp^2),\\
	N_{\tilde{\bar{E}}_{2T}^{\prime g}}=&-\{(1-x)^2 [(2 M \kappa_1 - (M+M_X) \kappa_2) (2 M (4 (M_X - M (1-x))^2 x - (2-x) (1-x)^2 \Delta_\perp^2) \kappa_1  \nonumber\\
	&- (4 (M+M_X)(M_X-M(1-x))^2 x + (1-x)^2 (M(x-2)+M_X(2+x)) \Delta_\perp^2) \kappa_2) + 4 k_\perp^2 \nonumber\\
	& \times((M_X^2 ((4-x)x-2)+ (1-x)^2 \Delta_\perp^2) \kappa_2^2 + 2 M M_X \kappa_2 (2(1-(3-x)x) \kappa_1 - x(1-2x) \kappa_2) -  (2\kappa_1 - \kappa_2) \nonumber\\
	& \times M^2 (2(x-2)x \kappa_1+ (2-x(6-5x)) \kappa_2)) + 4 k_\perp^2 (1-x) (4 M^2 (-2+x) \kappa_1^2 + 4 M (M+M_X - M_X x) \kappa_1 \kappa_2  \nonumber\\
	& + (\Delta_\perp^2 - x (M^2 - M_X^2+ \Delta_\perp^2))\kappa_2^2) \cos(2\theta)] [4 k_\perp^2 + 4 (M_X^2 + M^2 (-1+x)) x + \Delta_\perp^2 - 4 k_\perp \Delta_\perp \sin(\theta)]  \nonumber\\
	& \times[4 k_\perp^2 + 4 (M_X^2 + M^2 (-1+x)) x+ \Delta_\perp^2+ 4 k_\perp \Delta_\perp \sin(\theta)]\}/(2 M^2 \pi^3 x),
\end{align}
where $\bm{\Delta}_\perp^2 = -(1-\xi^2)t - 4\xi^2 M^2$, and $\theta$ is the angle between $\bm{\Delta}_\perp$ and $\bm{k}_\perp$.

The denominator is given by
\begin{align}
	D_\text{SM}=&\{ [4 k_\perp^2 + 4 M_X^2 x - 4 M^2 (1 - x) x + \Delta_\perp^2 + 4 (1 - x) \Lambda_X^2]^2 - 16 k_\perp^2 \Delta_\perp^2 \sin^2(\theta) \}^2 \nonumber\\
	&\times\{ [4 k_\perp^2 + 4 (M_X^2 - M^2 (1 - x)) x + (1 - x)^2 \Delta_\perp^2]^2 - 16 k_\perp^2 (1 - x)^2 \Delta_\perp^2 \sin^2(\theta) \}.
\end{align}

\end{document}